%% file: main.tex
\documentclass[sigconf]{acmart}
\usepackage[utf8]{inputenc}

\AtBeginDocument{%
  \providecommand\BibTeX{{%
    \normalfont B\kern-0.5em{\scshape i\kern-0.25em b}\kern-0.8em\TeX}}}

\copyrightyear{2021}
\acmYear{2021}
\setcopyright{rightsretained}
\acmConference[FAccT '21]{ACM Conference on Fairness, Accountability, and Transparency}{March 1--10, 2021}{Virtual Event, Canada}
\acmBooktitle{ACM Conference on Fairness, Accountability, and Transparency (FAccT '21), March 1--10, 2021, Virtual Event, Canada}\acmDOI{10.1145/3442188.3445896}
\acmISBN{978-1-4503-8309-7/21/03}

\usepackage{setspace}
\usepackage{tabularx}
\usepackage{enumitem}
\usepackage{footnote}
\makesavenoteenv{tabular}

\begin{document}
\newcommand{\etal}{\textit{et al. \@}}
\newcommand{\eg}{\textit{e.g., \@}}
\newcommand{\ie}{\textit{i.e., \@}}
\newcommand{\toadd}[1]{{\color{red}VP: #1}}
\newcommand{\vinod}[1]{{\color{red}VP: #1}}
\newcommand{\ben}[1]{{\color{blue}BH: #1}}
\newcommand{\nithya}[1]{{\color{red}NS: #1}}

\newcommand{\vpedit}[1]{{\color{red}#1}}

\newcommand{\myfigureshrinker}{\vspace{-1cm}}

\newcommand{\tabitem}{\hspace{2pt}\textbullet~~}

\setlength{\abovecaptionskip}{1ex}

\title{Re-imagining Algorithmic Fairness in India and Beyond}
\author{Nithya Sambasivan, Erin Arnesen, Ben Hutchinson, Tulsee Doshi, Vinodkumar Prabhakaran}
\email{(nithyasamba, erinarnesen, benhutch, tulsee, vinodkpg)@google.com}
\affiliation{%
  \institution{Google Research}
  \city{Mountain View}
  \state{CA}
  \postcode{94043}
}


\begin{abstract}
Conventional algorithmic fairness is West-centric, as seen in its sub-groups, values, and methods. In this paper, we de-center algorithmic fairness and analyse AI power in India. Based on 36 qualitative interviews and a discourse analysis of algorithmic deployments in India, we find that several assumptions of algorithmic fairness are challenged. We find that in India, data is not always reliable due to socio-economic factors, ML makers appear to follow double standards, and AI evokes unquestioning aspiration. We contend that localising model fairness alone can be window dressing in India, where the distance between models and oppressed communities is large. Instead, we re-imagine algorithmic fairness in India and provide a roadmap to re-contextualise data and models, empower oppressed communities, and enable Fair-ML ecosystems.
\end{abstract}

\vspace{-2mm}

\copyrightyear{2021}
\acmYear{2021}
\acmConference[FAccT '21]{ACM Conference on Fairness, Accountability, and Transparency}{March 1--10, 2021}{Virtual Event, Canada}
\acmBooktitle{ACM Conference on Fairness, Accountability, and Transparency (FAccT '21), March 1--10, 2021, Virtual Event, Canada}\acmDOI{10.1145/3442188.3445896}
\acmISBN{978-1-4503-8309-7/21/03}

\maketitle
\input{introduction}
\input{background}

\input{method}
\input{findings}

\input{discussion}

\input{conclusion}

\newpage

\begin{spacing}{0.9}
\bibliographystyle{ACM-Reference-Format}
\balance
\bibliography{references}
\end{spacing}

\end{document}

%% file: introduction.tex
\vspace{-2mm}
\section{Introduction}
\label{sec_intro}

Despite the exponential growth of fairness in Machine Learning (AI) research, it remains centred on Western concerns and histories---the structural injustices (\eg{}race and gender), the data (\eg{}ImageNet), the measurement scales (\eg{}Fitzpatrick scale), the legal tenets (\eg{}equal opportunity), and the enlightenment values. Conventional western AI fairness is becoming a \textit{universal} ethical framework for AI; consider the AI strategies from India \cite{NITI}, Tunisia \cite{TunisiaAI}, Mexico \cite{MexicoAI}, and Uruguay \cite{UruguayAI} that espouse fairness and transparency, but pay less attention to what is fair in local contexts.

Conventional measurements of algorithmic fairness make several assumptions based on Western institutions and infrastructures. To illustrate, consider Facial Recognition (FR), where demonstration of AI fairness failures and stakeholder coordination have resulted in bans and moratoria in the US. Several factors led to this outcome:
\begin{itemize}[leftmargin=*]
\item Decades of scientific empiricism on proxies and scales that corresponds to subgroups in the West \cite{fitzpatrick1988validity}.
\item Public datasets, APIs, and freedom of information acts are available to researchers to analyse model outcomes \cite{MachineB33:online, HowFacia39:online}.
\item AI research/industry is fairly responsive to bias reports from users and civil society \cite{Weareimp72:online, buolamwini2018gender}.
\item The existence of government representatives glued into technology policy, shaping AI regulation and accountability  \cite{JayapalJ44:online}. 
\item An active media systematically scrutinises and reports on downstream impacts of AI systems \cite{HowFacia39:online}
\end{itemize}

We argue that the above assumptions may not hold in much else of the world. While algorithmic fairness keeps AI within ethical and legal boundaries in the West, there is a real danger that naive generalisation of fairness will fail to keep AI deployments in check in the non-West. Scholars have pointed to how neoliberal AI follows the technical architecture of classic colonialism through data extraction, impairing indigenous innovation, and shipping manufactured services back to the data subjects---among communities already prone to exploitation, under-development, and inequality from centuries of imperialism \cite{wallerstein1991world, kwet2019digital, mohamed2020decolonial, birhane2020algorithmic, roy2014capitalism}.
Without engagement with the conditions, values, politics, and histories of the non-West, AI fairness can be a tokenism, at best---pernicious, at worst---for communities.
If algorithmic fairness is to serve as the ethical compass of AI, it is imperative that the field recognise its own defaults, biases, and blindspots to avoid exacerbating historical harms that it purports to mitigate.
We must take pains not to develop a general theory of algorithmic fairness based on the study of Western populations.
Could fairness, then, have structurally different meanings in the non-West? Could fairness frameworks that rely on Western infrastructures be counterproductive elsewhere? How do social, economic, and infrastructural factors influence Fair-ML?


In this paper, we study algorithmic power in contemporary India, and holistically re-imagine algorithmic fairness in India. Home to 1.38 billion people, India is a pluralistic nation of multiple languages, religions, cultural systems, and ethnicities. India is the site of a vibrant AI workforce. Hype and promise is palpable around AI---envisioned as a force-multiplier of socio-economic benefit for a large, under-privileged population \cite{NITI}. AI deployments are prolific, including in predictive policing \cite{PredictiveIndia} and facial recognition \cite{FacialRecIndia}. 
Despite the momentum on high-stakes AI systems, currently there is a lack of substantial policy or research on advancing algorithmic fairness for such a large population interfacing with AI. 

We report findings from 36 interviews with researchers and activists working in the grassroots with marginalised Indian communities, and from observations of current AI deployments in India. We use feminist, decolonial, and anti-caste lenses to analyze our data. We contend that India is on a unique path to AI, characterised by pluralism, socio-economic development, technocratic nation-building, and uneven AI capital---which requires us to confront many assumptions made in algorithmic fairness. Our findings point to three factors that need attention in Fair-ML in India:

\textit{Data and model distortions}: Infrastructures and social contracts in India challenge the assumption that datasets are faithful representations of people and phenomena.  
Models are over-fitted for digitally-rich profiles---typically middle-class men---further excluding the 50\% without Internet access.  
Sub-groups like caste (endogamous, ranked social identities, assigned at birth \cite{ambedkar1916castes, shanmugavelan}),\footnote{According to Shanmugavelan, ``caste is an inherited identity that can determine all aspects of one’s life opportunities, including personal rights, choices, freedom, dignity, access to capital and effective political participation in caste-affected societies'' \cite{shanmugavelan}. Dalits ('broken men' in Marathi) are the most inferiorised category of the people, are within the social hierarchy, but excluded in caste categories \cite{shanmugavelan, beteille1990race, ambedkar1916castes, bondcaste}.} gender, and religion require different fairness implementations; but AI systems in India are under-analyzed for biases, mirroring the limited mainstream public discourse on oppression. Indic social justice, like reservations, presents new fairness evaluations.\footnote{We use the term 'Indic' to refer to native Indian concepts.}

\textit{Double standards and distance by ML makers}: Indian users are perceived as `bottom billion' data subjects, petri dishes for intrusive models, and given poor recourse---thus, effectively limiting their agency.
While Indians are part of the AI workforce, a majority work in services, and engineers do not entirely represent marginalities, limiting re-mediation of distances. 

\textit{Unquestioning AI aspiration}: 
The AI imaginary is aspirational in the Indian state, media, and legislation. AI is readily adopted in high-stakes domains, often too early. Lack of an ecosystem of tools, policies, and stakeholders like journalists, researchers, and activists to interrogate high-stakes AI inhibits meaningful fairness in India.



In summary, we find that conventional Fair-ML may be inappropriate, insufficient, or even inimical in India if it does not engage with the local structures. In a societal context where the distance between models and dis-empowered communities is large---via technical distance, social distance, ethical distance, temporal distance, and physical distance---a myopic focus on localising fair model outputs alone can backfire. 
We call upon fairness researchers working in India to engage with \emph{end-to-end} factors impacting algorithms, like datasets and models, knowledge systems, the nation-state and justice systems, AI capital, and most importantly, the oppressed communities to whom we have ethical responsibilities. We present a holistic framework to operationalise algorithmic fairness in India, calling for: \textit{re-contextualising} data and model fairness; \textit{empowering} oppressed communities by participatory action; and \textit{enabling} an ecosystem for meaningful fairness. 

Our paper contributes by bringing to light how algorithmic power works in India, through a bottom-up analysis. Second, we present a holistic research agenda as a starting point to operationalise Fair-ML in India. The concerns we raise may certainly be true of other countries. The broader goal should be to develop global approaches to Fair-ML that reflect the needs of various contexts, while acknowledging that some principles are specific to context.



%% file: background.tex
\vspace{-2mm}
\section{Background}
\label{sec_background}

Recent years have seen the emergence of a rich body of literature on fairness and accountability in machine learning \eg \cite{barocas2017fairness,mehrabi2019survey}.
However, most of this research is framed in the Western context, by researchers situated in Western institutions, for mitigating social injustices prevalent in the West, using data and ontologies from the West, and implicitly imparting Western values, \eg in the premier FAccT conference, of the 138 papers published in 2019 and 2020, only a handful of papers even mention non-West countries, and only one of them---Marda's paper on New Delhi's predictive policing system\cite{marda2020data}---substantially engages with a non-Western context.


\vspace{-2mm}
\subsection{Western Orientation in Fair-ML}


\subsubsection{Axes of discrimination}
The majority of fairness research looks at racial \cite{lum2016predict,sap2019risk,davidson2019racial,manzini2019black,buolamwini2018gender} and gender biases \cite{bolukbasi2016man,buolamwini2018gender,sun2019mitigating,zhao2017men} in models---two dimensions that dominate the American public discourse. 
However these categories are culturally and historically situated \cite{hanna2020towards}. Even the categorisation of proxies in fairness analyses have Western biases and origins; \eg{} the Fitzpatrick skin type is often used by researchers as a phenotype \cite{buolamwini2018gender}, but was originally developed to categorise UV sensitivity) \cite{fitzpatrick1988validity}.
While other axes of discrimination and injustices such as disability status \cite{hutchinson2020social}, age \cite{diaz2018addressing}, and sexual orientation \cite{garg2019counterfactual} have gotten some attention, 
biases relevant to other geographies and cultures are not explored (\eg{}Adivasis (indigeneous tribes of South Asia) and Dalits). As \cite{mulligan2019thing} points out, tackling these issues require a deep understanding of the social structures and power dynamics therein, which points to a wide gap in literature.

\subsubsection{Legal framing}
Since early inquiries into algorithmic fairness largely dealt with US law enforcement (predictive policing and recidivism risk assessment) as well as state regulations (\eg{}in housing, loans, and education), the research framings often rely implicitly on US laws such as the Civil Rights Acts and Fair Housing Act, as well as on US legal concepts of discrimination. Indeed, researchers since the late 1960s have tried to translate US anti-discrimination law into statistical metrics \cite{hutchinson201950}. 
The community also often repurposes terminology from US legal domains, such as ``disparate impact'', ``disparate treatment'', and ``equal opportunity'',
or use them as points of triangulation, in order to compare technical properties of fairness through analogy with legal concepts \cite{green2020algorithmic, green2020false}.
%
%
%

\subsubsection{Philosophical roots}
Connections have been made between algorithmic fairness and Western concepts such as egalitarianism \cite{binns2018fairness}, consequentialism \cite{roff2020expected, mulligan2019thing}, deontic justice \cite{binns2018fairness, mulligan2019thing}, and Rawls' distributive justice \cite{mulligan2019thing, joseph2016rawlsian}.
Indeed, notions of algorithmic fairness seem to fit within a broad arc of enlightenment and post-enlightenment thinking, including in actuarial risk assessment \cite{ochigame2020}. Dr. B. R. Ambedkar's ((1891–1956), fondly called Babasaheb, the leader and dominant ideological source of today's Dalit politics) anti-caste movement was rooted in social justice, distinct from Rawl's distributive justice \cite{rodrigues2011justice} (also see Sen's critique of Rawl's idea of original position and inadequacies of impartiality-driven justice and fairness \cite{sen2009idea}).
Fairness' status as the {\em de facto} moral standard of choice and signifier of justice, is itself a sign of cultural situatedness.
%
%
Other moral foundations \cite{graham2013moral} of cultural importance may often be overlooked by the West, including purity/sanctity.
Traditional societies often value {\em restorative justice}, which emphasises repairing harms \cite{boyes2014suffolk}, rather than fairness, \eg{}contemporary Australian Aboriginal leaders emphasise reconciliation rather than fairness in their political goals \cite{inayatullah2006culture}. 
Furthermore, cultural relationships such as power distance, and temporal orientation, are known to mediate the importance placed on fairness  \cite{lund2013culture}.

\vspace{-2mm}
\subsection{Fairness perceptions across cultures}

Social psychologists have argued that justice and fairness require a lens that go beyond the Euro-American cultural confines \cite{leung2001social}.
While the concern for justice has a long history in the West (\eg{}Aristotle, Rawls) and the East (\eg{}Confucius, Chanakya), they show that the majority of empirical work on social justice has been situated in the US and Western Europe, grounding the understanding of justice in the Western cultural context. 
Summarising decades worth of research, \cite{leung2001social} says that the more collectivist and hierarchical societies in the East differs from the more individualistic and egalitarian cultures of the West in how different forms of justice---distributive, procedural, and retributive---are conceptualised and achieved. 
For instance, \cite{blake2015ontogeny} compared the acquisition of fairness behaviour in seven different societies: Canada, India, Mexico, Peru, Senegal, Uganda, and the USA, and found that while children from all cultures developed aversion towards disadvantageous inequity (avoid receiving less than a peer), advantageous inequity aversion (avoid receiving more than a peer) was more prevalent in the West.
Similarly, a study of children in three different cultures found that notions of distributive justice are not universal: ``children from a partially hunter-gatherer, egalitarian African culture distributed the spoils more equally than did the other two cultures, with merit playing only a limited role'' \cite{schafer2015fair}. 
See above point on Dr. B. R. Ambedkar's centring on priority-based social justice for caste inequalities.
The above works point to the dangers in defining fairness of algorithmic systems based solely on a Western lens.

\vspace{-2mm}
\subsection{Algorithmic fairness in the non-West}
The call for a global lens in AI accountability is not new \cite{paul2018reflecting,hagerty2019global}, but the ethical principles in AI are often interpreted, prioritised, contextualised, and implemented differently across the globe \cite{jobin2019global}. Recently, the IEEE Standards Association highlighted the monopoly of Western ethical traditions in AI ethics, and inquired how incorporating Buddhist, Ubuntu, and Shinto-inspired ethical traditions might change the processes of responsible AI \cite{ieeeglobal2019}.
Researchers have also challenged the normalisation of Western implicit beliefs, biases, and issues in specific geographic contexts; \eg{}India, Brazil and Nigeria \cite{sambasivan2018toward}, and China and Korea \cite{shin2019toward}.
Representational gaps in data is documented as one of the major challenges in achieving responsible AI from a global perspective \cite{arora2016bottom,shankar2017no}. For instance, \cite{shankar2017no} highlights the glaring gaps in geo-diversity of open datasets such as ImageNet and Open Images that drive much of the computer vision research. \cite{agarwal2020entity} shows that NLP models disproportionately fail to even detect names of people from non-Western backgrounds.

\vspace{-2mm}
\subsection{Accountability for unfairness}
Discussion of accountability is critical to any discussions of fairness, \ie{}how do we hold deployers of systems accountable for unfair outcomes?
Is it fair to deploy a system that lacks in accountability? Accountability is fundamentally about answerability for actions \cite{kohli2018translation}, and central to these are three phases by which an actor is made answerable to a forum: information-sharing, deliberation and discussion, and the imposition of consequences  \cite{wieringa2020account}. 
Since outcomes of ML deployments can be difficult to predict, proposals for accountability include participatory design \cite{katell2020toward} and participatory problem formulation \cite{martin2020participatory}, sharing the responsibility for designing solutions with the community. 
Nissenbaum distinguishes four barriers to responsibility in computer systems: (1) the problem of many hands, (2) bugs, (3) blaming the computer, and (4) ownership without liability \cite{nissenbaum1996accountability}.
These barriers become more complicated when technology spans cultures: more, and more remote, hands are involved; intended behaviours may not be defined; computer-blaming may meet computer-worship head on (see Section~\ref{sec_findings}); and questions of ownership and liability become more complicated.

Specific to the Indian context, scholars and activists have outlined opportunities for AI in India \cite{kalyanakrishnan2018opportunities}, proposed policy deliberation frameworks that take into account the unique policy landscape of India \cite{marda2018artificial}, and questioned the intrusive data collection practices through Aadhaar (biometric-based unique identity for Indians) in India \cite{ramanathan2014biometrics,ramanathan2015considering}. Researchers have documented societal biases in predictive policing in New Delhi \cite{marda2020data}, caste \cite{thorat2007legacy} and ethnicity \cite{ChutiaCh43:online} biases in job applications, call-center job callbacks \cite{banerjee2009labor}, caste-based wage-gaps \cite{madheswaran2007caste}, caste discrimination in agricultural loans decisions \cite{kumar2013does}, and even in online matrimonial ads \cite{rajadesingan2019smart}.

%% file: method.tex
\vspace{-2mm}

\section{Method}
\label{sec_methodology}
Our research results come from a critical synthesis of expert interviews and discourse analysis. Our methods were chosen in order to provide an expansive account of who is building ML for whom, what the on-the-ground experiences are, what the various processes of unfairness and exclusions are, and how they relate to social justice. 

We conducted qualitative interviews with 36 expert researchers, activists, and lawyers working closely with marginalised Indian communities at the grassroots. 
Expert interviews are a qualitative research technique used in exploratory phases, providing practical insider knowledge and surrogacy for a broader community \cite{bogner2009interviewing}. Importantly, experts helped us gain access to a nascent and difficult topic, considering the early algorithmic deployments in the public sector in India.
Our respondents were chosen from a wide range of areas to create a holistic analysis of algorithmic power in India. Respondents came from Computer Science (11), Activism (9), Law and Public Policy (6), Science and Technology Studies (5), Development Economics (2), Sociology (2), and Journalism (1). All respondents had 10-30 years of experience working with marginalised communities or on social justice. Specific expertise areas included caste, gender, labour, disability, surveillance, privacy, health, constitutional rights, and financial inclusion. 24 respondents were based in India, 2 in Europe, 1 in Southeast Asia, the rest in the USA; 25 of them self-identified as male, 10 as female, and 1 as non-binary.

In conjunction with qualitative interviews, we conducted an analysis of various algorithmic deployments and emerging policies in India, starting from Aadhaar (2009).
We identified and analysed various Indian news publications (\eg{}TheWire.in, Times of India), policy documents (\eg{}NITI Aayog, Srikrishna Bill), and community media (\eg{}Roundtable India, Feminism in India), and prior research. Due to secondary sources, our citations are on the higher side.

\textbf{Recruitment and moderation}
We recruited respondents via a combination of reaching out directly and personal contacts, using purposeful sampling \cite{palinkas2015purposeful}---\ie identifying and selecting experts with relevant experience---iterative until saturation. We conducted all interviews in English (preferred language of participants). The semi-structured interviews focused on 1) unfairness through discrimination in India; 2) technology production and consumption; 3) the historical and present role of fairness and ethics in India; 4) biases, stereotypes and proxies; 5) data; 6) laws and policy relating to fairness; and 7) canonical applications of fairness, evaluated in the Indian context. Respondents were compensated for the study (giftcards of 100 USD, 85 EUR, and 2000 INR), based on purchasing power parity and non-coercion. Employer restrictions prevented us from compensating government employees. Interviews lasted an hour each, and were conducted using video conferencing and captured via field notes and video recordings. 

\textbf{Analysis and coding}
Transcripts were coded and analyzed for patterns using an inductive approach~\cite{Thomas2006general}. From a careful reading of the transcripts, we developed categories and clustered excerpts, conveying key themes from the data. Two team members created a code book based on the themes, with seven top-level categories (sub-group discrimination, data and models, law and policy, ML biases and harms, AI applications, ML makers, and solutions) and several sub-categories (\eg{}caste, missing data, proxies, consent, algorithmic literacy, and so on). The themes that we describe in Section~\ref{sec_findings} were then developed and applied iteratively to the codes.

Our data is analysed using feminist, decolonial, and anti-caste lenses. A South Asian feminist stance allows us to examine oppressed communities as encountering and subverting forces of power, while locating in contextual specifics of family, caste, class, and religion.\footnote{We are sympathetic to Dalit feminist scholars, like Rege, who have critiqued postcolonial or subaltern feminist thoughts for the lack of anti-caste scrutiny \cite{rege1998dalit}} South Asian feminism is a critique of the white, Western feminism that saw non-western women as powerless victims that needed rescuing \cite{mohanty2005feminism, chaudhuri2004feminism}. 
Following Dr. B. R. Ambedkar's insight on how caste hierarchies and patriarchies are linked in India \cite{chakravarti1993conceptualising}, we echo that no social justice commitment in India can take place without examining caste, gender, and religion. A decolonial perspective (borrowed from Latin American and African scholars like \cite{escobar2011encountering, mignolo2011darker, wa1992decolonising,fanon2007wretched, dorfman1975read}) helps us confront inequalities from colonisation in India, providing new openings for knowledge and justice in AI fairness research. To Dr. B. R. Ambedkar and Periyar E. V. Ramasamy, colonialism predates the British era, and decolonisation is a continuum. For Dalit emancipatory politics, deconstructing colonial ideologies of the powerful, superior, and privileged begins by removing influences and privileges of dominant-caste members.\footnote{Thanks to Murali Shanmugavelan for contributing these points.}


\textbf{Research ethics}
We took great care to create a research ethics protocol to protect respondent privacy and safety, especially due to the sensitive nature of our inquiry. During recruitment, participants were informed of the purpose of the study, the question categories, and researcher affiliations. Participants signed informed consent acknowledging their awareness of the study purpose and researcher affiliation prior to the interview. At the beginning of each interview, the moderator additionally obtained verbal consent. We stored all data in a private Google Drive folder, with access limited to our team. To protect participant identity, we deleted all personally identifiable information in research files. We redact identifiable details when quoting participants. Every respondent was given the choice of default anonymity or being included in Acknowledgements. 

All co-authors of this paper work at the intersection of under-served communities and technology, with backgrounds in HCI, critical algorithmic studies, and ML fairness. The first author constructed the research approach and has had grassroots commitments with marginalised Indian communities for nearly 15 years. The first and second author moderated interviews. All authors were involved in the framing, analysis, and synthesis. Three of us are Indian and two of us are White. All of us come from privileged positions of class and/or caste. We acknowledge the above are our interpretations of research ethics, which may not be universal. 




%% file: findings.tex
 \vspace{-2mm}
\section{Findings}
\label{sec_findings}

\begin{figure}[t!]
\includegraphics[width=\linewidth]{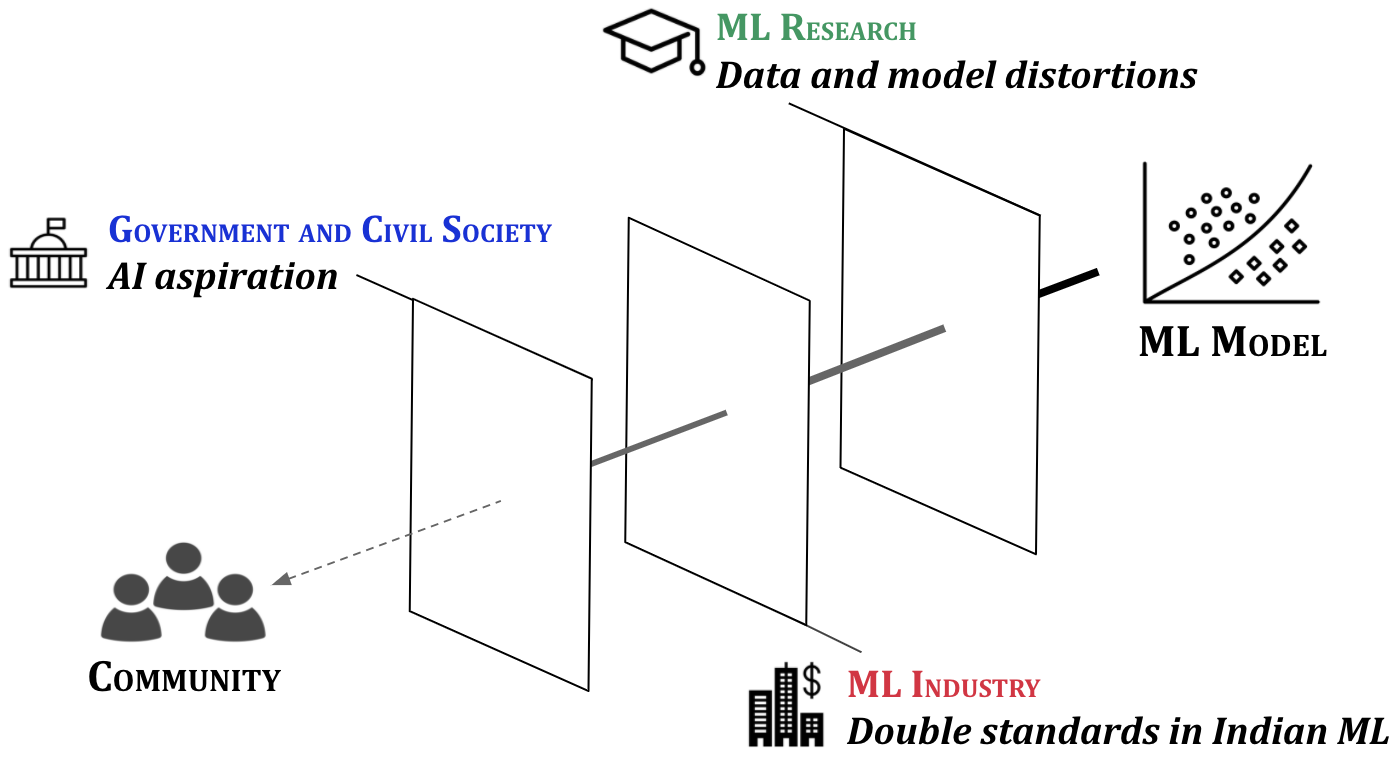}
\caption{\small Algorithmic powerful  in India, where the distance between models and oppressed communities is large}
\label{fig:Indiadistance}
 \Description{Three screens are shown in between communities and ML models. The screens are government and civil society for AI aspiration, technology industry for third world ML treatment, and ML research for flawed data and model assumptions. The screens limit the impact of models and create distance with communities.}
\end{figure}

We now present three themes (see Figure~\ref{fig:Indiadistance}) that we found to contrast views in conventional algorithmic fairness.

\vspace{-2mm}
\subsection{Data and model distortions}
Datasets are often seen as reliable representations of populations.
Biases in models are frequently attributed to biased datasets, presupposing the possibility of achieving fairness by ``fixing'' the data \cite{holstein2019improving}. However, social contracts, informal infrastructures, and population scale in India lead us to question the reliability of datasets. 

 \vspace{-.5em}
\subsubsection{Data considerations} \hfill\\
\textbf{Missing data and humans}
Our respondents discussed how data points were often missing because of social infrastructures and systemic disparities in India. Entire communities may be missing or misrepresented in datasets, exacerbated by digital divides, leading to wrong conclusions \cite{barocas2016big, lerman2013big, crawford2013hidden, crawford2013think} and residual unfairness \cite{kallus2018residual}. Half the Indian population lacks access to the Internet---the excluded half is primarily women, rural communities, and Adivasis \cite{women2020mobile, InIndiaA10:online, Jain2016Indias, MissionC54:online, COVID19l97:online}. Datasets derived from Internet connectivity will exclude a significant population, \eg many argued that India's mandatory \textsc{covid}-19 contact tracing app excluded hundreds of millions due to access constraints, pointing to the futility of digital nation-wide tracing (also see \cite{AarogyaS88:online}). 
Moreover, India's data footprint is relatively new, being a recent entrant to 4G mobile data.
Many respondents observed a bias towards upper-middle class problems, data, and deployments due to easier data access and revenue, as P8, CS/IS (computer science/information sciences researcher) put it, ``\textit{rich people problems like cardiac disease and cancer, not poor people's Tuberculosis, prioritised in AI [in India]}'', exacerbating inequities among those who benefit from AI and those who do not. 

Several respondents pointed to missing data due to class, gender, and caste inequities in accessing and creating online content, \eg{}many safety apps use data mapping to mark areas as unsafe, in order to calculate an area-wide safety score for use by law enforcement \cite{safetipin,citizencop} (women's safety is a pressing issue in India, in public consciousness since the 2012 Nirbhaya gang rape \cite{5inNewDe60:online}. P4 (anti-caste, communications researcher) described how rape is socially (caste, culture, and religion) contextualised and some incidents get more visibility than others, in turn becoming data, in turn getting fed into safety apps---a perpetual source of unfairness. Many respondents were concerned that the safety apps were populated by middle-class users and tended to mark Dalit, Muslim, and slum areas as unsafe, potentially leading to hyper-patrolling in these areas. 

Data was reported to be `missing' due to artful user practices to manipulate algorithms, motivated by privacy, abuse, and reputation concerns. \eg{}studies have shown how women users have `confused' algorithms, motivated by privacy needs \cite{sambasivan2018privacy, masika2015negotiating}). Another class of user practices that happened outside of applications led to `off data' traces. As an example, P17, CS/IS researcher, pointed to how auto rickshaw drivers created practices outside of ride-sharing apps, like calling passengers to verify landmarks (as Indian addresses are harder to specify \cite{Economic79:online}) or cancelling rides in-app (which used mobile payments) to carry out rides for a cash payment. Respondents described how data, like inferred gender, lacking an understanding of context was prone to inaccurate inferences.

Many respondents pointed to the frequent unavailability of socio-economic and demographic datasets at national, state, and municipal levels for public fairness research. Some respondents reported on how the state and industry apparati collected and retained valuable, large-scale data, but the datasets were not always made publicly available due to infrastructure and non-transparency issues. As P5, public policy researcher, described, ``\textit{The country has the ability to collect large amounts of data, but there is no access to it, and not in a machine-readable format.''} In particular, respondents shared how datasets featuring migration, incarceration, employment, or education, by sub-groups, were unavailable to the public. Scholarship like Bond's caste report \cite{bondcaste} argues that there is limited political will to collect and share socio-economic indicators by caste or religion. 

A rich human infrastructure \cite{sambasivan2010human} from India's public service delivery, \eg{}frontline data workers, call-center operators, and administrative staff extends into AI data collection. However, they face disproportionate work burden, sometimes leading to data collection errors \cite{datalabelling, Counting37:online, ismail2018engaging}. Many discussed how consent to a data worker stemmed from high interpersonal trust and holding them in high respect---relationships which may not be transitive to the downstream AI applications. In some cases, though, data workers have been shown to fudge data without actual conversations with affected people; efforts like \emph{jun sanwais} (public hearings) and Janata Information Systems organized by the Mazdoor Kisan Shakti Sangatan are examples to secure representation through data. \cite{singh2018, jenkins1999accounts}.

\textbf{Mis-recorded identities}
Statistical fairness makes a critical assumption so pervasively that it is rarely even stated: that user data corresponds one-to-one to people.\footnote{This issue is somewhat related to what \cite{olteanu2019social} call ``Non-individual agents''.} However the one-to-one correspondence in datasets often fails in India. Ground truth on full names, location, contact details, biometrics, and their usage patterns can be unstable, especially for marginalised groups. User identity can be mis-recorded by the data collection instrument, assuming a static individual correspondence or expected behaviour. 
Since conventional gender roles in Indian society lead to men having better access to devices, documentation, and mobility (see \cite{sambasivan2018privacy, donner2008express}), women often borrowed phones. A few respondents pointed to how household dynamics impacted data collection, especially when using the door-to-door data collection method, \eg how heads of households, typically men, often answered data-gathering surveys on behalf of women, but responses were recorded as women's. 

Several AI-based applications use phone numbers as a proxy for identity in account creation (and content sharing) in the non-West, where mobile-first usage is common and e-mail is not \cite{donner2015after}. However, this approach fails due to device sharing patterns \cite{sambasivan2018privacy}, increased use of multiple SIM cards, and the frequency with which people change their numbers. Several respondents mentioned how location may not be permanent or tethered to a home, \eg{}migrant workers regularly travel across porous nation-state boundaries.

\input{subgroups_table}

\textbf{Discriminated sub-groups and proxies}
AI systems in India remain under-analysed for biases, mirroring the limited public discourse on oppression. In Table 1, we present a summary of discriminated sub-groups in India, derived from our interviews and enriched through secondary research and statistics from authoritative sources, to substantiate attributes and proxies. Furthermore, we describe below some common discriminatory proxies and attributes that came up during our interviews. While the proxies may be similar to those in the West, the implementation and cultural logics may vary in India, \eg{}P19, STS researcher, pointed to how Hijra community members (a marginalised intersex or transgender community) may live together in one housing unit and be seen as fraudulent or invisible to models using family units. Proxies may not generalise even within the country, \eg{}asset ownership: ``\textit{If you live in Mumbai, having a motorbike is a nuisance. If rural, you're the richest in town.}'' (P9, CS/IS researcher).

\begin{itemize}[leftmargin=*]
    \item \textit{Names:} are revelatory proxies for caste, religion, gender, and ethnicity, and have contributed to discrimination in India \cite{thorat2007legacy, ChutiaCh43:online}, \eg{}Banerjee or Khan connote to caste or religion location. 
    \item \textit{Zip codes:} can correspond to caste, religion, ethnicity, or class. Many Indian zip codes are heterogeneous, with slums and posh houses abutting in the same area \cite{appadurai2000spectral,bharathi2018isolated}, unlike the rather homogeneous zip codes in the West from the segregated past \cite{ZIPCodeH5:online}.
    \item \textit{Occupation:} traditional occupations may correspond to caste, gender, or religion; \eg{}manual scavenging or butchery. 
    \item \textit{Expenditure:} on dietary and lifestyle items may be proxies for religion, caste, or ethnicity; \eg{}expenses on pork or beef. 
    \item \textit{Skin tone:} may indicate caste, ethnicity, and class; however, unlike correlations between race and skin tone, correspondences to Indian sub-groups is weaker. Dark skin tones can be discriminated against in India \cite{FairButN13:online}. Many respondents described how datasets under-collected darker skin tones and measurement scales like Fitzpatrick scale are 
   not calibrated on diverse Indian skin tones.
    \item \textit{Mobility:} can correlate to gender and disability. Indian women travel shorter distances and depend on public transport, for safety and cost \cite{WhyUrban26:online}. Persons with disabilities often have lower mobility in India, due to a lack of ramps and accessible infrastructure 
    \cite{Disabili59:online}.
    \item \textit{Language:} Language can correspond to religion, class, ethnicity, and caste. Many AI systems serve in English, which only 10\% of Indians understand \cite{InIndiaw91:online}. India has 30 languages with over a million speakers. Everyday slurs such as `dhobi', `kameena', `pariah', or `junglee' are reported to be rampant online \cite{Casteist69:online, ChutiaCh43:online}.
    \item \textit{Devices and infrastructure:} Internet access corresponds to gender, caste, and class, with 67\% Internet users being males \cite{Highgend58:online}. 
    \item \textit{Documentation}: Several AI applications require state-issued documentation like Aadhaar or birth certificates, \eg{}in finance. The economically poor are also reported to be document-poor \cite{Indiasp84:online}.\footnote{National IDs are contested in the non-West, where they are used to deliver welfare in an `objective' manner, but lead to populations falling through the cracks (see research on IDs in Aadhaar \cite{Howaglit16:online, singh2017margins}, post-apartheid South Africa \cite{donovan2015biometric} and Sri Lankan Tamils and C\^{o}te d'Ivoirians \cite{Womenand71:online}). Documentation has been known to be a weapon to dominate the non-literate in postcolonial societies \cite{gupta2012red} Also see administrative violence \cite{spade2015normal}.} 
\end{itemize}
\subsubsection{Model considerations} \hfill\\
\textbf{Over-fitting models to the privileged}
Respondents described how AI models in India overfitted to `good' data profiles of the digitally-rich, privileged communities, as a result of poor cultural understanding and exclusion on part of AI makers. Respondents noted that the sub-groups that had access to underlying variables for data-rich profiles, like money, mobility, and literacy, were typically middle-class men. 
Model inputs in India appear to be disproportionately skewed due to large disparities in digital access. For instance, P11, tech policy researcher, illustrated how lending apps instantly determined creditworthiness through alternative credit histories built based on the user's SMS messages, calls, and social networks (due to limited credit or banking history). Popular lending apps equate `good credit' with whether the user called their parents daily, had stored over 58 contacts, played car-racing games, and could repay in 30 days \cite{Mobilelo46:online}. Many respondents described how lending models imagined middle-class men as end-users---even with many microfinance studies showing that women have high loan repayment rates \cite{d2011women, swain2009does}. In some cases, those with `poor' data profiles subverted model predictions---as in P23's (STS researcher) research on financial lending, where women overwhelmingly availed and paid back loans in the names of male relatives to avoid perceived gender bias in credit scoring. Model re-training left new room for bias, though, due to a lack of Fair-ML standards for India, \eg{}an FR service used by police stations in eight Indian states retrained a western FR model on photos of Bollywood and regional film stars to mitigate the bias \cite{FacialRecIndia}---but Indian film stars are overwhelmingly fair-skinned, conventionally attractive, and able-bodied \cite{karan2008obsessions}, not fully representative of the larger society.

\textbf{Indic justice in models}
Popular fairness techniques, such as equal opportunity and equal odds, stem from epistemological and legal systems of the US (\eg{}\cite{dobbe2018broader, xiang2019legal}). India's own justice approaches present new and relevant ways to operationalise algorithmic fairness locally. Nearly all respondents discussed reservations as a technique for algorithmic fairness in India. One of the restorative justice measures to repair centuries of subordination---reservations are a type of affirmative action enshrined in the Indian constitution \cite{RichardsonFairness}. Reservations assigns quotas for marginalised communities at the national and state levels.\footnote{While the US Supreme court has banned various quotas \cite{joshi2018racial}, there is a history of quotas in the US, sometimes discriminatory, \eg{New Deal black worker quotas \cite{newdeal} and anti-semitic quotas in universities \cite{hollinger1998science}}. Quotas in India are legal and common. Thanks to Martin Wattenberg for this point.} Originally designed for Dalits and Adivasis, reservations have expanded to include other backward castes, women, persons with disabilities, and religious minorities. Depending on the policy, reservations can allocate quotas from 30\% to 80\%. 
Reservations in India have been described as one of the world's most radical policies \cite{baker_2001} (see \cite{RichardsonFairness} for more). Several studies have pointed to the successful redistribution of resources towards oppressed sub-groups, through reservations \cite{pande2003can, duflo2005political, borooah2007effectiveness}. 


 \vspace{-2mm}
\subsection{Double standards by ML makers}

\textbf{'Bottom billion' petri dishes}
Several respondents discussed how AI developers, both regional and international---private and state---treated Indian user communities as `petri dishes' for models. Many criticised how neo-liberal AI (including those from the state) tended to treat Indians as `bottom billion data subjects' in the periphery \cite{wallerstein1991world}---being subject to intrusive models, non-consensual automation, poor tech policies, inadequate user research, low-cost or free products that are low standard, and considered `unsaturated markets'. India's diversity of languages, scripts, and populace has been reported to be attractive for improving model robustness and training data \cite{India'sm28:online}. 
Many discussed how low quality designs, algorithms, and support were deployed for Indians, attributing to weak tech policy and enforcement of accountability in India. Several respondents described how AI makers had a transactional mindset towards Indians, seeing them as agency-less data subjects that generated large-scale behavioural traces to improve models. 

In contrast to how AI industry and research were moderately responsive to user bias reports in the West, recourse and redress for Indians were perceived to be non-existent. Respondents described that when recourse existed, it was often culturally-insensitive or dehumanising, \eg{}a respondent was violently questioned about their clothing by staff of a ride-sharing application, during redressal for an assault faced in a taxi (also see \cite{sambasivan2019they} for poor recourse). Several respondents described how lack of recourse was even more dire for marginalised users. \eg{}P14 (CS/IS researcher) described, ``\textit{[Ordering a ride-share] a person with a disability would choose electronic payment, but the driver insisted on cash. They said they are blind and wanted to pay electronically, but the driver declined and just moved on. No way to report it.}'' Even when feedback mechanisms were included, respondents shared that they were not always localised for India, and incidents were not always recognised unless an activist contacted the company staff directly. Many respondents shared how street-level bureaucrats, administrative offices, and front line workers---the human infrastructures \cite{sambasivan2010human} who played a crucial role in providing recourse to marginalised Indian communities---were removed in AI systems. Further, the high-tech illegibility of AI was noted to render recourse out of reach for groups marginalised by literacy, legal, and educational capital (see \cite{veeraraghavan2013dealing} for 'hiding behind a computer'). As P12 (STS researcher) explained, ``\textit{If decisions are made by a centralised server, communities don't even know what has gone wrong, why [welfare] has stopped, they don't know who to go to to fix the problem.}'' Many described how social audits and working with civil society created a better understanding and accountability.\footnote{Social audits like jan sanwais have long gauged effectiveness of civil programmes through village-level audits of documents, \eg to curb corrupt funds siphoning \cite{socialaudits}.}

Some respondents pointed to how Dalit and Muslim bodies were used as test subjects for AI surveillance, \eg{}pointing to how human efficiency trackers were increasingly deployed among Dalit sanitation workers in cities like Panchkula and Nagpur. Equipped with microphones, GPS, cameras, and a SIM, the trackers allowed detailed surveillance of movement and work, leading to some women workers avoiding restrooms for fear of camera capture, avoiding sensitive conversations for fear of snooping, and waiting for the tracker to die before going home \cite{Surveill67:online}. Such interventions were criticised for placing power in the hands of dominant-caste supervisors. P21 (legal researcher) pointed out that surveillance has historically been targeted at the working-poor, ``\textit{the house cleaner who is constantly suspected of stealing dried fruits or jewellery. Stepping out of their house means that their every move is tracked. Someone recording the face, the heartbeat..under the pretext of efficiency. Her job is to clean faeces in the morning and now she is a guinea pig for new AI.}'' 

\textbf{Entrenched privilege and distance}
Nearly all respondents described how AI makers and researchers, including regional makers, were heavily distant from the Indian communities they served. 
Several respondents discussed how Indian AI engineers were largely privileged class and caste males.\footnote{India fares slightly better than the US in gender representation in the tech workforce; however, gender roles and safety concerns lead to nearly 80\% of women leaving computing by their thirties (coinciding with family/parenting responsibilities) \cite{thakkar2018unexpected}.} 
For \eg{}P17 (CS/IS researcher) described, ``\textit{Who is designing AI? Incredibly entitled, Brahmin, certainly male. They've never encountered discrimination in their life. These guys are talking about primitive women. If they're designing AI, they haven't got a clue about the rest of the people. Then it becomes fairness for who?}'' Many respondents described how the Indian technology sector claimed to be `merit-based', open to anyone highly gifted in the technical sciences; but many have pointed to how merit is a function of caste privilege \cite{subramanian2015making, upadhya2007employment}. Many, though not all, graduates of Indian Institutes of Technology, founders of pioneering Indian software companies and nearly all Nobel prize winners of Indian origin have come from privileged castes and class backgrounds \cite{MostIndi27:online, subramanian2015making}. As P21 (legal researcher) explained the pervasive privilege in AI, ``\textit{Silicon Valley Brahmins [Indians] are not questioning the social structure they grew up in, and white tech workers do not understand caste to spot and mitigate obvious harms.
}'' 
%
While engineers and researchers are mostly privileged everywhere, the stark socio-economic disparities between Indian engineers and the marginalised communities may further amplify the distances.


 \vspace{-2mm}
\subsection{Unquestioning AI aspiration}
\textbf{AI euphoria}
Several respondents described how strong aspiration for AI for socio-economic upliftment was accompanied by high trust in automation, limited transparency, and the lack of an empowered Fair-ML ecosystem in India. Contrast with the West, where a large, active stakeholder ecosystem (of civil society, journalists, and law makers) is AI-literate and has access to open APIs and data. Many respondents described how public sector AI projects in India were viewed as modernising efforts to overcome age-old inefficiencies in resource delivery (also in \cite{sambasivan2019remarkable}). The AI embrace was attributed to follow the trajectory of recent high-tech interventions (such as Aadhaar, \textsc{MGNREGA} payments, and the National Register of Citizens \textsc{(NRC)}). Researchers have pointed to the aspirational role played by technology in India, signifying symbolic meanings of modernity and progress via technocracy \cite{pal2015banalities, pal2008computers, sambasivan2017imagined}. AI for societal benefit is a pivotal development thrust in India, with a focus on healthcare, agriculture, education, smart cities, and mobility \cite{NITI}---influencing citizen imaginaries of AI. In an international AI perceptions survey (2019), Indians ranked first in rating AI as `exciting', `futuristic' and `mostly good for society' \cite{kelley2019happy}.

Several respondents pointed to how automation solutions had fervent rhetoric; whereas in practice, accuracy and performance of systems were low. Many described how disproportionate confidence in high-tech solutions, combined with limited technology policy engagement among decision-makers, appeared to lead to sub-par high-stakes solutions, 
\eg{}the FR service used by Delhi Police was reported to have very low accuracy and failed to distinguish between boy and girl children \cite{techcrunch}.\footnote{A confidence threshold of 80-95\% is recommended for law enforcement AI \cite{HowAccur35:online}} 
Some respondents mentioned how a few automation solutions were announced following public sentiment, but turned into surveillance \eg{}how predictive policing in Delhi and FR in train stations was announced after Nirbhaya's gruesome gangrape in 2012 and women's safety incidents \cite{Nirbhaya74:online, Facialre30:online}. 

\textbf{Disputing AI4All}
Many respondents pointed to how emerging `4good' deployments tended to leave out minorities. \eg{}P29 (LGBTQ+ activist) discussed how AI was justified in the public domain, \eg{}surveillance for smart cities,\footnote{98 Indian cities are smart city sites, to be equipped with intelligent traffic, waste and energy management, and CCTV crime monitoring. \url{http://smartcities.gov.in/content}.} as women's safety measures, but tended to invisibilise transgender members or increase monitoring of Dalit and Muslim areas, \eg{}a FR was deployed outside women's restrooms to detect intrusion by non-female entrants, potentially leading to violence against transgender members. 

Many respondents expressed concern over AI advances in detecting sexuality, criminality, or terrorism (\eg{}\cite{seo2018partially, wang2018deep}) potentially being exported to India and harming minorities. P29 remarked on targeted attacks \cite{whatsapp}, ``\textit{part of the smart cities project is a Facial Recognition database where anyone can upload images. Imagine the vigilantism against dalit, poor, Muslims, trans persons if someone uploads a photo of them and it was used for sex offenders [arrests].}''

\textbf{Algorithmic opacity and authority}
In contrast to the `black box AI problem', \ie{}even the humans who design models do not always understand how variables are being combined to make inferences \cite{rudin2019we}, many respondents discussed an \textit{end-to-end} opacity of inputs, model behaviour, and outcomes in India. Fairness in India was reported to suffer from a lack of access to contributing datasets, APIs, and documentation, with several respondents describing how challenging it was for researchers and civil society to assess the high-stakes AI systems. As P11 described, ``\textit{Opacity is quite acute [in India]. People talk about blackboxes, reverse engineering inputs from outputs. What happens when you don't have the output? What happens when you can't reverse engineer at all?}''.

AI's `neutral' and `human-free' associations lent credence to its algorithmic authority. In January 2020, over a thousand protestors were arrested during protests in Delhi, aided by FR. The official statement was, ``\textit{This is a software. It does not see faith. It does not see clothes. It only sees the face and through the face the person is caught.}'' \cite{techcrunch}. While algorithms may not be trained on sub-group identification, proxies may correspond to Dalits, Adivasis, and Muslims disproportionately. \eg{}according to the National Crime Records Bureau (NCRB) in 2015, 34\% of undertrials were Dalits and Adivasis (25\% of the population); 20\% were Muslims (14\% of population); and 70\% were non-literate (26\% of the population) \cite{NCRBdata58:online}. 

Several respondents discussed a lack of inclusion of diverse stakeholders in decision-making processes, laws, and policies for public sector AI. Some talked about a colonial mindset of tight control in decision-making on automation laws, leading to reticent and monoscopic views by the judiciary and state. P5 (public policy researcher) pointed to how mission and vision statements for public sector AI tended to portray AI like magic, rather than contending with the realities of ``\textit{how things worked on-the-ground in a developing country}''. Additionally, respondents pointed to significant scope creep in high-stakes AI, \eg{}a few mentioned how the tender for an FR system was initially motivated by detecting criminals, later missing children, to then arresting protestors \cite{techcrunch}. 

\textbf{Questioning AI power}
Algorithmic fairness requires a buffer zone of journalists, activists, and researchers to keep AI system builders accountable. Many respondents described how limited debate and analysis of AI in India led to a weak implementation of Fair-ML in India. Issues of algorithmic bias were not widely raised in the public consciousness in India, at the time of the study. Respondents described how technology journalists in India---a keystone species for public discourse---covered app launches and investments, and less on algorithms and their impacts. P15 (journalist) pointed out that journalists may face disapproval for questioning certain narratives. \textit{``The broader issue is that AI biases do not come up in Indian press. Our definition of tech journalism has been about covering the business of tech [...] It is different from the US, where there is a more combative relationship. We don't ask tough questions of the state or tech companies.''} A seminal report by Newslaundry/Oxfam described how privileged castes comprised Indian news media, invisibilising the vast oppressed majority in public discourse \cite{Newslaundry}. 


%% file: subgroups_table.tex
\begin{savenotes}
\begin{table}
\small
\setlength{\tabcolsep}{2pt}
\begin{tabularx}{\linewidth}{X}
\toprule
\multicolumn{1}{c}{\bf Sub-groups, Proxies and Harms} \\
\midrule
\begin{minipage}[t]{\linewidth}
\textbf{Caste} 
\footnotesize{(17\% Dalits; 8\% Adivasi; 40\% Other Backward Class (OBC)
)\cite{CensusIndia2011}} 
\begin{itemize}[noitemsep,leftmargin=*,topsep=2pt]
  \item \footnotesize{Societal harms: Human rights atrocities. Poverty. Land, knowledge and language battles  \cite{xaxa2011tribes, ambedkar2014annihilation, Twothird31:online}. }
    \item \footnotesize{Proxies: Surname. Skin tone. Occupation. Neighborhood. Language.} 
    \item \footnotesize{Tech harms: Low literacy and phone ownership. Online misrepresentation \& exclusion. Accuracy gap of Facial Recognition {(FR)}. Limits of Fitzpatrick scale. Caste-based discrimination in tech ({}\cite{TheCisco63:online}). }
\end{itemize}
\end{minipage}\vspace{2pt}\\
\midrule

\begin{minipage}[t]{\linewidth}
\textbf{Gender}
\footnotesize{(48.5\% female)\cite{IndiaMinistryStatistics2018}} 
\begin{itemize}[noitemsep,leftmargin=*,topsep=2pt]
    \item \footnotesize{Societal harms: Sexual crimes. Dowry. Violence. Female infanticide.}
  \item \footnotesize{Proxies: Name. Type of labor. Mobility from home. }%
    \item \footnotesize{Tech harms: Accuracy gap in FR. Lower creditworthiness score. Recommendation algorithms favoring majority male users. Online abuse and 'racey’ content issues. Low Internet access.} 
\end{itemize}
\end{minipage}\vspace{2pt}\\
\midrule

\begin{minipage}[t]{\linewidth}
\textbf{Religion}
\footnotesize{(80\% Hindu, 14\% Muslim, 6\% Christians, Sikhs, Buddhists, Jains and indigeneous) \cite{CensusIndia2011}}  
\begin{itemize}[noitemsep,leftmargin=*,topsep=2pt]
    \item \footnotesize{Societal harms: Discrimination, lynching, vigilantism, and gang-rape against Muslims and others \cite{84DeadIn98:online}.}
    \item \footnotesize{Proxies: Name. Neighborhood. Expenses. Work. Language. Clothing. } 
    \item \footnotesize{Tech harms: Online stereotypes and hate speech, \eg Islamophobia. Discriminatory inferences due to lifestyle, location, appearance. Targeted Internet disruptions.} 
\end{itemize}
\end{minipage}\vspace{2pt}\\
\midrule

\begin{minipage}[t]{\linewidth}
\textbf{Ability}
\footnotesize{(5\%--8\%+ persons with disabilities) \cite{WorldBank2009Disabilities}} 
\begin{itemize}[noitemsep,leftmargin=*,topsep=2pt]
    \item \footnotesize{Societal harms: Stigma. Inaccessible education, transport \& work.} 
    \item \footnotesize{Proxies: Non-normative facial features, speech patterns, body shape \& movements. Use of assistive devices.} 
    \item \footnotesize{Tech harms: Assumed homogeneity in physical, mental presentation. Paternalistic words and images. No accessibility mandate. }
\end{itemize}
\end{minipage}\vspace{2pt}\\
\midrule

\begin{minipage}[t]{\linewidth}
\textbf{Class}
\footnotesize{(30\% live below poverty line; 48\% on \$2--\$10/day)\cite{Rangarajan2014Poverty}}
\begin{itemize}[noitemsep,leftmargin=*,topsep=2pt]
    \item \footnotesize{Societal harms: Poverty. Inadequate food, shelter, health, \& housing.}
    \item \footnotesize{Proxies: Spoken \& written language(s). Mother tongue. Literacy. Feature / Smart Phone Ownership. Rural vs. urban.}
    \item \footnotesize{Tech harms: Linguistic bias towards mainstream languages. Model bias towards middle class users. Limited or lack of internet access. }
\end{itemize}
\end{minipage}\vspace{2pt}\\
\midrule

\begin{minipage}[t]{\linewidth}
\textbf{Gender Identity \& Sexual Orientation}
\footnotesize{(No official LGBTQ+ data)} 
\begin{itemize}[noitemsep,leftmargin=*,topsep=2pt]
    \item \footnotesize{Societal harms: Discrimination and abuse. Lack of acceptance and visibility, despite the recent decriminalisation.\cite{tamang2020section}} 
    \item \footnotesize{Proxies: Gender declaration. Name. }%
    \item \footnotesize{Tech harms: FR "outing" and accuracy. Gender binary surveillance systems (\eg in dormitories). M/F ads targeting.  Catfishing and extortion abuse attacks.}
\end{itemize}
\end{minipage}\vspace{2pt}\\
\midrule

\begin{minipage}[t]{\linewidth}
\textbf{Ethnicity}
\footnotesize{(4\% NorthEast) \cite{CensusIndia2011}}
\begin{itemize}[noitemsep,leftmargin=*,topsep=2pt]
    \item \footnotesize{Societal harms: Racist slurs, discrimination, and physical attacks.}
    \item \footnotesize{Proxies: Skin tone. Facial features. Mother tongue. State. Name. }
    \item \footnotesize{Tech harms: Accuracy gap in FR. Online misrepresentation \& exclusion. Inaccurate inferences due to lifestyle, \eg migrant labor. }
\end{itemize} 
\end{minipage}\vspace{2pt}\\
\bottomrule
\end{tabularx}
\normalsize
\vspace{2pt}
\caption{\label{tab_subgroups} \small Axes of potential ML (un)fairness in India
}
\end{table}
\end{savenotes}

%% file: discussion.tex
\vspace{-2mm}

\section{Towards AI Fairness in India}
\begin{figure*}[ht!]
\includegraphics[width=.79\linewidth]{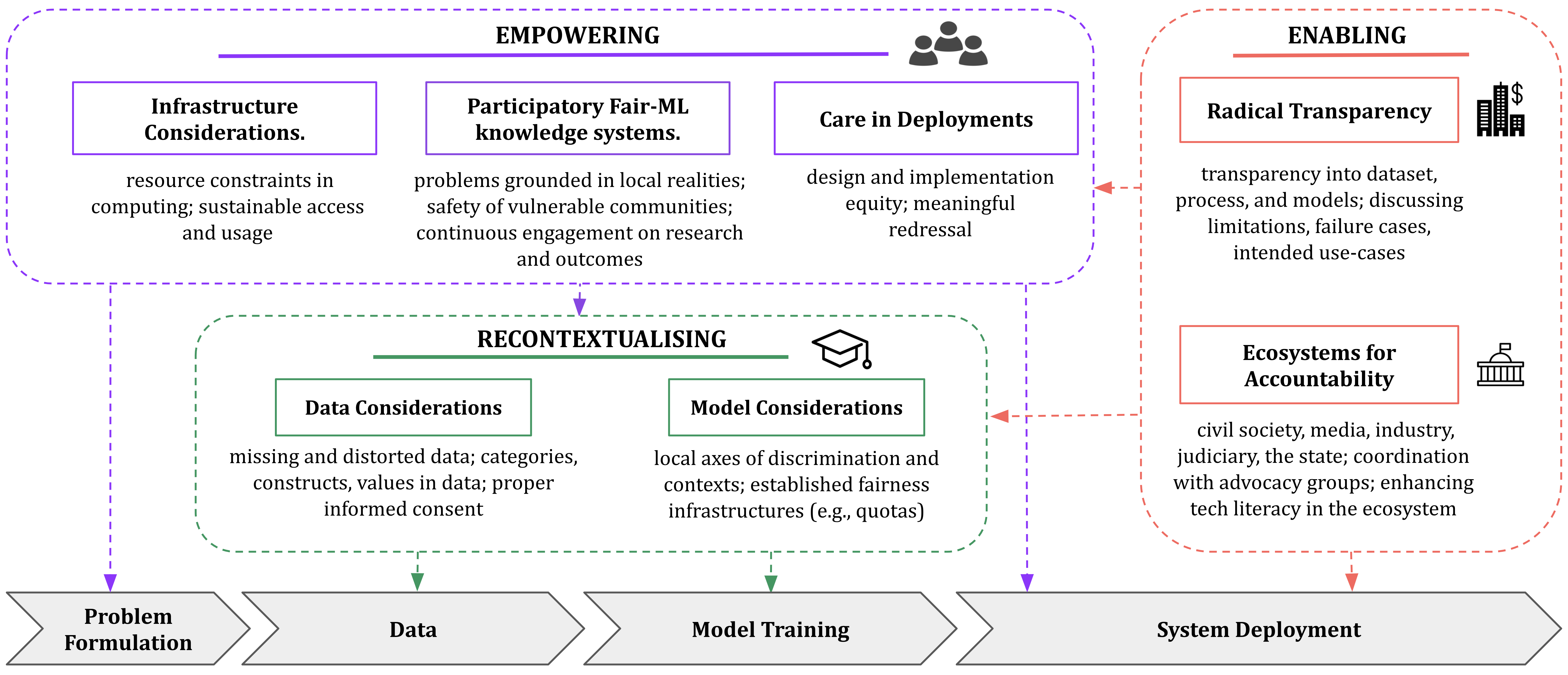}
\caption{Research pathways for Algorithmic Fairness in India.}
\label{fig:Indiaframework}
\end{figure*}

In Section~\ref{sec_findings}, we demonstrated that several underlying assumptions about algorithmic fairness and its enabling infrastructures fail in the Indian context. To meaningfully operationalise algorithmic fairness in India, we extend the spirit of Dr. B.R. Ambedkar's call to action, ``\textit{there cannot be political reform without social reform}'' \cite{ambedkar2014annihilation}. We need to substantively innovate on how to empower oppressed communities and enable justice within the surrounding socio-political infrastructures in India. Missing Indic factors and values in data and models, compounded by double standards and disjuncture in ML, deployed in an environment of unquestioning AI aspiration, face the risk of reinforcing injustices and harms. We need to understand and design for \textit{end-to-end} chains of algorithmic power, including how AI systems are conceived, produced, utilised, appropriated, assessed, and contested in India. We humbly submit that these are large, open-ended challenges that have perhaps not received much focus or are considered large in scope. However, a Fair-ML strategy for India needs to reflect its deeply plural, complex, and contradictory nature, and needs to go beyond model fairness. We propose an AI fairness research agenda in India, where we call for action along three critical and contingent pathways: \textit{Recontextualising}, \textit{Empowering}, and \textit{Enabling}. Figure~\ref{fig:Indiaframework} shows the core considerations of these pathways. The pathways present opportunities for cross-disciplinary and cross-institutional collaboration. While it is important to incorporate Indian concepts of fairness into AI that impacts Indian communities, the broader goal should be to develop general approaches to fairness that reflect the needs of local communities and are appropriate to the local contexts. 


\vspace{-2mm}
\subsection{Recontextualising Data and Models}

How might existing algorithmic fairness evaluation and mitigation approaches be \textit{recontextualised} to the Indian context, and what novel challenges does it give rise to? 

\subsubsection{Data considerations}\hfill\\
Data plays a critical role in measurements and mitigations of algorithmic bias. However, as seen in Section~\ref{sec_findings}, social, economic, and infrastructural factors challenge the reliance on datasets in India. Based on our findings, we outline some recommendations and put forth critical research questions regarding data and its uses in India.  

Due to the challenges to completeness and representation discussed in Section~\ref{sec_findings}, we must be (even more than usual) sceptical of Indian datasets until they are trust-worthy. For instance, how could fairness research handle the known data distortions guided by traditional gender roles? What are the risks in identifying caste in models? Should instances where models are deliberately `confused' (see Section~\ref{sec_findings}) be identified, and if so what should we do with such data? We must also account for data voids \cite{golebiewski2019data} for which statistical extrapolations might be invalid.

The vibrant role played by human infrastructures in providing, negotiating, collecting, and stewarding data points to new ways of looking at data as dialogue, rather than operations. Such data gathering via community relationships lead us to view data records as products of both beholder and the beheld. Building ties with community workers in the AI dataset collection process can be a starting point in creating high quality data, while respecting their situated knowledge. Combining observational research and dataset analysis will help us avoid misreadings of data. 
Normative frameworks (\eg{} perhaps ethics of care \cite{held2006ethics, asaro2019ai,zevenbergen2020}) may be relevant to take into account these social relations.
A related question is how data consent might work fairly in India. One approach could be to create transitive informed consent, built upon personal relationships and trust in data workers, with transparency on potential downstream applications. Ideas like collective consent \cite{MozillaF26:online}, data trusts \cite{Nesta}, and data co-ops may enhance community agency in datasets, while simultaneously increasing data reliability. 

Finally, at a fundamental level, we must question the categories and constructs we model in datasets, and how we measure them. As well as the situatedness of social categories such as gender (cf. \textit{Hijra}) and race \cite{hanna2020towards}, ontologies of affect (sentiment, inappropriateness, etc.), taboos (\textit{Halal}, revealing clothing, etc.), and social behaviours (headshakes, headwear, clothing, etc) are similarly contextual. How do we justify the ``knowing'' of social information by encoding it in data? We must also question if being ``data-driven'' is inconsistent with local values, goals and contexts. When data are appropriate for endemic goals (\eg{} caste membership for quotas), what form should their distributions and annotations take? Linguistically and culturally pluralistic communities should be given voices in these negotiations in ways that respect Indian norms of representation.

\subsubsection{Model and model (un)fairness considerations} \hfill\\
The prominent axes of historical injustices in India listed in Table~\ref{tab_subgroups} could be a starting point to detect and mitigate unfairness issues in trained models (\eg{}\cite{bolukbasi2016man,zhang2018mitigating}), alongside testing strategies, e.g. perturbation testing \cite{prabhakaran2019perturbation}, data augmentation \cite{zmigrod2019counterfactual}, adversarial testing \cite{kurakin2016adversarial}, 
and adherence to terminology guidelines for oppressed groups, such as \textsc{SIGACCESS}. However, it is important to note that operationalising fairness approaches from the West to these axes is often nontrivial. For instance, personal names act as a signifier for various socio-demographic attributes in India, however there are no large datasets of Indian names (like the US Census data, or the SSA data) that are readily available for fairness evaluations. In addition, disparities along the same axes may manifest very differently in India. For instance, gender disparities in the Indian workforce follow significantly different patterns compared to the West. How would an algorithm made fairer along gender based on datasets from the West be decontextualised and recontextualised for India? 

Another important consideration is how the algorithmic fairness interventions work with the existing infrastructures in India that surrounds decision making processes.
For instance, how do they work in the context of established fairness initiatives such as reservations/quotas? As an illustration, compared to the US undergraduate admission process of selection from a pool of candidates, the undergraduate admissions in India is done through various joint seat allocation processes, over hundreds of programmes, across dozens of universities and colleges that takes into account test scores, ordered preference lists provided by students, as well as various affirmative action quotas \cite{baswana2019centralized}. The quota system gives rise to the problem of matching under distributional constraints, a known problem in economics \cite{kamada2015efficient, goto2017designing, ashlagi2020assignment}, but has not received attention within the FAccT community (although \cite{cotter2019optimization} is related). First-order Fair-ML problems could include representational biases of caste and other sub-groups in NLP models, biases in Indic language NLP including challenges from code-mixing, Indian subgroup biases in computer vision, tackling online misinformation, benchmarking using Indic datasets, and fair allocation models in public welfare.

\vspace{-2mm}
\subsection{Empowering Communities}
Recontextualising data and models can only go so far without participatorily \textit{empowering} communities in identifying problems, specifying fairness expectations, and designing systems. 

\subsubsection{Participatory Fair-ML knowledge systems}\hfill{}\\
Context-free assumptions in Fair-ML, whether in homegrown or international AI systems, can not just fail, but produce harm inadvertently when applied to different infrastructural and cultural systems. As Mbembe describes the Western epistemic tradition, the knowing subject is enclosed in itself and produces supposedly objective knowledge of the world, ``without being part of that world, and he or she is by all accounts able to produce knowledge that is supposed to be universal and independent of context'' \cite{mbembe2015decolonizing}. How can we move away from the `common good' defined by the researcher---the supposedly all-knowing entity who has the expertise and experience necessary to identify what is of benefit to all \cite{nathan2017good}. Humbly creating grassroots commitments with vulnerable communities is an important first step. Our study discusses how caste, religion, and tribe are eluded even within the Indian technology discourse and policy. Modes of epistemic production in Fair-ML should enable marginalised communities to produce knowledge about themselves in the policies or designs. Grassroots efforts like Deep Learning Indaba \cite{dlindaba}
and Khipu \cite{khipu}
are exemplar of bootstrapping AI research in communities. Initiatives like Design Beku \cite{designbeku}
and SEWA \cite{sewa}
are excellent decolonial examples of participatorily co-designing with under-served communities. 

\subsubsection{Low-resource considerations} \hfill\\
India's heterogeneous literacies, economics, and infrastructures mean that Fair-ML researchers' commitment should go beyond model outputs, to deployments in accessible systems. Half the population of India is not online. Layers of the stack like interfaces, devices, connectivity, languages, and costs are important to ensure access. Learning from computing fields where constraints have been embraced as design material like \textsc{ICTD} \cite{toyama2015geek} and \textsc{HCI4D} \cite{dray2012human} can help, \eg{}via delay-tolerant connectivity, low cost devices, text-free interfaces, intermediaries, and NGO partnerships (see \cite{brewer2005case, heimerl2013local, kumar2015mobile, medhi2006text, sambasivan2010intermediated, gandhi2007digital, vivek2018technology}). 
Data infrastructures to build localised datasets would enhance access equity (\eg\cite{lacuna, sambasivan2021cascades}).


\subsubsection{First-world care in deployments}\hfill\\
Critiques were raised in our study on how neo-liberal AI followed a broader pattern of extraction from the `bottom billion' data subjects and labourers. Low costs, large and diverse populations, and policy infirmities have been cited as reasons for following double standards in India, \eg{}in non-consensual human trials and waste dumping \cite{macklin2004double} (also see \cite{mohamed2020decolonial}).  
Past disasters in India, like the fatal Union Carbide gas leak in 1984---one of the world's worst industrial accidents---point to faulty design and low quality, double standards for the `third world' \cite{Courttol47:online}.  
Similarly, unequal standards, inadequate safeguards, and dubious applications of AI in the non-West can lead to catastrophic effects (similar analogies have been made for content moderation \cite{roberts2016digital, sambasivan2019they}).
Fair-ML researchers should understand the systems into which they are embedding, engage with Indian realities and user feedback, and whether the recourse is meaningful. 

\vspace{-2mm}

\subsection{Enabling Fair-ML Ecosystems}

AI is increasingly perceived as a masculine, hype-filled, techno-utopian enterprise, with nations turning into AI superpowers \cite{Whoeverl94:online}. 
AI is even more aspirational and consequential in non-Western nations, where it performs an `efficiency' function in distributing scarce socio-economic resources and differentiating economies.
For Fair-ML research to be impactful and sustainable, it is crucial for researchers to \textit{enable} a critically conscious Fair-ML ecosystem. 

\subsubsection{Ecosystems for accountability}\hfill\\
Bootstrapping an ecosystem made up of civil society, media, industry, judiciary, and the state is necessary for accountability in Fair-ML (recall the US FR example). Moving from ivory tower research approaches to solidarity with various stakeholders through partnerships, evidence-based policy, and policy maker education can help create a sustainable Fair-ML ecosystem based on sound empirical and ethical norms, \eg{} we should consider research with algorithmic advocacy groups like Internet Freedom Foundation \cite{iff}, that have advanced landmark changes in net neutrality and privacy. Efforts like the AI Observatory to catalogue, understand harms, and demand accountability of automated decision support systems in India are crucial first steps \cite{AIobs}. Technology journalism is a keystone of equitable automation and needs to be fostered for AI. 

\subsubsection{Critical transparency}\hfill\\
Inscrutability suppresses algorithmic fairness. Besides the role played by ecosystem-wide regulations and standards, radical transparency should be espoused and enacted by Fair-ML researchers committed to India. 
Transparency on datasets, processes, and models (\eg{}\cite{mitchell2019model,gebru2018datasheets,bender2018data}), openly discussing limitations, failure cases, and lessons learnt 
can help move from the `magic pill' role of fairness as a checklist for ethical issues in India---to a more pragmatic, flawed, and evolving scientific function.

%% file: conclusion.tex
\vspace{-2mm}
\section{Conclusion}
As AI becomes global, algorithmic fairness naturally follows. Context matters. We must take care to not copy-paste the western-normative fairness everywhere.   
We presented a qualitative study and discourse analysis of algorithmic power in India, and found that algorithmic fairness assumptions are challenged in the Indian context. We found that data was not always reliable due to socio-economic factors, ML products for Indian users sufffer from double standards, and AI was seen with unquestioning aspiration. We called for an end-to-end re-imagining of algorithmic fairness that involves re-contextualising data and models, empowering oppressed communities, and enabling fairness ecosystems. The considerations we identified are certainly not limited to India; likewise, we call for inclusively evolving global approaches to Fair-ML. 

\vspace{-2mm}
\section{Acknowledgements}
Our thanks to the experts who shared their knowledge and wisdom with us: A. Aneesh, Aishwarya Lakshmiratan, Ameen Jauhar, Amit Sethi, Anil Joshi, Arindrajit Basu, Avinash Kumar, Chiranjeeb Bhattacharya, Dhruv Lakra, George Sebastian, Jacki O Neill, Mainack Mondal, Maya Indira Ganesh, Murali Shanmugavelan, Nandana Sengupta, Neha Kumar, Rahul De, Rahul Matthan, Rajesh Veeraraghavan, Ranjit Singh, Ryan Joseph Figueiredo (Equal Asia Foundation), Savita Bailur, Sayomdeb Mukerjee, Shanti Raghavan, Shyam Suri, Smita, Sriram Somanchi, Suraj Yengde, Vidushi Marda, and Vivek Srinivasan, and others who wish to stay anonymous. To Murali Shanmugavelan for educating us and connecting this paper to anti-caste emancipatory politics and theories. To Jose M. Faleiro, Daniel Russell, Jess Holbrook, Fernanda Viegas, Martin Wattenberg, Alex Hanna, and Reena Jana for their invaluable feedback.


%% file: main.bbl

\begin{thebibliography}{220}


\ifx \showCODEN    \undefined \def \showCODEN     #1{\unskip}     \fi
\ifx \showDOI      \undefined \def \showDOI       #1{#1}\fi
\ifx \showISBNx    \undefined \def \showISBNx     #1{\unskip}     \fi
\ifx \showISBNxiii \undefined \def \showISBNxiii  #1{\unskip}     \fi
\ifx \showISSN     \undefined \def \showISSN      #1{\unskip}     \fi
\ifx \showLCCN     \undefined \def \showLCCN      #1{\unskip}     \fi
\ifx \shownote     \undefined \def \shownote      #1{#1}          \fi
\ifx \showarticletitle \undefined \def \showarticletitle #1{#1}   \fi
\ifx \showURL      \undefined \def \showURL       {\relax}        \fi
\providecommand\bibfield[2]{#2}
\providecommand\bibinfo[2]{#2}
\providecommand\natexlab[1]{#1}
\providecommand\showeprint[2][]{arXiv:#2}

\bibitem[\protect\citeauthoryear{??}{NIT}{2018}]%
        {NITI}
 \bibinfo{year}{2018}\natexlab{}.
\newblock \bibinfo{booktitle}{\emph{National Strategy for Artificial
  Intelligence \#AI4ALL}}.
\newblock \bibinfo{publisher}{Niti Aayog}.
\newblock


\bibitem[\protect\citeauthoryear{??}{cit}{2020}]%
        {citizencop}
 \bibinfo{year}{2020}\natexlab{}.
\newblock \bibinfo{booktitle}{\emph{Citizen COP Foundation}}.
\newblock
\urldef\tempurl%
\url{https://www.citizencop.org}
\showURL{%
\tempurl}


\bibitem[\protect\citeauthoryear{??}{dli}{2020}]%
        {dlindaba}
 \bibinfo{year}{2020}\natexlab{}.
\newblock \bibinfo{booktitle}{\emph{Deep Learning Indaba}}.
\newblock
\urldef\tempurl%
\url{https://deeplearningindaba.com/2020/}
\showURL{%
\tempurl}


\bibitem[\protect\citeauthoryear{??}{des}{2020}]%
        {designbeku}
 \bibinfo{year}{2020}\natexlab{}.
\newblock \bibinfo{booktitle}{\emph{Design Beku}}.
\newblock
\urldef\tempurl%
\url{https://designbeku.in/}
\showURL{%
\tempurl}


\bibitem[\protect\citeauthoryear{??}{tec}{2020}]%
        {techcrunch}
 \bibinfo{year}{2020}\natexlab{}.
\newblock \bibinfo{title}{India used facial recognition tech to identify 1,100
  individuals at a recent riot | TechCrunch}.
\newblock
  \bibinfo{howpublished}{\url{https://techcrunch.com/2020/03/11/india-used-facial-recognition-tech-to-identify-1100-individuals-at-a-recent-riot}}.
\newblock
\newblock
\shownote{(Accessed on 07/28/2020).}


\bibitem[\protect\citeauthoryear{??}{iff}{2020}]%
        {iff}
 \bibinfo{year}{2020}\natexlab{}.
\newblock \bibinfo{booktitle}{\emph{Internet Freedom Foundation}}.
\newblock
\urldef\tempurl%
\url{https://internetfreedom.in/}
\showURL{%
\tempurl}


\bibitem[\protect\citeauthoryear{??}{khi}{2020}]%
        {khipu}
 \bibinfo{year}{2020}\natexlab{}.
\newblock \bibinfo{booktitle}{\emph{Khipu AI}}.
\newblock
\urldef\tempurl%
\url{https://github.com/khipu-ai}
\showURL{%
\tempurl}


\bibitem[\protect\citeauthoryear{??}{lac}{2020}]%
        {lacuna}
 \bibinfo{year}{2020}\natexlab{}.
\newblock \bibinfo{booktitle}{\emph{Lacuna Fund}}.
\newblock
\urldef\tempurl%
\url{https://lacunafund.org/}
\showURL{%
\tempurl}


\bibitem[\protect\citeauthoryear{??}{saf}{2020}]%
        {safetipin}
 \bibinfo{year}{2020}\natexlab{}.
\newblock \bibinfo{booktitle}{\emph{Safetipin}}.
\newblock
\urldef\tempurl%
\url{https://safetipin.com/}
\showURL{%
\tempurl}


\bibitem[\protect\citeauthoryear{??}{sew}{2020}]%
        {sewa}
 \bibinfo{year}{2020}\natexlab{}.
\newblock \bibinfo{booktitle}{\emph{SEWA}}.
\newblock
\urldef\tempurl%
\url{http://www.sewa.org/}
\showURL{%
\tempurl}


\bibitem[\protect\citeauthoryear{Abraham and Rao}{Abraham and Rao}{[n.d.]}]%
        {84DeadIn98:online}
\bibfield{author}{\bibinfo{person}{Delna Abraham} {and} \bibinfo{person}{Ojaswi
  Rao}.} \bibinfo{year}{[n.d.]}\natexlab{}.
\newblock \bibinfo{title}{84\% Dead In Cow-Related Violence Since 2010 Are
  Muslim; 97\% Attacks After 2014 | IndiaSpend}.
\newblock
  \bibinfo{howpublished}{\url{https://archive.indiaspend.com/cover-story/86-dead-in-cow-related-violence-since-2010-are-muslim-97-attacks-after-2014-2014}}.
\newblock
\newblock
\shownote{(Accessed on 08/16/2020).}


\bibitem[\protect\citeauthoryear{Agarwal, Yang, Wallace, and Nenkova}{Agarwal
  et~al\mbox{.}}{2020}]%
        {agarwal2020entity}
\bibfield{author}{\bibinfo{person}{Oshin Agarwal}, \bibinfo{person}{Yinfei
  Yang}, \bibinfo{person}{Byron~C Wallace}, {and} \bibinfo{person}{Ani
  Nenkova}.} \bibinfo{year}{2020}\natexlab{}.
\newblock \showarticletitle{Entity-Switched Datasets: An Approach to Auditing
  the In-Domain Robustness of Named Entity Recognition Models}.
\newblock \bibinfo{journal}{\emph{arXiv preprint arXiv:2004.04123}}
  (\bibinfo{year}{2020}).
\newblock


\bibitem[\protect\citeauthoryear{(Agesic)}{(Agesic)}{2019}]%
        {UruguayAI}
\bibfield{author}{\bibinfo{person}{Digital Government~Agency (Agesic)}.}
  \bibinfo{year}{2019}\natexlab{}.
\newblock \showarticletitle{Artificial Intelligence for the digital government
  | English version}.
  \bibinfo{howpublished}{\url{https://www.gub.uy/agencia-gobierno-electronico-sociedad-informacion-conocimiento/sites/agencia-gobierno-electronico-sociedad-informacion-conocimiento/files/documentos/publicaciones/IA\%20Strategy\%20-20english\%20version.pdf}}.
  In \bibinfo{booktitle}{\emph{AI whitepaper.}}
\newblock


\bibitem[\protect\citeauthoryear{Aggarwal}{Aggarwal}{2018}]%
        {India'sm28:online}
\bibfield{author}{\bibinfo{person}{Varun Aggarwal}.}
  \bibinfo{year}{2018}\natexlab{}.
\newblock \bibinfo{title}{India’s mess of complexity is just what AI needs |
  MIT Technology Review}.
\newblock
  \bibinfo{howpublished}{\url{https://www.technologyreview.com/2018/06/27/240474/indias-mess-of-complexity-is-just-what-ai-needs/}}.
\newblock
\newblock
\shownote{(Accessed on 09/18/2020).}


\bibitem[\protect\citeauthoryear{Agrawal}{Agrawal}{2020}]%
        {ChutiaCh43:online}
\bibfield{author}{\bibinfo{person}{Saumya Agrawal}.}
  \bibinfo{year}{2020}\natexlab{}.
\newblock \bibinfo{title}{Chutia| 'Chutia not slang, but community where I
  belong': Assam woman's online job application rejected due to surname |
  Trending \& Viral News}.
\newblock
  \bibinfo{howpublished}{\url{https://www.timesnownews.com/the-buzz/article/chutia-not-slang-but-community-where-i-belong-assam-womans-online-job-application-rejected-due-to-surname/625556}}.
\newblock
\newblock
\shownote{(Accessed on 09/28/2020).}


\bibitem[\protect\citeauthoryear{Amazon}{Amazon}{2020}]%
        {Weareimp72:online}
\bibfield{author}{\bibinfo{person}{Amazon}.} \bibinfo{year}{2020}\natexlab{}.
\newblock \bibinfo{title}{We are implementing a one-year moratorium on police
  use of Rekognition}.
\newblock
  \bibinfo{howpublished}{\url{https://blog.aboutamazon.com/policy/we-are-implementing-a-one-year-moratorium-on-police-use-of-rekognition}}.
\newblock
\newblock
\shownote{(Accessed on 08/29/2020).}


\bibitem[\protect\citeauthoryear{Ambedkar}{Ambedkar}{1916}]%
        {ambedkar1916castes}
\bibfield{author}{\bibinfo{person}{BR Ambedkar}.}
  \bibinfo{year}{1916}\natexlab{}.
\newblock \showarticletitle{Castes in India: Their mechanism, genesis and
  development (Vol. 1)}.
\newblock \bibinfo{journal}{\emph{Columbia: Indian Antiquary. Ambedkar, BR
  (1936). Annihilation of Caste. Jullundur: Bheem Patrika Publications}}
  (\bibinfo{year}{1916}).
\newblock


\bibitem[\protect\citeauthoryear{Ambedkar}{Ambedkar}{2014}]%
        {ambedkar2014annihilation}
\bibfield{author}{\bibinfo{person}{Bhimrao~Ramji Ambedkar}.}
  \bibinfo{year}{2014}\natexlab{}.
\newblock \bibinfo{booktitle}{\emph{Annihilation of caste: The annotated
  critical edition}}.
\newblock \bibinfo{publisher}{Verso Books}.
\newblock


\bibitem[\protect\citeauthoryear{Angwin, Larson, Mattu, and Kirchner}{Angwin
  et~al\mbox{.}}{2016}]%
        {MachineB33:online}
\bibfield{author}{\bibinfo{person}{Julia Angwin}, \bibinfo{person}{Jeff
  Larson}, \bibinfo{person}{Surya Mattu}, {and} \bibinfo{person}{Lauren
  Kirchner}.} \bibinfo{year}{2016}\natexlab{}.
\newblock \bibinfo{title}{Machine Bias — ProPublica}.
\newblock
  \bibinfo{howpublished}{\url{https://www.propublica.org/article/machine-bias-risk-assessments-in-criminal-sentencing}}.
\newblock
\newblock
\shownote{(Accessed on 07/30/2020).}


\bibitem[\protect\citeauthoryear{Appadurai}{Appadurai}{2000}]%
        {appadurai2000spectral}
\bibfield{author}{\bibinfo{person}{Arjun Appadurai}.}
  \bibinfo{year}{2000}\natexlab{}.
\newblock \showarticletitle{Spectral housing and urban cleansing: notes on
  millennial Mumbai}.
\newblock \bibinfo{journal}{\emph{Public culture}} \bibinfo{volume}{12},
  \bibinfo{number}{3} (\bibinfo{year}{2000}), \bibinfo{pages}{627--651}.
\newblock


\bibitem[\protect\citeauthoryear{Arora}{Arora}{2016}]%
        {arora2016bottom}
\bibfield{author}{\bibinfo{person}{Payal Arora}.}
  \bibinfo{year}{2016}\natexlab{}.
\newblock \showarticletitle{Bottom of the data pyramid: Big data and the global
  south}.
\newblock \bibinfo{journal}{\emph{International Journal of Communication}}
  \bibinfo{volume}{10} (\bibinfo{year}{2016}), \bibinfo{pages}{19}.
\newblock


\bibitem[\protect\citeauthoryear{Asaro}{Asaro}{2019}]%
        {asaro2019ai}
\bibfield{author}{\bibinfo{person}{Peter~M Asaro}.}
  \bibinfo{year}{2019}\natexlab{}.
\newblock \showarticletitle{AI ethics in predictive policing: From models of
  threat to an ethics of care}.
\newblock \bibinfo{journal}{\emph{IEEE Technology and Society Magazine}}
  \bibinfo{volume}{38}, \bibinfo{number}{2} (\bibinfo{year}{2019}),
  \bibinfo{pages}{40--53}.
\newblock


\bibitem[\protect\citeauthoryear{Ashlagi, Saberi, and Shameli}{Ashlagi
  et~al\mbox{.}}{2020}]%
        {ashlagi2020assignment}
\bibfield{author}{\bibinfo{person}{Itai Ashlagi}, \bibinfo{person}{Amin
  Saberi}, {and} \bibinfo{person}{Ali Shameli}.}
  \bibinfo{year}{2020}\natexlab{}.
\newblock \showarticletitle{Assignment mechanisms under distributional
  constraints}.
\newblock \bibinfo{journal}{\emph{Operations Research}} \bibinfo{volume}{68},
  \bibinfo{number}{2} (\bibinfo{year}{2020}), \bibinfo{pages}{467--479}.
\newblock


\bibitem[\protect\citeauthoryear{Bailur, Srivastava, and Smertnik}{Bailur
  et~al\mbox{.}}{2019}]%
        {Womenand71:online}
\bibfield{author}{\bibinfo{person}{Savita Bailur}, \bibinfo{person}{Devina
  Srivastava}, {and} \bibinfo{person}{H\'{e}l\`{e}ne (Caribou~Digital)
  Smertnik}.} \bibinfo{year}{2019}\natexlab{}.
\newblock \bibinfo{title}{Women and ID in a digital age: Five fundamental
  barriers and new design questions}.
\newblock
  \bibinfo{howpublished}{\url{https://savitabailur.com/2019/09/09/women-and-id-in-a-digital-age-five-fundamental-barriers-and-new-design-questions/}}.
\newblock
\newblock
\shownote{(Accessed on 08/02/2020).}


\bibitem[\protect\citeauthoryear{Baker}{Baker}{2001}]%
        {baker_2001}
\bibfield{author}{\bibinfo{person}{Robert Baker}.}
  \bibinfo{year}{2001}\natexlab{}.
\newblock \showarticletitle{Bioethics and Human Rights: A Historical
  Perspective}.
\newblock \bibinfo{journal}{\emph{Cambridge Quarterly of Healthcare Ethics}}
  \bibinfo{volume}{10}, \bibinfo{number}{3} (\bibinfo{year}{2001}),
  \bibinfo{pages}{241–252}.
\newblock
\urldef\tempurl%
\url{https://doi.org/10.1017/S0963180101003048}
\showDOI{\tempurl}


\bibitem[\protect\citeauthoryear{Banaji and Bhat}{Banaji and Bhat}{2019}]%
        {whatsapp}
\bibfield{author}{\bibinfo{person}{Shakuntala Banaji} {and}
  \bibinfo{person}{Ram Bhat}.} \bibinfo{year}{2019}\natexlab{}.
\newblock \bibinfo{booktitle}{\emph{WhatsApp Vigilantes: An exploration of
  citizen reception and circulation of WhatsApp misinformation linked to mob
  violence in India}}.
\newblock \bibinfo{publisher}{Department of Media and Communications, LSE}.
\newblock


\bibitem[\protect\citeauthoryear{Banerjee, Bertrand, Datta, and
  Mullainathan}{Banerjee et~al\mbox{.}}{2009}]%
        {banerjee2009labor}
\bibfield{author}{\bibinfo{person}{Abhijit Banerjee}, \bibinfo{person}{Marianne
  Bertrand}, \bibinfo{person}{Saugato Datta}, {and} \bibinfo{person}{Sendhil
  Mullainathan}.} \bibinfo{year}{2009}\natexlab{}.
\newblock \showarticletitle{Labor market discrimination in Delhi: Evidence from
  a field experiment}.
\newblock \bibinfo{journal}{\emph{Journal of comparative Economics}}
  \bibinfo{volume}{37}, \bibinfo{number}{1} (\bibinfo{year}{2009}),
  \bibinfo{pages}{14--27}.
\newblock


\bibitem[\protect\citeauthoryear{Barik}{Barik}{2020}]%
        {Facialre30:online}
\bibfield{author}{\bibinfo{person}{Soumyarendra Barik}.}
  \bibinfo{year}{2020}\natexlab{}.
\newblock \bibinfo{title}{Facial recognition based surveillance systems to be
  installed at 983 railway stations across India}.
\newblock
  \bibinfo{howpublished}{\url{https://www.medianama.com/2020/01/223-facial-recognition-system-indian-railways-facial-recognition/}}.
\newblock
\newblock
\shownote{(Accessed on 10/03/2020).}


\bibitem[\protect\citeauthoryear{Barocas, Hardt, and Narayanan}{Barocas
  et~al\mbox{.}}{2017}]%
        {barocas2017fairness}
\bibfield{author}{\bibinfo{person}{Solon Barocas}, \bibinfo{person}{Moritz
  Hardt}, {and} \bibinfo{person}{Arvind Narayanan}.}
  \bibinfo{year}{2017}\natexlab{}.
\newblock \showarticletitle{Fairness in machine learning}.
\newblock \bibinfo{journal}{\emph{NIPS Tutorial}}  \bibinfo{volume}{1}
  (\bibinfo{year}{2017}).
\newblock


\bibitem[\protect\citeauthoryear{Barocas and Selbst}{Barocas and
  Selbst}{2016}]%
        {barocas2016big}
\bibfield{author}{\bibinfo{person}{Solon Barocas} {and}
  \bibinfo{person}{Andrew~D Selbst}.} \bibinfo{year}{2016}\natexlab{}.
\newblock \showarticletitle{Big data's disparate impact}.
\newblock \bibinfo{journal}{\emph{Calif. L. Rev.}}  \bibinfo{volume}{104}
  (\bibinfo{year}{2016}), \bibinfo{pages}{671}.
\newblock


\bibitem[\protect\citeauthoryear{Baswana, Chakrabarti, Chandran, Kanoria, and
  Patange}{Baswana et~al\mbox{.}}{2019}]%
        {baswana2019centralized}
\bibfield{author}{\bibinfo{person}{Surender Baswana},
  \bibinfo{person}{Partha~Pratim Chakrabarti}, \bibinfo{person}{Sharat
  Chandran}, \bibinfo{person}{Yashodhan Kanoria}, {and}
  \bibinfo{person}{Utkarsh Patange}.} \bibinfo{year}{2019}\natexlab{}.
\newblock \showarticletitle{Centralized admissions for engineering colleges in
  India}.
\newblock \bibinfo{journal}{\emph{INFORMS Journal on Applied Analytics}}
  \bibinfo{volume}{49}, \bibinfo{number}{5} (\bibinfo{year}{2019}),
  \bibinfo{pages}{338--354}.
\newblock


\bibitem[\protect\citeauthoryear{Baxi}{Baxi}{2018}]%
        {PredictiveIndia}
\bibfield{author}{\bibinfo{person}{Abhishek Baxi}.}
  \bibinfo{year}{2018}\natexlab{}.
\newblock \bibinfo{title}{Law Enforcement Agencies In India Are Using
  Artificial Intelligence To Nab Criminals}.
\newblock
  \bibinfo{howpublished}{\url{https://www.forbes.com/sites/baxiabhishek/2018/09/28/law-enforcement-agencies-in-india-are-using-artificial-intelligence-to-nab-criminals-heres-how}}.
\newblock
\newblock
\shownote{(Accessed on 08/30/2020).}


\bibitem[\protect\citeauthoryear{BBC}{BBC}{2019}]%
        {Nirbhaya74:online}
\bibfield{author}{\bibinfo{person}{BBC}.} \bibinfo{year}{2019}\natexlab{}.
\newblock \bibinfo{title}{Nirbhaya case: Four Indian men executed for 2012
  Delhi bus rape and murder - BBC News}.
\newblock
  \bibinfo{howpublished}{\url{https://www.bbc.com/news/world-asia-india-51969961}}.
\newblock
\newblock
\shownote{(Accessed on 09/01/2020).}


\bibitem[\protect\citeauthoryear{Bender and Friedman}{Bender and
  Friedman}{2018}]%
        {bender2018data}
\bibfield{author}{\bibinfo{person}{Emily~M Bender} {and} \bibinfo{person}{Batya
  Friedman}.} \bibinfo{year}{2018}\natexlab{}.
\newblock \showarticletitle{Data statements for natural language processing:
  Toward mitigating system bias and enabling better science}.
\newblock \bibinfo{journal}{\emph{Transactions of the Association for
  Computational Linguistics}}  \bibinfo{volume}{6} (\bibinfo{year}{2018}),
  \bibinfo{pages}{587--604}.
\newblock


\bibitem[\protect\citeauthoryear{Beteille}{Beteille}{1990}]%
        {beteille1990race}
\bibfield{author}{\bibinfo{person}{Andre Beteille}.}
  \bibinfo{year}{1990}\natexlab{}.
\newblock \showarticletitle{Race, caste and gender}.
\newblock \bibinfo{journal}{\emph{Man}} (\bibinfo{year}{1990}),
  \bibinfo{pages}{489--504}.
\newblock


\bibitem[\protect\citeauthoryear{Bharathi, Malghan, and Rahman}{Bharathi
  et~al\mbox{.}}{2018}]%
        {bharathi2018isolated}
\bibfield{author}{\bibinfo{person}{Naveen Bharathi}, \bibinfo{person}{Deepak~V
  Malghan}, {and} \bibinfo{person}{Andaleeb Rahman}.}
  \bibinfo{year}{2018}\natexlab{}.
\newblock \showarticletitle{Isolated by caste: Neighbourhood-scale residential
  segregation in Indian metros}.
\newblock \bibinfo{journal}{\emph{IIM Bangalore Research Paper}}
  \bibinfo{number}{572} (\bibinfo{year}{2018}).
\newblock


\bibitem[\protect\citeauthoryear{Bhonsle and Prasad}{Bhonsle and
  Prasad}{2020}]%
        {Counting37:online}
\bibfield{author}{\bibinfo{person}{Anubha Bhonsle} {and}
  \bibinfo{person}{Pallavi Prasad}.} \bibinfo{year}{2020}\natexlab{}.
\newblock \bibinfo{title}{Counting cows, not rural health indicators}.
\newblock
  \bibinfo{howpublished}{\url{https://ruralindiaonline.org/articles/counting-cows-not-rural-health-indicators/}}.
\newblock
\newblock
\shownote{(Accessed on 08/02/2020).}


\bibitem[\protect\citeauthoryear{Binns}{Binns}{2018}]%
        {binns2018fairness}
\bibfield{author}{\bibinfo{person}{Reuben Binns}.}
  \bibinfo{year}{2018}\natexlab{}.
\newblock \showarticletitle{Fairness in machine learning: Lessons from
  political philosophy}. In \bibinfo{booktitle}{\emph{Conference on Fairness,
  Accountability and Transparency}}. \bibinfo{pages}{149--159}.
\newblock


\bibitem[\protect\citeauthoryear{Birhane}{Birhane}{2020}]%
        {birhane2020algorithmic}
\bibfield{author}{\bibinfo{person}{Abeba Birhane}.}
  \bibinfo{year}{2020}\natexlab{}.
\newblock \showarticletitle{Algorithmic colonization of Africa}.
\newblock \bibinfo{journal}{\emph{SCRIPTed}}  \bibinfo{volume}{17}
  (\bibinfo{year}{2020}), \bibinfo{pages}{389}.
\newblock


\bibitem[\protect\citeauthoryear{Blake, McAuliffe, Corbit, Callaghan, Barry,
  Bowie, Kleutsch, Kramer, Ross, Vongsachang, et~al\mbox{.}}{Blake
  et~al\mbox{.}}{2015}]%
        {blake2015ontogeny}
\bibfield{author}{\bibinfo{person}{PR Blake}, \bibinfo{person}{K McAuliffe},
  \bibinfo{person}{J Corbit}, \bibinfo{person}{TC Callaghan},
  \bibinfo{person}{O Barry}, \bibinfo{person}{A Bowie}, \bibinfo{person}{L
  Kleutsch}, \bibinfo{person}{KL Kramer}, \bibinfo{person}{E Ross},
  \bibinfo{person}{H Vongsachang}, {et~al\mbox{.}}}
  \bibinfo{year}{2015}\natexlab{}.
\newblock \showarticletitle{The ontogeny of fairness in seven societies}.
\newblock \bibinfo{journal}{\emph{Nature}} \bibinfo{volume}{528},
  \bibinfo{number}{7581} (\bibinfo{year}{2015}), \bibinfo{pages}{258--261}.
\newblock


\bibitem[\protect\citeauthoryear{Bogner, Littig, and Menz}{Bogner
  et~al\mbox{.}}{2009}]%
        {bogner2009interviewing}
\bibfield{author}{\bibinfo{person}{Alexander Bogner}, \bibinfo{person}{Beate
  Littig}, {and} \bibinfo{person}{Wolfgang Menz}.}
  \bibinfo{year}{2009}\natexlab{}.
\newblock \bibinfo{booktitle}{\emph{Interviewing experts}}.
\newblock \bibinfo{publisher}{Springer}.
\newblock


\bibitem[\protect\citeauthoryear{Bolukbasi, Chang, Zou, Saligrama, and
  Kalai}{Bolukbasi et~al\mbox{.}}{2016}]%
        {bolukbasi2016man}
\bibfield{author}{\bibinfo{person}{Tolga Bolukbasi}, \bibinfo{person}{Kai-Wei
  Chang}, \bibinfo{person}{James~Y Zou}, \bibinfo{person}{Venkatesh Saligrama},
  {and} \bibinfo{person}{Adam~T Kalai}.} \bibinfo{year}{2016}\natexlab{}.
\newblock \showarticletitle{Man is to computer programmer as woman is to
  homemaker? debiasing word embeddings}. In \bibinfo{booktitle}{\emph{Advances
  in neural information processing systems}}. \bibinfo{pages}{4349--4357}.
\newblock


\bibitem[\protect\citeauthoryear{Borooah, Dubey, and Iyer}{Borooah
  et~al\mbox{.}}{2007}]%
        {borooah2007effectiveness}
\bibfield{author}{\bibinfo{person}{Vani~K Borooah}, \bibinfo{person}{Amaresh
  Dubey}, {and} \bibinfo{person}{Sriya Iyer}.} \bibinfo{year}{2007}\natexlab{}.
\newblock \showarticletitle{The effectiveness of jobs reservation: caste,
  religion and economic status in India}.
\newblock \bibinfo{journal}{\emph{Development and change}}
  \bibinfo{volume}{38}, \bibinfo{number}{3} (\bibinfo{year}{2007}),
  \bibinfo{pages}{423--445}.
\newblock


\bibitem[\protect\citeauthoryear{Boyes-Watson}{Boyes-Watson}{2014}]%
        {boyes2014suffolk}
\bibfield{author}{\bibinfo{person}{C Boyes-Watson}.}
  \bibinfo{year}{2014}\natexlab{}.
\newblock \showarticletitle{Suffolk University, College of Arts \& Sciences}.
\newblock \bibinfo{journal}{\emph{Center for Restorative Justice. Retrieved on
  November}}  \bibinfo{volume}{28} (\bibinfo{year}{2014}),
  \bibinfo{pages}{2015}.
\newblock


\bibitem[\protect\citeauthoryear{Brewer, Demmer, Du, Ho, Kam, Nedevschi, Pal,
  Patra, Surana, and Fall}{Brewer et~al\mbox{.}}{2005}]%
        {brewer2005case}
\bibfield{author}{\bibinfo{person}{Eric Brewer}, \bibinfo{person}{Michael
  Demmer}, \bibinfo{person}{Bowei Du}, \bibinfo{person}{Melissa Ho},
  \bibinfo{person}{Matthew Kam}, \bibinfo{person}{Sergiu Nedevschi},
  \bibinfo{person}{Joyojeet Pal}, \bibinfo{person}{Rabin Patra},
  \bibinfo{person}{Sonesh Surana}, {and} \bibinfo{person}{Kevin Fall}.}
  \bibinfo{year}{2005}\natexlab{}.
\newblock \showarticletitle{The case for technology in developing regions}.
\newblock \bibinfo{journal}{\emph{Computer}} \bibinfo{volume}{38},
  \bibinfo{number}{6} (\bibinfo{year}{2005}), \bibinfo{pages}{25--38}.
\newblock


\bibitem[\protect\citeauthoryear{Buolamwini and Gebru}{Buolamwini and
  Gebru}{2018}]%
        {buolamwini2018gender}
\bibfield{author}{\bibinfo{person}{Joy Buolamwini} {and}
  \bibinfo{person}{Timnit Gebru}.} \bibinfo{year}{2018}\natexlab{}.
\newblock \showarticletitle{Gender shades: Intersectional accuracy disparities
  in commercial gender classification}. In \bibinfo{booktitle}{\emph{Conference
  on fairness, accountability and transparency}}. \bibinfo{pages}{77--91}.
\newblock


\bibitem[\protect\citeauthoryear{Central Statistics~Office and
  Implementation}{Central Statistics~Office and Implementation}{2018}]%
        {IndiaMinistryStatistics2018}
\bibfield{author}{\bibinfo{person}{Ministry of~Statistics Central
  Statistics~Office} {and} \bibinfo{person}{Programme Implementation}.}
  \bibinfo{year}{2018}\natexlab{}.
\newblock \bibinfo{booktitle}{\emph{Women and Men in India: A statistical
  compilation of Gender related Indicators in India}}.
\newblock \bibinfo{type}{{T}echnical {R}eport}.
  \bibinfo{institution}{Government of India}.
\newblock


\bibitem[\protect\citeauthoryear{Chakravarti}{Chakravarti}{1993}]%
        {chakravarti1993conceptualising}
\bibfield{author}{\bibinfo{person}{Uma Chakravarti}.}
  \bibinfo{year}{1993}\natexlab{}.
\newblock \showarticletitle{Conceptualising Brahmanical patriarchy in early
  India: Gender, caste, class and state}.
\newblock \bibinfo{journal}{\emph{Economic and Political Weekly}}
  (\bibinfo{year}{1993}), \bibinfo{pages}{579--585}.
\newblock


\bibitem[\protect\citeauthoryear{Chaudhuri}{Chaudhuri}{2004}]%
        {chaudhuri2004feminism}
\bibfield{author}{\bibinfo{person}{Maitrayee Chaudhuri}.}
  \bibinfo{year}{2004}\natexlab{}.
\newblock \showarticletitle{Feminism in India}.
\newblock  (\bibinfo{year}{2004}).
\newblock


\bibitem[\protect\citeauthoryear{Clark}{Clark}{2013}]%
        {ZIPCodeH5:online}
\bibfield{author}{\bibinfo{person}{Anna Clark}.}
  \bibinfo{year}{2013}\natexlab{}.
\newblock \bibinfo{title}{ZIP Code History: How They Define Us | The New
  Republic}.
\newblock
  \bibinfo{howpublished}{\url{https://newrepublic.com/article/112558/zip-code-history-how-they-define-us}}.
\newblock
\newblock
\shownote{(Accessed on 09/24/2020).}


\bibitem[\protect\citeauthoryear{Cotter, Jiang, Gupta, Wang, Narayan, You, and
  Sridharan}{Cotter et~al\mbox{.}}{2019}]%
        {cotter2019optimization}
\bibfield{author}{\bibinfo{person}{Andrew Cotter}, \bibinfo{person}{Heinrich
  Jiang}, \bibinfo{person}{Maya~R Gupta}, \bibinfo{person}{Serena Wang},
  \bibinfo{person}{Taman Narayan}, \bibinfo{person}{Seungil You}, {and}
  \bibinfo{person}{Karthik Sridharan}.} \bibinfo{year}{2019}\natexlab{}.
\newblock \showarticletitle{Optimization with Non-Differentiable Constraints
  with Applications to Fairness, Recall, Churn, and Other Goals.}
\newblock \bibinfo{journal}{\emph{Journal of Machine Learning Research}}
  \bibinfo{volume}{20}, \bibinfo{number}{172} (\bibinfo{year}{2019}),
  \bibinfo{pages}{1--59}.
\newblock


\bibitem[\protect\citeauthoryear{Crawford}{Crawford}{2013a}]%
        {crawford2013hidden}
\bibfield{author}{\bibinfo{person}{Kate Crawford}.}
  \bibinfo{year}{2013}\natexlab{a}.
\newblock \showarticletitle{The hidden biases in big data}.
\newblock \bibinfo{journal}{\emph{Harvard business review}}
  \bibinfo{volume}{1}, \bibinfo{number}{1} (\bibinfo{year}{2013}),
  \bibinfo{pages}{814}.
\newblock


\bibitem[\protect\citeauthoryear{Crawford}{Crawford}{2013b}]%
        {crawford2013think}
\bibfield{author}{\bibinfo{person}{Kate Crawford}.}
  \bibinfo{year}{2013}\natexlab{b}.
\newblock \showarticletitle{Think again: Big data}.
\newblock \bibinfo{journal}{\emph{Foreign Policy}}  \bibinfo{volume}{9}
  (\bibinfo{year}{2013}).
\newblock


\bibitem[\protect\citeauthoryear{Crumpler}{Crumpler}{2020}]%
        {HowAccur35:online}
\bibfield{author}{\bibinfo{person}{William Crumpler}.}
  \bibinfo{year}{2020}\natexlab{}.
\newblock \bibinfo{title}{How Accurate are Facial Recognition Systems – and
  Why Does It Matter? | Center for Strategic and International Studies}.
\newblock
\newblock
\newblock
\shownote{(Accessed on 07/28/2020).}


\bibitem[\protect\citeauthoryear{Culture}{Culture}{2018}]%
        {Economic79:online}
\bibfield{author}{\bibinfo{person}{Camera Culture}.}
  \bibinfo{year}{2018}\natexlab{}.
\newblock \bibinfo{title}{Economic Impact of Discoverability of Localities and
  Addresses in India — Emerging Worlds}.
\newblock
  \bibinfo{howpublished}{\url{http://mitemergingworlds.com/blog/2018/2/12/economic-impact-of-discoverability-of-localities-and-addresses-in-india}}.
\newblock
\newblock
\shownote{(Accessed on 09/24/2020).}


\bibitem[\protect\citeauthoryear{Dahir}{Dahir}{2019}]%
        {Mobilelo46:online}
\bibfield{author}{\bibinfo{person}{Abdi~Lahir Dahir}.}
  \bibinfo{year}{2019}\natexlab{}.
\newblock \bibinfo{title}{Mobile loans apps Tala, Branch, Okash face scrutiny
  in Kenya — Quartz Africa}.
\newblock
  \bibinfo{howpublished}{\url{https://qz.com/africa/1712796/mobile-loans-apps-tala-branch-okash-face-scrutiny-in-kenya/}}.
\newblock
\newblock
\shownote{(Accessed on 08/04/2020).}


\bibitem[\protect\citeauthoryear{Davidson, Bhattacharya, and Weber}{Davidson
  et~al\mbox{.}}{2019}]%
        {davidson2019racial}
\bibfield{author}{\bibinfo{person}{Thomas Davidson}, \bibinfo{person}{Debasmita
  Bhattacharya}, {and} \bibinfo{person}{Ingmar Weber}.}
  \bibinfo{year}{2019}\natexlab{}.
\newblock \showarticletitle{Racial bias in hate speech and abusive language
  detection datasets}.
\newblock \bibinfo{journal}{\emph{arXiv preprint arXiv:1905.12516}}
  (\bibinfo{year}{2019}).
\newblock


\bibitem[\protect\citeauthoryear{Deal}{Deal}{[n.d.]}]%
        {newdeal}
\bibfield{author}{\bibinfo{person}{The Living~New Deal}.}
  \bibinfo{year}{[n.d.]}\natexlab{}.
\newblock \bibinfo{title}{African Americans}.
\newblock
  \bibinfo{howpublished}{\url{https://livingnewdeal.org/what-was-the-new-deal/new-deal-inclusion/african-americans-2/}}.
\newblock
\newblock
\shownote{(Accessed on 08/29/2020).}


\bibitem[\protect\citeauthoryear{Diaz, Johnson, Lazar, Piper, and Gergle}{Diaz
  et~al\mbox{.}}{2018}]%
        {diaz2018addressing}
\bibfield{author}{\bibinfo{person}{Mark Diaz}, \bibinfo{person}{Isaac Johnson},
  \bibinfo{person}{Amanda Lazar}, \bibinfo{person}{Anne~Marie Piper}, {and}
  \bibinfo{person}{Darren Gergle}.} \bibinfo{year}{2018}\natexlab{}.
\newblock \showarticletitle{Addressing Age-Related Bias in Sentiment Analysis}.
  In \bibinfo{booktitle}{\emph{Proceedings of the 2018 CHI Conference on Human
  Factors in Computing Systems}} (Montreal QC, Canada)
  \emph{(\bibinfo{series}{CHI ’18})}. \bibinfo{publisher}{Association for
  Computing Machinery}, \bibinfo{address}{New York, NY, USA},
  \bibinfo{pages}{1–14}.
\newblock
\showISBNx{9781450356206}
\urldef\tempurl%
\url{https://doi.org/10.1145/3173574.3173986}
\showDOI{\tempurl}


\bibitem[\protect\citeauthoryear{Dixit}{Dixit}{July}]%
        {FairButN13:online}
\bibfield{author}{\bibinfo{person}{Neha Dixit}.}
  \bibinfo{year}{July}\natexlab{}.
\newblock \bibinfo{title}{Fair, But Not So Lovely: India’s Obsession With
  Skin Whitening | by Neha Dixit | BRIGHT Magazine}.
\newblock
  \bibinfo{howpublished}{\url{https://brightthemag.com/fair-but-not-so-lovely-indias-obsession-with-skin-whitening-beauty-body-image-bleaching-4d6ba9c9743d}}.
\newblock
\newblock
\shownote{(Accessed on 09/25/2020).}


\bibitem[\protect\citeauthoryear{Dixit}{Dixit}{2019}]%
        {FacialRecIndia}
\bibfield{author}{\bibinfo{person}{Pranav Dixit}.}
  \bibinfo{year}{2019}\natexlab{}.
\newblock \bibinfo{title}{India Is Creating A National Facial Recognition
  System}.
\newblock
  \bibinfo{howpublished}{\url{https://www.buzzfeednews.com/article/pranavdixit/india-is-creating-a-national-facial-recognition-system-and}}.
\newblock
\newblock
\shownote{(Accessed on 08/30/2020).}


\bibitem[\protect\citeauthoryear{Dobbe, Dean, Gilbert, and Kohli}{Dobbe
  et~al\mbox{.}}{2018}]%
        {dobbe2018broader}
\bibfield{author}{\bibinfo{person}{Roel Dobbe}, \bibinfo{person}{Sarah Dean},
  \bibinfo{person}{Thomas Gilbert}, {and} \bibinfo{person}{Nitin Kohli}.}
  \bibinfo{year}{2018}\natexlab{}.
\newblock \showarticletitle{A broader view on bias in automated
  decision-making: Reflecting on epistemology and dynamics}.
\newblock \bibinfo{journal}{\emph{arXiv preprint arXiv:1807.00553}}
  (\bibinfo{year}{2018}).
\newblock


\bibitem[\protect\citeauthoryear{Donner}{Donner}{2015}]%
        {donner2015after}
\bibfield{author}{\bibinfo{person}{Jonathan Donner}.}
  \bibinfo{year}{2015}\natexlab{}.
\newblock \bibinfo{booktitle}{\emph{After access: Inclusion, development, and a
  more mobile Internet}}.
\newblock \bibinfo{publisher}{MIT press}.
\newblock


\bibitem[\protect\citeauthoryear{Donner, Rangaswamy, Steenson, and Wei}{Donner
  et~al\mbox{.}}{2008}]%
        {donner2008express}
\bibfield{author}{\bibinfo{person}{Jonathan Donner}, \bibinfo{person}{Nimmi
  Rangaswamy}, \bibinfo{person}{M Steenson}, {and} \bibinfo{person}{Carolyn
  Wei}.} \bibinfo{year}{2008}\natexlab{}.
\newblock \showarticletitle{``Express yourself '' / ``Stay together'': Tensions
  surrounding mobile communication in the middle-class Indian family}.
\newblock \bibinfo{journal}{\emph{J. Katz (Ed.), Handbook of mobile
  communication studies}} (\bibinfo{year}{2008}), \bibinfo{pages}{325--337}.
\newblock


\bibitem[\protect\citeauthoryear{Donovan}{Donovan}{2015}]%
        {donovan2015biometric}
\bibfield{author}{\bibinfo{person}{Kevin~P Donovan}.}
  \bibinfo{year}{2015}\natexlab{}.
\newblock \showarticletitle{The biometric imaginary: Bureaucratic
  technopolitics in post-apartheid welfare}.
\newblock \bibinfo{journal}{\emph{Journal of Southern African Studies}}
  \bibinfo{volume}{41}, \bibinfo{number}{4} (\bibinfo{year}{2015}),
  \bibinfo{pages}{815--833}.
\newblock


\bibitem[\protect\citeauthoryear{Dorfman and Mattelart}{Dorfman and
  Mattelart}{1975}]%
        {dorfman1975read}
\bibfield{author}{\bibinfo{person}{Ariel Dorfman} {and} \bibinfo{person}{Armand
  Mattelart}.} \bibinfo{year}{1975}\natexlab{}.
\newblock \bibinfo{booktitle}{\emph{How to Read Donald Duck}}.
\newblock \bibinfo{publisher}{International General New York}.
\newblock


\bibitem[\protect\citeauthoryear{Dray, Light, Dearden, Evers, Densmore,
  Ramachandran, Kam, Marsden, Sambasivan, Smyth, et~al\mbox{.}}{Dray
  et~al\mbox{.}}{2012}]%
        {dray2012human}
\bibfield{author}{\bibinfo{person}{Susan Dray}, \bibinfo{person}{Ann Light},
  \bibinfo{person}{A Dearden}, \bibinfo{person}{Vanessa Evers},
  \bibinfo{person}{Melissa Densmore}, \bibinfo{person}{D Ramachandran},
  \bibinfo{person}{M Kam}, \bibinfo{person}{G Marsden}, \bibinfo{person}{N
  Sambasivan}, \bibinfo{person}{T Smyth}, {et~al\mbox{.}}}
  \bibinfo{year}{2012}\natexlab{}.
\newblock \showarticletitle{Human--Computer Interaction for Development:
  Changing Human--Computer Interaction to Change the World}.
\newblock In \bibinfo{booktitle}{\emph{The Human-Computer Interaction Handbook:
  Fundamentals, Evolving Technologies, and Emerging Applications, Third
  Edition}}. \bibinfo{publisher}{CRC press}, \bibinfo{pages}{1369--1394}.
\newblock


\bibitem[\protect\citeauthoryear{Duflo}{Duflo}{2005}]%
        {duflo2005political}
\bibfield{author}{\bibinfo{person}{Esther Duflo}.}
  \bibinfo{year}{2005}\natexlab{}.
\newblock \showarticletitle{Why political reservations?}
\newblock \bibinfo{journal}{\emph{Journal of the European Economic
  Association}} \bibinfo{volume}{3}, \bibinfo{number}{2-3}
  (\bibinfo{year}{2005}), \bibinfo{pages}{668--678}.
\newblock


\bibitem[\protect\citeauthoryear{D’espallier, Gu{\'e}rin, and
  Mersland}{D’espallier et~al\mbox{.}}{2011}]%
        {d2011women}
\bibfield{author}{\bibinfo{person}{Bert D’espallier},
  \bibinfo{person}{Isabelle Gu{\'e}rin}, {and} \bibinfo{person}{Roy Mersland}.}
  \bibinfo{year}{2011}\natexlab{}.
\newblock \showarticletitle{Women and repayment in microfinance: A global
  analysis}.
\newblock \bibinfo{journal}{\emph{World development}} \bibinfo{volume}{39},
  \bibinfo{number}{5} (\bibinfo{year}{2011}), \bibinfo{pages}{758--772}.
\newblock


\bibitem[\protect\citeauthoryear{Escobar}{Escobar}{2011}]%
        {escobar2011encountering}
\bibfield{author}{\bibinfo{person}{Arturo Escobar}.}
  \bibinfo{year}{2011}\natexlab{}.
\newblock \bibinfo{booktitle}{\emph{Encountering development: The making and
  unmaking of the Third World}}. Vol.~\bibinfo{volume}{1}.
\newblock \bibinfo{publisher}{Princeton University Press}.
\newblock


\bibitem[\protect\citeauthoryear{Express}{Express}{2020}]%
        {MostIndi27:online}
\bibfield{author}{\bibinfo{person}{Indian Express}.}
  \bibinfo{year}{2020}\natexlab{}.
\newblock \bibinfo{title}{Most Indian Nobel winners Brahmins: Gujarat Speaker
  Rajendra Trivedi}.
\newblock
  \bibinfo{howpublished}{\url{https://indianexpress.com/article/cities/ahmedabad/most-indian-nobel-winners-brahmins-gujarat-speaker-rajendra-trivedi-6198741/}}.
\newblock
\newblock
\shownote{(Accessed on 09/04/2020).}


\bibitem[\protect\citeauthoryear{Fanon}{Fanon}{2007}]%
        {fanon2007wretched}
\bibfield{author}{\bibinfo{person}{Frantz Fanon}.}
  \bibinfo{year}{2007}\natexlab{}.
\newblock \bibinfo{booktitle}{\emph{The wretched of the earth}}.
\newblock \bibinfo{publisher}{Grove/Atlantic, Inc.}
\newblock


\bibitem[\protect\citeauthoryear{Fitzpatrick}{Fitzpatrick}{1988}]%
        {fitzpatrick1988validity}
\bibfield{author}{\bibinfo{person}{Thomas~B Fitzpatrick}.}
  \bibinfo{year}{1988}\natexlab{}.
\newblock \showarticletitle{The validity and practicality of sun-reactive skin
  types I through VI}.
\newblock \bibinfo{journal}{\emph{Archives of dermatology}}
  \bibinfo{volume}{124}, \bibinfo{number}{6} (\bibinfo{year}{1988}),
  \bibinfo{pages}{869--871}.
\newblock


\bibitem[\protect\citeauthoryear{Gandhi, Veeraraghavan, Toyama, and
  Ramprasad}{Gandhi et~al\mbox{.}}{2007}]%
        {gandhi2007digital}
\bibfield{author}{\bibinfo{person}{Rikin Gandhi}, \bibinfo{person}{Rajesh
  Veeraraghavan}, \bibinfo{person}{Kentaro Toyama}, {and}
  \bibinfo{person}{Vanaja Ramprasad}.} \bibinfo{year}{2007}\natexlab{}.
\newblock \showarticletitle{Digital green: Participatory video for agricultural
  extension}. In \bibinfo{booktitle}{\emph{2007 International conference on
  information and communication technologies and development}}. IEEE,
  \bibinfo{pages}{1--10}.
\newblock


\bibitem[\protect\citeauthoryear{Gardiner}{Gardiner}{2013}]%
        {5inNewDe60:online}
\bibfield{author}{\bibinfo{person}{Harris Gardiner}.}
  \bibinfo{year}{2013}\natexlab{}.
\newblock \bibinfo{title}{5 in New Delhi Rape Case Face Murder Charges - The
  New York Times}.
\newblock
  \bibinfo{howpublished}{\url{https://www.nytimes.com/2013/01/04/world/asia/murder-charges-filed-against-5-men-in-india-gang-rape.html}}.
\newblock
\newblock
\shownote{(Accessed on 09/13/2020).}


\bibitem[\protect\citeauthoryear{Garg, Perot, Limtiaco, Taly, Chi, and
  Beutel}{Garg et~al\mbox{.}}{2019}]%
        {garg2019counterfactual}
\bibfield{author}{\bibinfo{person}{Sahaj Garg}, \bibinfo{person}{Vincent
  Perot}, \bibinfo{person}{Nicole Limtiaco}, \bibinfo{person}{Ankur Taly},
  \bibinfo{person}{Ed~H Chi}, {and} \bibinfo{person}{Alex Beutel}.}
  \bibinfo{year}{2019}\natexlab{}.
\newblock \showarticletitle{Counterfactual fairness in text classification
  through robustness}. In \bibinfo{booktitle}{\emph{Proceedings of the 2019
  AAAI/ACM Conference on AI, Ethics, and Society}}. \bibinfo{pages}{219--226}.
\newblock


\bibitem[\protect\citeauthoryear{Gebru, Morgenstern, Vecchione, Vaughan,
  Wallach, Daum{\'e}~III, and Crawford}{Gebru et~al\mbox{.}}{2018}]%
        {gebru2018datasheets}
\bibfield{author}{\bibinfo{person}{Timnit Gebru}, \bibinfo{person}{Jamie
  Morgenstern}, \bibinfo{person}{Briana Vecchione},
  \bibinfo{person}{Jennifer~Wortman Vaughan}, \bibinfo{person}{Hanna Wallach},
  \bibinfo{person}{Hal Daum{\'e}~III}, {and} \bibinfo{person}{Kate Crawford}.}
  \bibinfo{year}{2018}\natexlab{}.
\newblock \showarticletitle{Datasheets for datasets}.
\newblock \bibinfo{journal}{\emph{arXiv preprint arXiv:1803.09010}}
  (\bibinfo{year}{2018}).
\newblock


\bibitem[\protect\citeauthoryear{Golebiewski and Boyd}{Golebiewski and
  Boyd}{2019}]%
        {golebiewski2019data}
\bibfield{author}{\bibinfo{person}{Michael Golebiewski} {and}
  \bibinfo{person}{Danah Boyd}.} \bibinfo{year}{2019}\natexlab{}.
\newblock \showarticletitle{Data voids: Where missing data can easily be
  exploited}.
\newblock \bibinfo{journal}{\emph{Data \& Society}} (\bibinfo{year}{2019}).
\newblock


\bibitem[\protect\citeauthoryear{Goto, Kojima, Kurata, Tamura, and Yokoo}{Goto
  et~al\mbox{.}}{2017}]%
        {goto2017designing}
\bibfield{author}{\bibinfo{person}{Masahiro Goto}, \bibinfo{person}{Fuhito
  Kojima}, \bibinfo{person}{Ryoji Kurata}, \bibinfo{person}{Akihisa Tamura},
  {and} \bibinfo{person}{Makoto Yokoo}.} \bibinfo{year}{2017}\natexlab{}.
\newblock \showarticletitle{Designing matching mechanisms under general
  distributional constraints}.
\newblock \bibinfo{journal}{\emph{American Economic Journal: Microeconomics}}
  \bibinfo{volume}{9}, \bibinfo{number}{2} (\bibinfo{year}{2017}),
  \bibinfo{pages}{226--62}.
\newblock


\bibitem[\protect\citeauthoryear{Graham, Haidt, Koleva, Motyl, Iyer, Wojcik,
  and Ditto}{Graham et~al\mbox{.}}{2013}]%
        {graham2013moral}
\bibfield{author}{\bibinfo{person}{Jesse Graham}, \bibinfo{person}{Jonathan
  Haidt}, \bibinfo{person}{Sena Koleva}, \bibinfo{person}{Matt Motyl},
  \bibinfo{person}{Ravi Iyer}, \bibinfo{person}{Sean~P Wojcik}, {and}
  \bibinfo{person}{Peter~H Ditto}.} \bibinfo{year}{2013}\natexlab{}.
\newblock \showarticletitle{Moral foundations theory: The pragmatic validity of
  moral pluralism}.
\newblock In \bibinfo{booktitle}{\emph{Advances in experimental social
  psychology}}. Vol.~\bibinfo{volume}{47}. \bibinfo{publisher}{Elsevier},
  \bibinfo{pages}{55--130}.
\newblock


\bibitem[\protect\citeauthoryear{Green}{Green}{2020}]%
        {green2020false}
\bibfield{author}{\bibinfo{person}{Ben Green}.}
  \bibinfo{year}{2020}\natexlab{}.
\newblock \showarticletitle{The false promise of risk assessments: epistemic
  reform and the limits of fairness}. In \bibinfo{booktitle}{\emph{Proceedings
  of the 2020 Conference on Fairness, Accountability, and Transparency}}.
  \bibinfo{pages}{594--606}.
\newblock


\bibitem[\protect\citeauthoryear{Green and Viljoen}{Green and Viljoen}{2020}]%
        {green2020algorithmic}
\bibfield{author}{\bibinfo{person}{Ben Green} {and} \bibinfo{person}{Salom{\'e}
  Viljoen}.} \bibinfo{year}{2020}\natexlab{}.
\newblock \showarticletitle{Algorithmic realism: expanding the boundaries of
  algorithmic thought}. In \bibinfo{booktitle}{\emph{Proceedings of the 2020
  Conference on Fairness, Accountability, and Transparency}}.
  \bibinfo{pages}{19--31}.
\newblock


\bibitem[\protect\citeauthoryear{Gupta}{Gupta}{2012}]%
        {gupta2012red}
\bibfield{author}{\bibinfo{person}{Akhil Gupta}.}
  \bibinfo{year}{2012}\natexlab{}.
\newblock \bibinfo{booktitle}{\emph{Red tape: Bureaucracy, structural violence,
  and poverty in India}}.
\newblock \bibinfo{publisher}{Duke University Press}.
\newblock


\bibitem[\protect\citeauthoryear{Hagerty and Rubinov}{Hagerty and
  Rubinov}{2019}]%
        {hagerty2019global}
\bibfield{author}{\bibinfo{person}{Alexa Hagerty} {and} \bibinfo{person}{Igor
  Rubinov}.} \bibinfo{year}{2019}\natexlab{}.
\newblock \showarticletitle{Global AI Ethics: A Review of the Social Impacts
  and Ethical Implications of Artificial Intelligence}.
\newblock \bibinfo{journal}{\emph{arXiv}} (\bibinfo{year}{2019}),
  \bibinfo{pages}{arXiv--1907}.
\newblock


\bibitem[\protect\citeauthoryear{Hanna, Denton, Smart, and Smith-Loud}{Hanna
  et~al\mbox{.}}{2020}]%
        {hanna2020towards}
\bibfield{author}{\bibinfo{person}{Alex Hanna}, \bibinfo{person}{Emily Denton},
  \bibinfo{person}{Andrew Smart}, {and} \bibinfo{person}{Jamila Smith-Loud}.}
  \bibinfo{year}{2020}\natexlab{}.
\newblock \showarticletitle{Towards a critical race methodology in algorithmic
  fairness}. In \bibinfo{booktitle}{\emph{Proceedings of the 2020 Conference on
  Fairness, Accountability, and Transparency}}. \bibinfo{pages}{501--512}.
\newblock


\bibitem[\protect\citeauthoryear{Heimerl, Hasan, Ali, Brewer, and
  Parikh}{Heimerl et~al\mbox{.}}{2013}]%
        {heimerl2013local}
\bibfield{author}{\bibinfo{person}{Kurtis Heimerl}, \bibinfo{person}{Shaddi
  Hasan}, \bibinfo{person}{Kashif Ali}, \bibinfo{person}{Eric Brewer}, {and}
  \bibinfo{person}{Tapan Parikh}.} \bibinfo{year}{2013}\natexlab{}.
\newblock \showarticletitle{Local, sustainable, small-scale cellular networks}.
  In \bibinfo{booktitle}{\emph{Proceedings of the Sixth International
  Conference on Information and Communication Technologies and Development:
  Full Papers-Volume 1}}. \bibinfo{pages}{2--12}.
\newblock


\bibitem[\protect\citeauthoryear{Held et~al\mbox{.}}{Held
  et~al\mbox{.}}{2006}]%
        {held2006ethics}
\bibfield{author}{\bibinfo{person}{Virginia Held} {et~al\mbox{.}}}
  \bibinfo{year}{2006}\natexlab{}.
\newblock \bibinfo{booktitle}{\emph{The ethics of care: Personal, political,
  and global}}.
\newblock \bibinfo{publisher}{Oxford University Press on Demand}.
\newblock


\bibitem[\protect\citeauthoryear{Hollinger}{Hollinger}{1998}]%
        {hollinger1998science}
\bibfield{author}{\bibinfo{person}{David~A Hollinger}.}
  \bibinfo{year}{1998}\natexlab{}.
\newblock \bibinfo{booktitle}{\emph{Science, Jews, and secular culture: studies
  in mid-twentieth-century American intellectual history}}.
\newblock \bibinfo{publisher}{Princeton University Press}.
\newblock


\bibitem[\protect\citeauthoryear{Holstein, Wortman~Vaughan, Daum{\'e}~III,
  Dudik, and Wallach}{Holstein et~al\mbox{.}}{2019}]%
        {holstein2019improving}
\bibfield{author}{\bibinfo{person}{Kenneth Holstein}, \bibinfo{person}{Jennifer
  Wortman~Vaughan}, \bibinfo{person}{Hal Daum{\'e}~III}, \bibinfo{person}{Miro
  Dudik}, {and} \bibinfo{person}{Hanna Wallach}.}
  \bibinfo{year}{2019}\natexlab{}.
\newblock \showarticletitle{Improving fairness in machine learning systems:
  What do industry practitioners need?}. In
  \bibinfo{booktitle}{\emph{Proceedings of the 2019 CHI Conference on Human
  Factors in Computing Systems}}. \bibinfo{pages}{1--16}.
\newblock


\bibitem[\protect\citeauthoryear{Hutchinson and Mitchell}{Hutchinson and
  Mitchell}{2019}]%
        {hutchinson201950}
\bibfield{author}{\bibinfo{person}{Ben Hutchinson} {and}
  \bibinfo{person}{Margaret Mitchell}.} \bibinfo{year}{2019}\natexlab{}.
\newblock \showarticletitle{50 years of test (un) fairness: Lessons for machine
  learning}. In \bibinfo{booktitle}{\emph{Proceedings of the Conference on
  Fairness, Accountability, and Transparency}}. \bibinfo{pages}{49--58}.
\newblock


\bibitem[\protect\citeauthoryear{Hutchinson, Prabhakaran, Denton, Webster,
  Zhong, and Denuyl}{Hutchinson et~al\mbox{.}}{2020}]%
        {hutchinson2020social}
\bibfield{author}{\bibinfo{person}{Ben Hutchinson}, \bibinfo{person}{Vinodkumar
  Prabhakaran}, \bibinfo{person}{Emily Denton}, \bibinfo{person}{Kellie
  Webster}, \bibinfo{person}{Yu Zhong}, {and} \bibinfo{person}{Stephen
  Denuyl}.} \bibinfo{year}{2020}\natexlab{}.
\newblock \showarticletitle{Social Biases in NLP Models as Barriers for Persons
  with Disabilities}.
\newblock \bibinfo{journal}{\emph{ACL}} (\bibinfo{year}{2020}).
\newblock


\bibitem[\protect\citeauthoryear{IDSN}{IDSN}{2010}]%
        {Twothird31:online}
\bibfield{author}{\bibinfo{person}{IDSN}.} \bibinfo{year}{2010}\natexlab{}.
\newblock \bibinfo{title}{Two thirds of India’s Dalits are poor -
  International Dalit Solidarity Network}.
\newblock
  \bibinfo{howpublished}{\url{https://idsn.org/two-thirds-of-indias-dalits-are-poor/}}.
\newblock
\newblock
\shownote{(Accessed on 08/13/2020).}


\bibitem[\protect\citeauthoryear{{IEEE}}{{IEEE}}{2019}]%
        {ieeeglobal2019}
\bibfield{author}{\bibinfo{person}{{IEEE}}.} \bibinfo{year}{2019}\natexlab{}.
\newblock \showarticletitle{{The IEEE Global Initiative on Ethics of Autonomous
  and Intelligent Systems. “Classical Ethics in A/IS”}}.
\newblock In \bibinfo{booktitle}{\emph{{Ethically Aligned Design: A Vision for
  Prioritizing Human Well-being with Autonomous and Intelligent Systems, First
  Edition}}}. \bibinfo{pages}{36--67}.
\newblock


\bibitem[\protect\citeauthoryear{Inayatullah}{Inayatullah}{2006}]%
        {inayatullah2006culture}
\bibfield{author}{\bibinfo{person}{S Inayatullah}.}
  \bibinfo{year}{2006}\natexlab{}.
\newblock \showarticletitle{Culture and Fairness: The Idea of Civilization
  Fairness}.
\newblock In \bibinfo{booktitle}{\emph{Fairness, Globalization and Public
  Institutions}}. \bibinfo{publisher}{University of Hawaii Press},
  \bibinfo{pages}{31--33}.
\newblock


\bibitem[\protect\citeauthoryear{Ismail and Kumar}{Ismail and Kumar}{2018}]%
        {ismail2018engaging}
\bibfield{author}{\bibinfo{person}{Azra Ismail} {and} \bibinfo{person}{Neha
  Kumar}.} \bibinfo{year}{2018}\natexlab{}.
\newblock \showarticletitle{Engaging solidarity in data collection practices
  for community health}.
\newblock \bibinfo{journal}{\emph{Proceedings of the ACM on Human-Computer
  Interaction}} \bibinfo{volume}{2}, \bibinfo{number}{CSCW}
  (\bibinfo{year}{2018}), \bibinfo{pages}{1--24}.
\newblock


\bibitem[\protect\citeauthoryear{Jain}{Jain}{2016}]%
        {Jain2016Indias}
\bibfield{author}{\bibinfo{person}{Mayank Jain}.}
  \bibinfo{year}{2016}\natexlab{}.
\newblock \showarticletitle{{India's internet population is exploding but women
  are not logging in}}.
\newblock \bibinfo{journal}{\emph{Scroll.in}} (\bibinfo{date}{26 9}
  \bibinfo{year}{2016}).
\newblock
\urldef\tempurl%
\url{https://scroll.in/article/816892/indias-internet-population-is-exploding-but-women-are-not-logging-inia}
\showURL{%
\tempurl}


\bibitem[\protect\citeauthoryear{Jenkins and Goetz}{Jenkins and Goetz}{1999}]%
        {jenkins1999accounts}
\bibfield{author}{\bibinfo{person}{Rob Jenkins} {and}
  \bibinfo{person}{Anne~Marie Goetz}.} \bibinfo{year}{1999}\natexlab{}.
\newblock \showarticletitle{Accounts and accountability: theoretical
  implications of the right-to-information movement in India}.
\newblock \bibinfo{journal}{\emph{Third world quarterly}} \bibinfo{volume}{20},
  \bibinfo{number}{3} (\bibinfo{year}{1999}), \bibinfo{pages}{603--622}.
\newblock


\bibitem[\protect\citeauthoryear{Jobin, Ienca, and Vayena}{Jobin
  et~al\mbox{.}}{2019}]%
        {jobin2019global}
\bibfield{author}{\bibinfo{person}{Anna Jobin}, \bibinfo{person}{Marcello
  Ienca}, {and} \bibinfo{person}{Effy Vayena}.}
  \bibinfo{year}{2019}\natexlab{}.
\newblock \showarticletitle{The global landscape of AI ethics guidelines}.
\newblock \bibinfo{journal}{\emph{Nature Machine Intelligence}}
  \bibinfo{volume}{1}, \bibinfo{number}{9} (\bibinfo{year}{2019}),
  \bibinfo{pages}{389--399}.
\newblock


\bibitem[\protect\citeauthoryear{Joseph, Kearns, Morgenstern, Neel, and
  Roth}{Joseph et~al\mbox{.}}{2016}]%
        {joseph2016rawlsian}
\bibfield{author}{\bibinfo{person}{Matthew Joseph}, \bibinfo{person}{Michael
  Kearns}, \bibinfo{person}{Jamie Morgenstern}, \bibinfo{person}{Seth Neel},
  {and} \bibinfo{person}{Aaron Roth}.} \bibinfo{year}{2016}\natexlab{}.
\newblock \showarticletitle{Rawlsian fairness for machine learning}.
\newblock \bibinfo{journal}{\emph{arXiv preprint arXiv:1610.09559}}
  \bibinfo{volume}{1}, \bibinfo{number}{2} (\bibinfo{year}{2016}).
\newblock


\bibitem[\protect\citeauthoryear{Joshi}{Joshi}{2020}]%
        {AIobs}
\bibfield{author}{\bibinfo{person}{Divij Joshi}.}
  \bibinfo{year}{2020}\natexlab{}.
\newblock \bibinfo{title}{AI Observatory}.
\newblock \bibinfo{howpublished}{\url{http://ai-observatory.in/}}.
\newblock
\newblock
\shownote{(Accessed on 12/30/2020).}


\bibitem[\protect\citeauthoryear{Joshi}{Joshi}{2018}]%
        {joshi2018racial}
\bibfield{author}{\bibinfo{person}{Yuvraj Joshi}.}
  \bibinfo{year}{2018}\natexlab{}.
\newblock \showarticletitle{Racial Indirection}.
\newblock \bibinfo{journal}{\emph{UCDL Rev.}}  \bibinfo{volume}{52}
  (\bibinfo{year}{2018}), \bibinfo{pages}{2495}.
\newblock


\bibitem[\protect\citeauthoryear{Kala}{Kala}{2019}]%
        {Highgend58:online}
\bibfield{author}{\bibinfo{person}{Rishi~Ranjan Kala}.}
  \bibinfo{year}{2019}\natexlab{}.
\newblock \bibinfo{title}{High gender disparity among internet users in India -
  The Financial Express}.
\newblock
  \bibinfo{howpublished}{\url{https://www.financialexpress.com/industry/high-gender-disparity-among-internet-users-in-india/1718951/}}.
\newblock
\newblock
\shownote{(Accessed on 10/06/2020).}


\bibitem[\protect\citeauthoryear{Kallus and Zhou}{Kallus and Zhou}{2018}]%
        {kallus2018residual}
\bibfield{author}{\bibinfo{person}{Nathan Kallus} {and} \bibinfo{person}{Angela
  Zhou}.} \bibinfo{year}{2018}\natexlab{}.
\newblock \showarticletitle{Residual unfairness in fair machine learning from
  prejudiced data}.
\newblock \bibinfo{journal}{\emph{arXiv preprint arXiv:1806.02887}}
  (\bibinfo{year}{2018}).
\newblock


\bibitem[\protect\citeauthoryear{Kalyanakrishnan, Panicker, Natarajan, and
  Rao}{Kalyanakrishnan et~al\mbox{.}}{2018}]%
        {kalyanakrishnan2018opportunities}
\bibfield{author}{\bibinfo{person}{Shivaram Kalyanakrishnan},
  \bibinfo{person}{Rahul~Alex Panicker}, \bibinfo{person}{Sarayu Natarajan},
  {and} \bibinfo{person}{Shreya Rao}.} \bibinfo{year}{2018}\natexlab{}.
\newblock \showarticletitle{Opportunities and Challenges for Artificial
  Intelligence in India}. In \bibinfo{booktitle}{\emph{Proceedings of the 2018
  AAAI/ACM Conference on AI, Ethics, and Society}}. \bibinfo{pages}{164--170}.
\newblock


\bibitem[\protect\citeauthoryear{Kamada and Kojima}{Kamada and Kojima}{2015}]%
        {kamada2015efficient}
\bibfield{author}{\bibinfo{person}{Yuichiro Kamada} {and}
  \bibinfo{person}{Fuhito Kojima}.} \bibinfo{year}{2015}\natexlab{}.
\newblock \showarticletitle{Efficient matching under distributional
  constraints: Theory and applications}.
\newblock \bibinfo{journal}{\emph{American Economic Review}}
  \bibinfo{volume}{105}, \bibinfo{number}{1} (\bibinfo{year}{2015}),
  \bibinfo{pages}{67--99}.
\newblock


\bibitem[\protect\citeauthoryear{Kamath and Kumar}{Kamath and Kumar}{2017}]%
        {InIndiaA10:online}
\bibfield{author}{\bibinfo{person}{Anant Kamath} {and} \bibinfo{person}{Vinay
  Kumar}.} \bibinfo{year}{2017}\natexlab{}.
\newblock \bibinfo{title}{In India, Accessible Phones Lead to Inaccessible
  Opportunities}.
\newblock
  \bibinfo{howpublished}{\url{https://thewire.in/caste/india-accessible-phones-still-lead-inaccessible-opportunities}}.
\newblock
\newblock
\shownote{(Accessed on 01/14/2021).}


\bibitem[\protect\citeauthoryear{Kandukuri}{Kandukuri}{2018}]%
        {Casteist69:online}
\bibfield{author}{\bibinfo{person}{Divya Kandukuri}.}
  \bibinfo{year}{2018}\natexlab{}.
\newblock \bibinfo{title}{Casteist Slurs You Need To Know - YouTube}.
\newblock
  \bibinfo{howpublished}{\url{https://www.youtube.com/watch?v=wJwkIxOpqZA}}.
\newblock
\newblock
\shownote{(Accessed on 09/25/2020).}


\bibitem[\protect\citeauthoryear{Karan}{Karan}{2008}]%
        {karan2008obsessions}
\bibfield{author}{\bibinfo{person}{Kavita Karan}.}
  \bibinfo{year}{2008}\natexlab{}.
\newblock \showarticletitle{Obsessions with fair skin: Color discourses in
  Indian advertising}.
\newblock \bibinfo{journal}{\emph{Advertising \& society review}}
  \bibinfo{volume}{9}, \bibinfo{number}{2} (\bibinfo{year}{2008}).
\newblock


\bibitem[\protect\citeauthoryear{Katell, Young, Dailey, Herman, Guetler, Tam,
  Bintz, Raz, and Krafft}{Katell et~al\mbox{.}}{2020}]%
        {katell2020toward}
\bibfield{author}{\bibinfo{person}{Michael Katell}, \bibinfo{person}{Meg
  Young}, \bibinfo{person}{Dharma Dailey}, \bibinfo{person}{Bernease Herman},
  \bibinfo{person}{Vivian Guetler}, \bibinfo{person}{Aaron Tam},
  \bibinfo{person}{Corinne Bintz}, \bibinfo{person}{Daniella Raz}, {and}
  \bibinfo{person}{PM Krafft}.} \bibinfo{year}{2020}\natexlab{}.
\newblock \showarticletitle{Toward situated interventions for algorithmic
  equity: lessons from the field}. In \bibinfo{booktitle}{\emph{Proceedings of
  the 2020 Conference on Fairness, Accountability, and Transparency}}.
  \bibinfo{pages}{45--55}.
\newblock


\bibitem[\protect\citeauthoryear{Kelley, Yang, Heldreth, Moessner, Sedley,
  Kramm, Newman, and Woodruff}{Kelley et~al\mbox{.}}{2019}]%
        {kelley2019happy}
\bibfield{author}{\bibinfo{person}{Patrick~Gage Kelley},
  \bibinfo{person}{Yongwei Yang}, \bibinfo{person}{Courtney Heldreth},
  \bibinfo{person}{Christopher Moessner}, \bibinfo{person}{Aaron Sedley},
  \bibinfo{person}{Andreas Kramm}, \bibinfo{person}{David Newman}, {and}
  \bibinfo{person}{Allison Woodruff}.} \bibinfo{year}{2019}\natexlab{}.
\newblock \showarticletitle{"Happy and Assured that life will be easy 10years
  from now.": Perceptions of Artificial Intelligence in 8 Countries}.
\newblock \bibinfo{journal}{\emph{arXiv preprint arXiv:2001.00081}}
  (\bibinfo{year}{2019}).
\newblock


\bibitem[\protect\citeauthoryear{Khaira}{Khaira}{2020}]%
        {Surveill67:online}
\bibfield{author}{\bibinfo{person}{Rachna Khaira}.}
  \bibinfo{year}{2020}\natexlab{}.
\newblock \bibinfo{title}{Surveillance Slavery: Swachh Bharat Tags Sanitation
  Workers To Live-Track Their Every Move | HuffPost India}.
\newblock
  \bibinfo{howpublished}{\url{https://www.huffingtonpost.in/entry/swacch-bharat-tags-sanitation-workers-to-live-track-their-every-move_in_5e4c98a9c5b6b0f6bff11f9b?guccounter=1}}.
\newblock
\newblock
\shownote{(Accessed on 07/28/2020).}


\bibitem[\protect\citeauthoryear{Kodali}{Kodali}{2020}]%
        {AarogyaS88:online}
\bibfield{author}{\bibinfo{person}{Srinivas Kodali}.}
  \bibinfo{year}{2020}\natexlab{}.
\newblock \bibinfo{title}{Aarogya Setu: A bridge too far? | Deccan Herald}.
\newblock
  \bibinfo{howpublished}{\url{https://www.deccanherald.com/specials/sunday-spotlight/aarogya-setu-a-bridge-too-far-835691.html}}.
\newblock
\newblock
\shownote{(Accessed on 08/01/2020).}


\bibitem[\protect\citeauthoryear{Kofman}{Kofman}{2016}]%
        {HowFacia39:online}
\bibfield{author}{\bibinfo{person}{Ava Kofman}.}
  \bibinfo{year}{2016}\natexlab{}.
\newblock \bibinfo{title}{How Facial Recognition Can Ruin Your Life --
  Intercept}.
\newblock
  \bibinfo{howpublished}{\url{https://theintercept.com/2016/10/13/how-a-facial-recognition-mismatch-can-ruin-your-life/}}.
\newblock
\newblock
\shownote{(Accessed on 07/30/2020).}


\bibitem[\protect\citeauthoryear{Kohli, Barreto, and Kroll}{Kohli
  et~al\mbox{.}}{2018}]%
        {kohli2018translation}
\bibfield{author}{\bibinfo{person}{Nitin Kohli}, \bibinfo{person}{Renata
  Barreto}, {and} \bibinfo{person}{Joshua~A Kroll}.}
  \bibinfo{year}{2018}\natexlab{}.
\newblock \showarticletitle{Translation tutorial: a shared lexicon for research
  and practice in human-centered software systems}. In
  \bibinfo{booktitle}{\emph{1st Conference on Fairness, Accountability, and
  Transparancy. New York, NY, USA}}, Vol.~\bibinfo{volume}{7}.
\newblock


\bibitem[\protect\citeauthoryear{Kumar and Anderson}{Kumar and
  Anderson}{2015}]%
        {kumar2015mobile}
\bibfield{author}{\bibinfo{person}{Neha Kumar} {and} \bibinfo{person}{Richard~J
  Anderson}.} \bibinfo{year}{2015}\natexlab{}.
\newblock \showarticletitle{Mobile phones for maternal health in rural India}.
  In \bibinfo{booktitle}{\emph{Proceedings of the 33rd Annual ACM Conference on
  Human Factors in Computing Systems}}. \bibinfo{pages}{427--436}.
\newblock


\bibitem[\protect\citeauthoryear{Kumar}{Kumar}{2013}]%
        {kumar2013does}
\bibfield{author}{\bibinfo{person}{Sunil~Mitra Kumar}.}
  \bibinfo{year}{2013}\natexlab{}.
\newblock \showarticletitle{Does access to formal agricultural credit depend on
  caste?}
\newblock \bibinfo{journal}{\emph{World Development}}  \bibinfo{volume}{43}
  (\bibinfo{year}{2013}), \bibinfo{pages}{315--328}.
\newblock


\bibitem[\protect\citeauthoryear{Kurakin, Goodfellow, and Bengio}{Kurakin
  et~al\mbox{.}}{2016}]%
        {kurakin2016adversarial}
\bibfield{author}{\bibinfo{person}{Alexey Kurakin}, \bibinfo{person}{Ian
  Goodfellow}, {and} \bibinfo{person}{Samy Bengio}.}
  \bibinfo{year}{2016}\natexlab{}.
\newblock \showarticletitle{Adversarial machine learning at scale}.
\newblock \bibinfo{journal}{\emph{arXiv preprint arXiv:1611.01236}}
  (\bibinfo{year}{2016}).
\newblock


\bibitem[\protect\citeauthoryear{Kwet}{Kwet}{2019}]%
        {kwet2019digital}
\bibfield{author}{\bibinfo{person}{Michael Kwet}.}
  \bibinfo{year}{2019}\natexlab{}.
\newblock \showarticletitle{Digital colonialism: US empire and the new
  imperialism in the Global South}.
\newblock \bibinfo{journal}{\emph{Race \& Class}} \bibinfo{volume}{60},
  \bibinfo{number}{4} (\bibinfo{year}{2019}), \bibinfo{pages}{3--26}.
\newblock


\bibitem[\protect\citeauthoryear{Lerman}{Lerman}{2013}]%
        {lerman2013big}
\bibfield{author}{\bibinfo{person}{Jonas Lerman}.}
  \bibinfo{year}{2013}\natexlab{}.
\newblock \showarticletitle{Big data and its exclusions}.
\newblock \bibinfo{journal}{\emph{Stan. L. Rev. Online}}  \bibinfo{volume}{66}
  (\bibinfo{year}{2013}), \bibinfo{pages}{55}.
\newblock


\bibitem[\protect\citeauthoryear{Leung and Stephan}{Leung and Stephan}{2001}]%
        {leung2001social}
\bibfield{author}{\bibinfo{person}{Kwok Leung} {and} \bibinfo{person}{Walter~G
  Stephan}.} \bibinfo{year}{2001}\natexlab{}.
\newblock \showarticletitle{Social Justice from a Cultural Perspective.}
\newblock  (\bibinfo{year}{2001}).
\newblock


\bibitem[\protect\citeauthoryear{Lum and Isaac}{Lum and Isaac}{2016}]%
        {lum2016predict}
\bibfield{author}{\bibinfo{person}{Kristian Lum} {and} \bibinfo{person}{William
  Isaac}.} \bibinfo{year}{2016}\natexlab{}.
\newblock \showarticletitle{To predict and serve?}
\newblock \bibinfo{journal}{\emph{Significance}} \bibinfo{volume}{13},
  \bibinfo{number}{5} (\bibinfo{year}{2016}), \bibinfo{pages}{14--19}.
\newblock


\bibitem[\protect\citeauthoryear{Lund, Scheer, and Kozlenkova}{Lund
  et~al\mbox{.}}{2013}]%
        {lund2013culture}
\bibfield{author}{\bibinfo{person}{Donald~J Lund}, \bibinfo{person}{Lisa~K
  Scheer}, {and} \bibinfo{person}{Irina~V Kozlenkova}.}
  \bibinfo{year}{2013}\natexlab{}.
\newblock \showarticletitle{Culture's impact on the importance of fairness in
  interorganizational relationships}.
\newblock \bibinfo{journal}{\emph{Journal of International Marketing}}
  \bibinfo{volume}{21}, \bibinfo{number}{4} (\bibinfo{year}{2013}),
  \bibinfo{pages}{21--43}.
\newblock


\bibitem[\protect\citeauthoryear{Macklin}{Macklin}{2004}]%
        {macklin2004double}
\bibfield{author}{\bibinfo{person}{Ruth Macklin}.}
  \bibinfo{year}{2004}\natexlab{}.
\newblock \bibinfo{booktitle}{\emph{Double standards in medical research in
  developing countries}}. Vol.~\bibinfo{volume}{2}.
\newblock \bibinfo{publisher}{Cambridge University Press}.
\newblock


\bibitem[\protect\citeauthoryear{Madheswaran and Attewell}{Madheswaran and
  Attewell}{2007}]%
        {madheswaran2007caste}
\bibfield{author}{\bibinfo{person}{Subramaniam Madheswaran} {and}
  \bibinfo{person}{Paul Attewell}.} \bibinfo{year}{2007}\natexlab{}.
\newblock \showarticletitle{Caste discrimination in the Indian urban labour
  market: Evidence from the National Sample Survey}.
\newblock \bibinfo{journal}{\emph{Economic and political Weekly}}
  (\bibinfo{year}{2007}), \bibinfo{pages}{4146--4153}.
\newblock


\bibitem[\protect\citeauthoryear{Manzini, Chong, Black, and Tsvetkov}{Manzini
  et~al\mbox{.}}{2019}]%
        {manzini2019black}
\bibfield{author}{\bibinfo{person}{Thomas Manzini}, \bibinfo{person}{Lim~Yao
  Chong}, \bibinfo{person}{Alan~W Black}, {and} \bibinfo{person}{Yulia
  Tsvetkov}.} \bibinfo{year}{2019}\natexlab{}.
\newblock \showarticletitle{Black is to Criminal as Caucasian is to Police:
  Detecting and Removing Multiclass Bias in Word Embeddings}. In
  \bibinfo{booktitle}{\emph{Proceedings of the 2019 Conference of the North
  American Chapter of the Association for Computational Linguistics: Human
  Language Technologies, Volume 1 (Long and Short Papers)}}.
  \bibinfo{pages}{615--621}.
\newblock


\bibitem[\protect\citeauthoryear{Marda}{Marda}{2018}]%
        {marda2018artificial}
\bibfield{author}{\bibinfo{person}{Vidushi Marda}.}
  \bibinfo{year}{2018}\natexlab{}.
\newblock \showarticletitle{Artificial intelligence policy in India: a
  framework for engaging the limits of data-driven decision-making}.
\newblock \bibinfo{journal}{\emph{Philosophical Transactions of the Royal
  Society A: Mathematical, Physical and Engineering Sciences}}
  \bibinfo{volume}{376}, \bibinfo{number}{2133} (\bibinfo{year}{2018}),
  \bibinfo{pages}{20180087}.
\newblock


\bibitem[\protect\citeauthoryear{Marda and Narayan}{Marda and Narayan}{2020}]%
        {marda2020data}
\bibfield{author}{\bibinfo{person}{Vidushi Marda} {and}
  \bibinfo{person}{Shivangi Narayan}.} \bibinfo{year}{2020}\natexlab{}.
\newblock \showarticletitle{Data in New Delhi's predictive policing system}. In
  \bibinfo{booktitle}{\emph{Proceedings of the 2020 Conference on Fairness,
  Accountability, and Transparency}}. \bibinfo{pages}{317--324}.
\newblock


\bibitem[\protect\citeauthoryear{Martin~Jr, Prabhakaran, Kuhlberg, Smart, and
  Isaac}{Martin~Jr et~al\mbox{.}}{2020}]%
        {martin2020participatory}
\bibfield{author}{\bibinfo{person}{Donald Martin~Jr}, \bibinfo{person}{Vinod
  Prabhakaran}, \bibinfo{person}{Jill Kuhlberg}, \bibinfo{person}{Andrew
  Smart}, {and} \bibinfo{person}{William~S Isaac}.}
  \bibinfo{year}{2020}\natexlab{}.
\newblock \showarticletitle{Participatory Problem Formulation for Fairer
  Machine Learning Through Community Based System Dynamics}.
\newblock \bibinfo{journal}{\emph{ICLR Workshop on Machine Learning in Real
  Life (ML-IRL)}} (\bibinfo{year}{2020}).
\newblock


\bibitem[\protect\citeauthoryear{Martinho-Truswell, Miller, Asare, Petheram,
  Stirling, G\'{o}mez~Mont, and Martinez}{Martinho-Truswell
  et~al\mbox{.}}{2018}]%
        {MexicoAI}
\bibfield{author}{\bibinfo{person}{Emma. Martinho-Truswell},
  \bibinfo{person}{Hannah. Miller}, \bibinfo{person}{Isak~Nti Asare},
  \bibinfo{person}{Andre Petheram}, \bibinfo{person}{Richard (Oxford~Insights)
  Stirling}, \bibinfo{person}{Constanza G\'{o}mez~Mont}, {and}
  \bibinfo{person}{Cristina (C~Minds) Martinez}.}
  \bibinfo{year}{2018}\natexlab{}.
\newblock \showarticletitle{Towards an AI strategy in Mexico: Harnessing the AI
  revolution}. In \bibinfo{booktitle}{\emph{AI whitepaper.}}
\newblock


\bibitem[\protect\citeauthoryear{Masika and Bailur}{Masika and Bailur}{2015}]%
        {masika2015negotiating}
\bibfield{author}{\bibinfo{person}{Rachel Masika} {and} \bibinfo{person}{Savita
  Bailur}.} \bibinfo{year}{2015}\natexlab{}.
\newblock \showarticletitle{Negotiating women’s agency through ICTs: A
  comparative study of Uganda and India}.
\newblock \bibinfo{journal}{\emph{Gender, Technology and Development}}
  \bibinfo{volume}{19}, \bibinfo{number}{1} (\bibinfo{year}{2015}),
  \bibinfo{pages}{43--69}.
\newblock


\bibitem[\protect\citeauthoryear{Mbembe}{Mbembe}{2015}]%
        {mbembe2015decolonizing}
\bibfield{author}{\bibinfo{person}{Achille Mbembe}.}
  \bibinfo{year}{2015}\natexlab{}.
\newblock \bibinfo{title}{Decolonizing knowledge and the question of the
  archive}.
\newblock
\newblock


\bibitem[\protect\citeauthoryear{Medhi, Sagar, and Toyama}{Medhi
  et~al\mbox{.}}{2006}]%
        {medhi2006text}
\bibfield{author}{\bibinfo{person}{Indrani Medhi}, \bibinfo{person}{Aman
  Sagar}, {and} \bibinfo{person}{Kentaro Toyama}.}
  \bibinfo{year}{2006}\natexlab{}.
\newblock \showarticletitle{Text-free user interfaces for illiterate and
  semi-literate users}. In \bibinfo{booktitle}{\emph{2006 international
  conference on information and communication technologies and development}}.
  IEEE, \bibinfo{pages}{72--82}.
\newblock


\bibitem[\protect\citeauthoryear{Mehrabi, Morstatter, Saxena, Lerman, and
  Galstyan}{Mehrabi et~al\mbox{.}}{2019}]%
        {mehrabi2019survey}
\bibfield{author}{\bibinfo{person}{Ninareh Mehrabi}, \bibinfo{person}{Fred
  Morstatter}, \bibinfo{person}{Nripsuta Saxena}, \bibinfo{person}{Kristina
  Lerman}, {and} \bibinfo{person}{Aram Galstyan}.}
  \bibinfo{year}{2019}\natexlab{}.
\newblock \showarticletitle{A survey on bias and fairness in machine learning}.
\newblock \bibinfo{journal}{\emph{arXiv preprint arXiv:1908.09635}}
  (\bibinfo{year}{2019}).
\newblock


\bibitem[\protect\citeauthoryear{Mignolo}{Mignolo}{2011}]%
        {mignolo2011darker}
\bibfield{author}{\bibinfo{person}{Walter Mignolo}.}
  \bibinfo{year}{2011}\natexlab{}.
\newblock \bibinfo{booktitle}{\emph{The darker side of western modernity:
  Global futures, decolonial options}}.
\newblock \bibinfo{publisher}{Duke University Press}.
\newblock


\bibitem[\protect\citeauthoryear{Ministry~of Home~Affairs}{Ministry~of
  Home~Affairs}{[n.d.]}]%
        {CensusIndia2011}
\bibfield{author}{\bibinfo{person}{Government of~India Ministry~of
  Home~Affairs}.} \bibinfo{year}{[n.d.]}\natexlab{}.
\newblock \bibinfo{title}{2011 Census Data}.
\newblock
  \bibinfo{howpublished}{\url{https://www.censusindia.gov.in/2011-Common/CensusData2011.html}}.
\newblock
\newblock
\shownote{(Accessed on 08/26/2020).}


\bibitem[\protect\citeauthoryear{Mitchell, Wu, Zaldivar, Barnes, Vasserman,
  Hutchinson, Spitzer, Raji, and Gebru}{Mitchell et~al\mbox{.}}{2019}]%
        {mitchell2019model}
\bibfield{author}{\bibinfo{person}{Margaret Mitchell}, \bibinfo{person}{Simone
  Wu}, \bibinfo{person}{Andrew Zaldivar}, \bibinfo{person}{Parker Barnes},
  \bibinfo{person}{Lucy Vasserman}, \bibinfo{person}{Ben Hutchinson},
  \bibinfo{person}{Elena Spitzer}, \bibinfo{person}{Inioluwa~Deborah Raji},
  {and} \bibinfo{person}{Timnit Gebru}.} \bibinfo{year}{2019}\natexlab{}.
\newblock \showarticletitle{Model cards for model reporting}. In
  \bibinfo{booktitle}{\emph{Proceedings of the conference on fairness,
  accountability, and transparency}}. \bibinfo{pages}{220--229}.
\newblock


\bibitem[\protect\citeauthoryear{Mohamed, Png, and Isaac}{Mohamed
  et~al\mbox{.}}{2020}]%
        {mohamed2020decolonial}
\bibfield{author}{\bibinfo{person}{Shakir Mohamed},
  \bibinfo{person}{Marie-Therese Png}, {and} \bibinfo{person}{William Isaac}.}
  \bibinfo{year}{2020}\natexlab{}.
\newblock \showarticletitle{Decolonial AI: Decolonial Theory as Sociotechnical
  Foresight in Artificial Intelligence}.
\newblock \bibinfo{journal}{\emph{Philosophy \& Technology}}
  (\bibinfo{year}{2020}), \bibinfo{pages}{1--26}.
\newblock


\bibitem[\protect\citeauthoryear{Mohan}{Mohan}{2018}]%
        {WhyUrban26:online}
\bibfield{author}{\bibinfo{person}{Angel Mohan}.}
  \bibinfo{year}{2018}\natexlab{}.
\newblock \bibinfo{title}{Why Urban Indian Women Turn Down Job Opportunities
  Away From Home |}.
\newblock
  \bibinfo{howpublished}{\url{https://www.indiaspend.com/why-urban-indian-women-turn-down-job-opportunities-away-from-home-94002/}}.
\newblock
\newblock
\shownote{(Accessed on 09/25/2020).}


\bibitem[\protect\citeauthoryear{Mohanty}{Mohanty}{2005}]%
        {mohanty2005feminism}
\bibfield{author}{\bibinfo{person}{Chandra~Talpade Mohanty}.}
  \bibinfo{year}{2005}\natexlab{}.
\newblock \bibinfo{booktitle}{\emph{Feminism without borders: Decolonizing
  theory, practicing solidarity}}.
\newblock \bibinfo{publisher}{Zubaan}.
\newblock


\bibitem[\protect\citeauthoryear{Mukherji}{Mukherji}{[n.d.]}]%
        {TheCisco63:online}
\bibfield{author}{\bibinfo{person}{Anahita Mukherji}.}
  \bibinfo{year}{[n.d.]}\natexlab{}.
\newblock \bibinfo{title}{The Cisco Case Could Expose Rampant Prejudice Against
  Dalits in Silicon Valley}.
\newblock
  \bibinfo{howpublished}{\url{https://thewire.in/caste/cisco-caste-discrimination-silicon-valley-dalit-prejudice}}.
\newblock
\newblock
\shownote{(Accessed on 08/14/2020).}


\bibitem[\protect\citeauthoryear{Mulgan and Straub}{Mulgan and
  Straub}{[n.d.]}]%
        {Nesta}
\bibfield{author}{\bibinfo{person}{Geoff Mulgan} {and} \bibinfo{person}{Vincent
  Straub}.} \bibinfo{year}{[n.d.]}\natexlab{}.
\newblock \bibinfo{title}{The new ecosystem of trust: how data trusts,
  collaboratives and coops can help govern data for the maximum public benefit
  | Nesta}.
\newblock
  \bibinfo{howpublished}{\url{https://www.nesta.org.uk/blog/new-ecosystem-trust/}}.
\newblock
\newblock
\shownote{(Accessed on 08/21/2020).}


\bibitem[\protect\citeauthoryear{Mulligan, Kroll, Kohli, and Wong}{Mulligan
  et~al\mbox{.}}{2019}]%
        {mulligan2019thing}
\bibfield{author}{\bibinfo{person}{Deirdre~K Mulligan},
  \bibinfo{person}{Joshua~A Kroll}, \bibinfo{person}{Nitin Kohli}, {and}
  \bibinfo{person}{Richmond~Y Wong}.} \bibinfo{year}{2019}\natexlab{}.
\newblock \showarticletitle{This Thing Called Fairness: Disciplinary Confusion
  Realizing a Value in Technology}.
\newblock \bibinfo{journal}{\emph{Proceedings of the ACM on Human-Computer
  Interaction}} \bibinfo{volume}{3}, \bibinfo{number}{CSCW}
  (\bibinfo{year}{2019}), \bibinfo{pages}{1--36}.
\newblock


\bibitem[\protect\citeauthoryear{Murali}{Murali}{2019}]%
        {datalabelling}
\bibfield{author}{\bibinfo{person}{Anand Murali}.}
  \bibinfo{year}{2019}\natexlab{}.
\newblock \bibinfo{title}{How India’s data labellers are powering the global
  AI race | FactorDaily}.
\newblock
  \bibinfo{howpublished}{\url{https://factordaily.com/indian-data-labellers-powering-the-global-ai-race/}}.
\newblock
\newblock
\shownote{(Accessed on 09/13/2020).}


\bibitem[\protect\citeauthoryear{Nathan, Kaczmarek, Castor, Cheng, and
  Mann}{Nathan et~al\mbox{.}}{2017}]%
        {nathan2017good}
\bibfield{author}{\bibinfo{person}{Lisa~P Nathan}, \bibinfo{person}{Michelle
  Kaczmarek}, \bibinfo{person}{Maggie Castor}, \bibinfo{person}{Shannon Cheng},
  {and} \bibinfo{person}{Raquel Mann}.} \bibinfo{year}{2017}\natexlab{}.
\newblock \showarticletitle{Good for Whom? Unsettling Research Practice}. In
  \bibinfo{booktitle}{\emph{Proceedings of the 8th International Conference on
  Communities and Technologies}}. \bibinfo{pages}{290--297}.
\newblock


\bibitem[\protect\citeauthoryear{Newslaundry and India}{Newslaundry and
  India}{2019}]%
        {Newslaundry}
\bibfield{author}{\bibinfo{person}{Newslaundry} {and} \bibinfo{person}{Oxfam
  India}.} \bibinfo{year}{2019}\natexlab{}.
\newblock \showarticletitle{{Who Tells Our Stories Matters: Representation of
  Marginalised Caste Groups in Indian Newsrooms}}.
\newblock  (\bibinfo{date}{8} \bibinfo{year}{2019}).
\newblock


\bibitem[\protect\citeauthoryear{Nissenbaum}{Nissenbaum}{1996}]%
        {nissenbaum1996accountability}
\bibfield{author}{\bibinfo{person}{Helen Nissenbaum}.}
  \bibinfo{year}{1996}\natexlab{}.
\newblock \showarticletitle{Accountability in a computerized society}.
\newblock \bibinfo{journal}{\emph{Science and engineering ethics}}
  \bibinfo{volume}{2}, \bibinfo{number}{1} (\bibinfo{year}{1996}),
  \bibinfo{pages}{25--42}.
\newblock


\bibitem[\protect\citeauthoryear{Ochigame}{Ochigame}{2020}]%
        {ochigame2020}
\bibfield{author}{\bibinfo{person}{Rodrigo Ochigame}.}
  \bibinfo{year}{2020}\natexlab{}.
\newblock \showarticletitle{The Long History of Algorithmic Fairness}.
\newblock \bibinfo{journal}{\emph{Phenomenal World}} (\bibinfo{year}{2020}).
\newblock


\bibitem[\protect\citeauthoryear{Olteanu, Castillo, Diaz, and Kiciman}{Olteanu
  et~al\mbox{.}}{2019}]%
        {olteanu2019social}
\bibfield{author}{\bibinfo{person}{Alexandra Olteanu}, \bibinfo{person}{Carlos
  Castillo}, \bibinfo{person}{Fernando Diaz}, {and} \bibinfo{person}{Emre
  Kiciman}.} \bibinfo{year}{2019}\natexlab{}.
\newblock \showarticletitle{Social data: Biases, methodological pitfalls, and
  ethical boundaries}.
\newblock \bibinfo{journal}{\emph{Frontiers in Big Data}}  \bibinfo{volume}{2}
  (\bibinfo{year}{2019}), \bibinfo{pages}{13}.
\newblock


\bibitem[\protect\citeauthoryear{Pal}{Pal}{2008}]%
        {pal2008computers}
\bibfield{author}{\bibinfo{person}{Joyojeet Pal}.}
  \bibinfo{year}{2008}\natexlab{}.
\newblock \showarticletitle{Computers and the promise of development:
  aspiration, neoliberalism and ``technolity'' in India's ICTD enterprise}.
\newblock \bibinfo{journal}{\emph{A paper presented at confronting the
  Challenge of Technology for Development: Experiences from the BRICS}}
  (\bibinfo{year}{2008}), \bibinfo{pages}{29--30}.
\newblock


\bibitem[\protect\citeauthoryear{Pal}{Pal}{2015}]%
        {pal2015banalities}
\bibfield{author}{\bibinfo{person}{Joyojeet Pal}.}
  \bibinfo{year}{2015}\natexlab{}.
\newblock \showarticletitle{Banalities turned viral: Narendra Modi and the
  political tweet}.
\newblock \bibinfo{journal}{\emph{Television \& New Media}}
  \bibinfo{volume}{16}, \bibinfo{number}{4} (\bibinfo{year}{2015}),
  \bibinfo{pages}{378--387}.
\newblock


\bibitem[\protect\citeauthoryear{Palinkas, Horwitz, Green, Wisdom, Duan, and
  Hoagwood}{Palinkas et~al\mbox{.}}{2015}]%
        {palinkas2015purposeful}
\bibfield{author}{\bibinfo{person}{Lawrence~A Palinkas},
  \bibinfo{person}{Sarah~M Horwitz}, \bibinfo{person}{Carla~A Green},
  \bibinfo{person}{Jennifer~P Wisdom}, \bibinfo{person}{Naihua Duan}, {and}
  \bibinfo{person}{Kimberly Hoagwood}.} \bibinfo{year}{2015}\natexlab{}.
\newblock \showarticletitle{Purposeful sampling for qualitative data collection
  and analysis in mixed method implementation research}.
\newblock \bibinfo{journal}{\emph{Administration and policy in mental health
  and mental health services research}} \bibinfo{volume}{42},
  \bibinfo{number}{5} (\bibinfo{year}{2015}), \bibinfo{pages}{533--544}.
\newblock


\bibitem[\protect\citeauthoryear{Pande}{Pande}{2003}]%
        {pande2003can}
\bibfield{author}{\bibinfo{person}{Rohini Pande}.}
  \bibinfo{year}{2003}\natexlab{}.
\newblock \showarticletitle{Can mandated political representation increase
  policy influence for disadvantaged minorities? Theory and evidence from
  India}.
\newblock \bibinfo{journal}{\emph{American Economic Review}}
  \bibinfo{volume}{93}, \bibinfo{number}{4} (\bibinfo{year}{2003}),
  \bibinfo{pages}{1132--1151}.
\newblock


\bibitem[\protect\citeauthoryear{Pandey}{Pandey}{2020}]%
        {COVID19l97:online}
\bibfield{author}{\bibinfo{person}{Kundan Pandey}.}
  \bibinfo{year}{2020}\natexlab{}.
\newblock \bibinfo{title}{COVID-19 lockdown highlights India’s great digital
  divide}.
\newblock
  \bibinfo{howpublished}{\url{https://www.downtoearth.org.in/news/governance/covid-19-lockdown-highlights-india-s-great-digital-divide-72514}}.
\newblock
\newblock
\shownote{(Accessed on 01/14/2021).}


\bibitem[\protect\citeauthoryear{Patnaik}{Patnaik}{2012}]%
        {socialaudits}
\bibfield{author}{\bibinfo{person}{Priti Patnaik}.}
  \bibinfo{year}{2012}\natexlab{}.
\newblock \bibinfo{title}{Social audits in India – a slow but sure way to
  fight corruption}.
\newblock
  \bibinfo{howpublished}{\url{https://www.theguardian.com/global-development/poverty-matters/2012/jan/13/india-social-audits-fight-corruption}}.
\newblock
\newblock
\shownote{(Accessed on 08/21/2020).}


\bibitem[\protect\citeauthoryear{Paul, Jolley, and Anthony}{Paul
  et~al\mbox{.}}{2018}]%
        {paul2018reflecting}
\bibfield{author}{\bibinfo{person}{Amy Paul}, \bibinfo{person}{C Jolley}, {and}
  \bibinfo{person}{Aubra Anthony}.} \bibinfo{year}{2018}\natexlab{}.
\newblock \showarticletitle{Reflecting the Past, Shaping the Future: Making AI
  Work for International Development}.
\newblock \bibinfo{journal}{\emph{USAID. gov}} (\bibinfo{year}{2018}).
\newblock


\bibitem[\protect\citeauthoryear{Prabhakaran, Hutchinson, and
  Mitchell}{Prabhakaran et~al\mbox{.}}{2019}]%
        {prabhakaran2019perturbation}
\bibfield{author}{\bibinfo{person}{Vinodkumar Prabhakaran},
  \bibinfo{person}{Ben Hutchinson}, {and} \bibinfo{person}{Margaret Mitchell}.}
  \bibinfo{year}{2019}\natexlab{}.
\newblock \showarticletitle{Perturbation Sensitivity Analysis to Detect
  Unintended Model Biases}. In \bibinfo{booktitle}{\emph{Proceedings of the
  2019 Conference on Empirical Methods in Natural Language Processing and the
  9th International Joint Conference on Natural Language Processing
  (EMNLP-IJCNLP)}}. \bibinfo{pages}{5744--5749}.
\newblock


\bibitem[\protect\citeauthoryear{Rajadesingan, Mahalingam, and
  Jurgens}{Rajadesingan et~al\mbox{.}}{2019}]%
        {rajadesingan2019smart}
\bibfield{author}{\bibinfo{person}{Ashwin Rajadesingan},
  \bibinfo{person}{Ramaswami Mahalingam}, {and} \bibinfo{person}{David
  Jurgens}.} \bibinfo{year}{2019}\natexlab{}.
\newblock \showarticletitle{Smart, Responsible, and Upper Caste Only: Measuring
  Caste Attitudes through Large-Scale Analysis of Matrimonial Profiles}. In
  \bibinfo{booktitle}{\emph{Proceedings of the International AAAI Conference on
  Web and Social Media}}, Vol.~\bibinfo{volume}{13}. \bibinfo{pages}{393--404}.
\newblock


\bibitem[\protect\citeauthoryear{Ramanathan}{Ramanathan}{2014}]%
        {ramanathan2014biometrics}
\bibfield{author}{\bibinfo{person}{Usha Ramanathan}.}
  \bibinfo{year}{2014}\natexlab{}.
\newblock \showarticletitle{Biometrics use for social protection programmes in
  India--Risk: Violating human rights of the poor}.
\newblock \bibinfo{journal}{\emph{United Nations Research Institute for Social
  Development}}  \bibinfo{volume}{2} (\bibinfo{year}{2014}).
\newblock


\bibitem[\protect\citeauthoryear{Ramanathan}{Ramanathan}{2015}]%
        {ramanathan2015considering}
\bibfield{author}{\bibinfo{person}{Usha Ramanathan}.}
  \bibinfo{year}{2015}\natexlab{}.
\newblock \showarticletitle{Considering Social Implications of Biometric
  Registration: A Database Intended for Every Citizen in India [Commentary]}.
\newblock \bibinfo{journal}{\emph{IEEE Technology and Society Magazine}}
  \bibinfo{volume}{34}, \bibinfo{number}{1} (\bibinfo{year}{2015}),
  \bibinfo{pages}{10--16}.
\newblock


\bibitem[\protect\citeauthoryear{Rangarajan, Mahendra~Dev, Sundaram, Vyas, and
  Datta}{Rangarajan et~al\mbox{.}}{2014}]%
        {Rangarajan2014Poverty}
\bibfield{author}{\bibinfo{person}{C. Rangarajan}, \bibinfo{person}{S.
  Mahendra~Dev}, \bibinfo{person}{K. Sundaram}, \bibinfo{person}{Mahesh Vyas},
  {and} \bibinfo{person}{K.L Datta}.} \bibinfo{year}{2014}\natexlab{}.
\newblock \bibinfo{booktitle}{\emph{Report of the Expert Group to Review the
  Methodology for Measurement of Poverty}}.
\newblock \bibinfo{type}{{T}echnical {R}eport}.
  \bibinfo{institution}{Government of India Planning Commission}.
\newblock


\bibitem[\protect\citeauthoryear{Ratclifee}{Ratclifee}{2019}]%
        {Howaglit16:online}
\bibfield{author}{\bibinfo{person}{Rebecca Ratclifee}.}
  \bibinfo{year}{2019}\natexlab{}.
\newblock \bibinfo{title}{How a glitch in India's biometric welfare system can
  be lethal | India | The Guardian}.
\newblock
  \bibinfo{howpublished}{\url{https://www.theguardian.com/technology/2019/oct/16/glitch-india-biometric-welfare-system-starvation}}.
\newblock
\newblock
\shownote{(Accessed on 07/29/2020).}


\bibitem[\protect\citeauthoryear{Rege}{Rege}{1998}]%
        {rege1998dalit}
\bibfield{author}{\bibinfo{person}{Sharmila Rege}.}
  \bibinfo{year}{1998}\natexlab{}.
\newblock \showarticletitle{Dalit women talk differently: A critique of
  'difference' and towards a Dalit feminist standpoint position}.
\newblock \bibinfo{journal}{\emph{Economic and Political Weekly}}
  (\bibinfo{year}{1998}), \bibinfo{pages}{WS39--WS46}.
\newblock


\bibitem[\protect\citeauthoryear{Region}{Region}{2009}]%
        {WorldBank2009Disabilities}
\bibfield{author}{\bibinfo{person}{World Bank Human Development Unit South~Asia
  Region}.} \bibinfo{year}{2009}\natexlab{}.
\newblock \bibinfo{title}{People with Disabilities in India: From Commitments
  to Outcomes}.
\newblock
  \bibinfo{howpublished}{\url{http://documents1.worldbank.org/curated/en/577801468259486686/pdf/502090WP0Peopl1Box0342042B01PUBLIC1.pdf}}.
\newblock
\newblock
\shownote{(Accessed on 08/26/2020).}


\bibitem[\protect\citeauthoryear{Richardson}{Richardson}{2012}]%
        {RichardsonFairness}
\bibfield{author}{\bibinfo{person}{S.~Henry Richardson}.}
  \bibinfo{year}{2012}\natexlab{}.
\newblock \bibinfo{title}{Fairness and Political Equality: India and the U.S.}
\newblock
  \bibinfo{howpublished}{\url{https://law.utah.edu/event/fairness-and-political-equality-india-and-the-u-s/}}.
\newblock


\bibitem[\protect\citeauthoryear{Roberts}{Roberts}{2016}]%
        {roberts2016digital}
\bibfield{author}{\bibinfo{person}{Sarah~T Roberts}.}
  \bibinfo{year}{2016}\natexlab{}.
\newblock \showarticletitle{Digital refuse: Canadian garbage, commercial
  content moderation and the global circulation of social media’s waste}.
\newblock \bibinfo{journal}{\emph{Wi: journal of mobile media}}
  (\bibinfo{year}{2016}).
\newblock


\bibitem[\protect\citeauthoryear{Rodrigues}{Rodrigues}{2011}]%
        {rodrigues2011justice}
\bibfield{author}{\bibinfo{person}{Valerian Rodrigues}.}
  \bibinfo{year}{2011}\natexlab{}.
\newblock \showarticletitle{Justice as the Lens: Interrogating Rawls through
  Sen and Ambedkar}.
\newblock \bibinfo{journal}{\emph{Indian Journal of Human Development}}
  \bibinfo{volume}{5}, \bibinfo{number}{1} (\bibinfo{year}{2011}),
  \bibinfo{pages}{153--174}.
\newblock


\bibitem[\protect\citeauthoryear{Roff}{Roff}{2020}]%
        {roff2020expected}
\bibfield{author}{\bibinfo{person}{Heather~M Roff}.}
  \bibinfo{year}{2020}\natexlab{}.
\newblock \showarticletitle{Expected utilitarianism}.
\newblock \bibinfo{journal}{\emph{arXiv preprint arXiv:2008.07321}}
  (\bibinfo{year}{2020}).
\newblock


\bibitem[\protect\citeauthoryear{Rowntree}{Rowntree}{2020}]%
        {women2020mobile}
\bibfield{author}{\bibinfo{person}{Oliver Rowntree}.}
  \bibinfo{year}{2020}\natexlab{}.
\newblock \bibinfo{title}{The mobile gender gap report 2020}.
\newblock
\newblock


\bibitem[\protect\citeauthoryear{Roy}{Roy}{2014}]%
        {roy2014capitalism}
\bibfield{author}{\bibinfo{person}{Arundhati Roy}.}
  \bibinfo{year}{2014}\natexlab{}.
\newblock \bibinfo{booktitle}{\emph{Capitalism: A ghost story}}.
\newblock \bibinfo{publisher}{Haymarket Books}.
\newblock


\bibitem[\protect\citeauthoryear{RT}{RT}{2017}]%
        {Whoeverl94:online}
\bibfield{author}{\bibinfo{person}{RT}.} \bibinfo{year}{2017}\natexlab{}.
\newblock \bibinfo{title}{'Whoever leads in AI will rule the world’: Putin to
  Russian children on Knowledge Day — RT World News}.
\newblock
  \bibinfo{howpublished}{\url{https://www.rt.com/news/401731-ai-rule-world-putin/}}.
\newblock
\newblock
\shownote{(Accessed on 09/20/2020).}


\bibitem[\protect\citeauthoryear{Rudin and Radin}{Rudin and Radin}{2019}]%
        {rudin2019we}
\bibfield{author}{\bibinfo{person}{Cynthia Rudin} {and} \bibinfo{person}{Joanna
  Radin}.} \bibinfo{year}{2019}\natexlab{}.
\newblock \showarticletitle{Why are we using black box models in AI when we
  don’t need to? A lesson from an explainable AI competition}.
\newblock \bibinfo{journal}{\emph{Harvard Data Science Review}}
  \bibinfo{volume}{1}, \bibinfo{number}{2} (\bibinfo{year}{2019}).
\newblock


\bibitem[\protect\citeauthoryear{Ruhaak}{Ruhaak}{[n.d.]}]%
        {MozillaF26:online}
\bibfield{author}{\bibinfo{person}{Anouk Ruhaak}.}
  \bibinfo{year}{[n.d.]}\natexlab{}.
\newblock \bibinfo{title}{Mozilla Foundation - When One Affects Many: The Case
  For Collective Consent}.
\newblock
  \bibinfo{howpublished}{\url{https://foundation.mozilla.org/en/blog/when-one-affects-many-case-collective-consent/}}.
\newblock
\newblock
\shownote{(Accessed on 08/21/2020).}


\bibitem[\protect\citeauthoryear{S}{S}{2020}]%
        {Indiasp84:online}
\bibfield{author}{\bibinfo{person}{Rukmini S}.}
  \bibinfo{year}{2020}\natexlab{}.
\newblock \bibinfo{title}{India’s poor are also document-poor}.
\newblock
  \bibinfo{howpublished}{\url{https://www.livemint.com/news/india/india-s-poor-are-also-document-poor-11578300732736.html}}.
\newblock
\newblock
\shownote{(Accessed on 09/13/2020).}


\bibitem[\protect\citeauthoryear{S}{S}{May}]%
        {InIndiaw91:online}
\bibfield{author}{\bibinfo{person}{Rukmini S}.} \bibinfo{year}{May}\natexlab{}.
\newblock \bibinfo{title}{In India, who speaks in English, and where?}
\newblock
  \bibinfo{howpublished}{\url{https://www.livemint.com/news/india/in-india-who-speaks-in-english-and-where-1557814101428.html}}.
\newblock
\newblock
\shownote{(Accessed on 09/25/2020).}


\bibitem[\protect\citeauthoryear{Sambasivan}{Sambasivan}{2019}]%
        {sambasivan2019remarkable}
\bibfield{author}{\bibinfo{person}{Nithya Sambasivan}.}
  \bibinfo{year}{2019}\natexlab{}.
\newblock \showarticletitle{The remarkable illusions of technology for social
  good}.
\newblock \bibinfo{journal}{\emph{interactions}} \bibinfo{volume}{26},
  \bibinfo{number}{3} (\bibinfo{year}{2019}), \bibinfo{pages}{64--66}.
\newblock


\bibitem[\protect\citeauthoryear{Sambasivan and Aoki}{Sambasivan and
  Aoki}{2017}]%
        {sambasivan2017imagined}
\bibfield{author}{\bibinfo{person}{Nithya Sambasivan} {and}
  \bibinfo{person}{Paul~M Aoki}.} \bibinfo{year}{2017}\natexlab{}.
\newblock \showarticletitle{Imagined Connectivities: Synthesized Conceptions of
  Public Wi-Fi in Urban India}. In \bibinfo{booktitle}{\emph{Proceedings of the
  2017 CHI Conference on Human Factors in Computing Systems}}.
  \bibinfo{pages}{5917--5928}.
\newblock


\bibitem[\protect\citeauthoryear{Sambasivan, Batool, Ahmed, Matthews, Thomas,
  Gayt{\'a}n-Lugo, Nemer, Bursztein, Churchill, and Consolvo}{Sambasivan
  et~al\mbox{.}}{2019}]%
        {sambasivan2019they}
\bibfield{author}{\bibinfo{person}{Nithya Sambasivan}, \bibinfo{person}{Amna
  Batool}, \bibinfo{person}{Nova Ahmed}, \bibinfo{person}{Tara Matthews},
  \bibinfo{person}{Kurt Thomas}, \bibinfo{person}{Laura~Sanely
  Gayt{\'a}n-Lugo}, \bibinfo{person}{David Nemer}, \bibinfo{person}{Elie
  Bursztein}, \bibinfo{person}{Elizabeth Churchill}, {and}
  \bibinfo{person}{Sunny Consolvo}.} \bibinfo{year}{2019}\natexlab{}.
\newblock \showarticletitle{" They Don't Leave Us Alone Anywhere We Go" Gender
  and Digital Abuse in South Asia}. In \bibinfo{booktitle}{\emph{proceedings of
  the 2019 CHI Conference on Human Factors in Computing Systems}}.
  \bibinfo{pages}{1--14}.
\newblock


\bibitem[\protect\citeauthoryear{Sambasivan, Checkley, Batool, Ahmed, Nemer,
  Gayt{\'a}n-Lugo, Matthews, Consolvo, and Churchill}{Sambasivan
  et~al\mbox{.}}{2018}]%
        {sambasivan2018privacy}
\bibfield{author}{\bibinfo{person}{Nithya Sambasivan}, \bibinfo{person}{Garen
  Checkley}, \bibinfo{person}{Amna Batool}, \bibinfo{person}{Nova Ahmed},
  \bibinfo{person}{David Nemer}, \bibinfo{person}{Laura~Sanely
  Gayt{\'a}n-Lugo}, \bibinfo{person}{Tara Matthews}, \bibinfo{person}{Sunny
  Consolvo}, {and} \bibinfo{person}{Elizabeth Churchill}.}
  \bibinfo{year}{2018}\natexlab{}.
\newblock \showarticletitle{" Privacy is not for me, it's for those rich
  women": Performative Privacy Practices on Mobile Phones by Women in South
  Asia}. In \bibinfo{booktitle}{\emph{Fourteenth Symposium on Usable Privacy
  and Security ($\{$SOUPS$\}$ 2018)}}. \bibinfo{pages}{127--142}.
\newblock


\bibitem[\protect\citeauthoryear{Sambasivan, Cutrell, Toyama, and
  Nardi}{Sambasivan et~al\mbox{.}}{2010}]%
        {sambasivan2010intermediated}
\bibfield{author}{\bibinfo{person}{Nithya Sambasivan}, \bibinfo{person}{Ed
  Cutrell}, \bibinfo{person}{Kentaro Toyama}, {and} \bibinfo{person}{Bonnie
  Nardi}.} \bibinfo{year}{2010}\natexlab{}.
\newblock \showarticletitle{Intermediated technology use in developing
  communities}. In \bibinfo{booktitle}{\emph{Proceedings of the SIGCHI
  Conference on Human Factors in Computing Systems}}.
  \bibinfo{pages}{2583--2592}.
\newblock


\bibitem[\protect\citeauthoryear{Sambasivan and Holbrook}{Sambasivan and
  Holbrook}{2018}]%
        {sambasivan2018toward}
\bibfield{author}{\bibinfo{person}{Nithya Sambasivan} {and}
  \bibinfo{person}{Jess Holbrook}.} \bibinfo{year}{2018}\natexlab{}.
\newblock \showarticletitle{Toward responsible AI for the next billion users}.
\newblock \bibinfo{journal}{\emph{interactions}} \bibinfo{volume}{26},
  \bibinfo{number}{1} (\bibinfo{year}{2018}), \bibinfo{pages}{68--71}.
\newblock


\bibitem[\protect\citeauthoryear{Sambasivan, Kapania, Highfill, Akrong,
  Paritosh, and Aroyo}{Sambasivan et~al\mbox{.}}{2021}]%
        {sambasivan2021cascades}
\bibfield{author}{\bibinfo{person}{Nithya Sambasivan}, \bibinfo{person}{Shivani
  Kapania}, \bibinfo{person}{Hannah Highfill}, \bibinfo{person}{Diana Akrong},
  \bibinfo{person}{Praveen Paritosh}, {and} \bibinfo{person}{Lora Aroyo}.}
  \bibinfo{year}{2021}\natexlab{}.
\newblock \showarticletitle{''Everyone wants to do the model work, not the data
  work'': Data Cascades in High-Stakes AI}. In
  \bibinfo{booktitle}{\emph{proceedings of the 2021 CHI Conference on Human
  Factors in Computing Systems}}.
\newblock


\bibitem[\protect\citeauthoryear{Sambasivan and Smyth}{Sambasivan and
  Smyth}{2010}]%
        {sambasivan2010human}
\bibfield{author}{\bibinfo{person}{Nithya Sambasivan} {and}
  \bibinfo{person}{Thomas Smyth}.} \bibinfo{year}{2010}\natexlab{}.
\newblock \showarticletitle{The human infrastructure of ICTD}. In
  \bibinfo{booktitle}{\emph{Proceedings of the 4th ACM/IEEE international
  conference on information and communication technologies and development}}.
  \bibinfo{pages}{1--9}.
\newblock


\bibitem[\protect\citeauthoryear{Sap, Card, Gabriel, Choi, and Smith}{Sap
  et~al\mbox{.}}{2019}]%
        {sap2019risk}
\bibfield{author}{\bibinfo{person}{Maarten Sap}, \bibinfo{person}{Dallas Card},
  \bibinfo{person}{Saadia Gabriel}, \bibinfo{person}{Yejin Choi}, {and}
  \bibinfo{person}{Noah~A Smith}.} \bibinfo{year}{2019}\natexlab{}.
\newblock \showarticletitle{The risk of racial bias in hate speech detection}.
  In \bibinfo{booktitle}{\emph{Proceedings of the 57th Annual Meeting of the
  Association for Computational Linguistics}}. \bibinfo{pages}{1668--1678}.
\newblock


\bibitem[\protect\citeauthoryear{Saracini and Shanmugavelan}{Saracini and
  Shanmugavelan}{2019}]%
        {bondcaste}
\bibfield{author}{\bibinfo{person}{Nadia Saracini} {and}
  \bibinfo{person}{Murali Shanmugavelan}.} \bibinfo{year}{2019}\natexlab{}.
\newblock \showarticletitle{BOND: Caste and Development}.
\newblock  (\bibinfo{year}{2019}).
\newblock


\bibitem[\protect\citeauthoryear{Sch{\"a}fer, Haun, and Tomasello}{Sch{\"a}fer
  et~al\mbox{.}}{2015}]%
        {schafer2015fair}
\bibfield{author}{\bibinfo{person}{Marie Sch{\"a}fer},
  \bibinfo{person}{Daniel~BM Haun}, {and} \bibinfo{person}{Michael Tomasello}.}
  \bibinfo{year}{2015}\natexlab{}.
\newblock \showarticletitle{Fair is not fair everywhere}.
\newblock \bibinfo{journal}{\emph{Psychological science}} \bibinfo{volume}{26},
  \bibinfo{number}{8} (\bibinfo{year}{2015}), \bibinfo{pages}{1252--1260}.
\newblock


\bibitem[\protect\citeauthoryear{Sen}{Sen}{2009}]%
        {sen2009idea}
\bibfield{author}{\bibinfo{person}{Amartya~Kumar Sen}.}
  \bibinfo{year}{2009}\natexlab{}.
\newblock \bibinfo{booktitle}{\emph{The idea of justice}}.
\newblock \bibinfo{publisher}{Harvard University Press}.
\newblock


\bibitem[\protect\citeauthoryear{Seo, Chan, Brantingham, Leap, Vayanos, Tambe,
  and Liu}{Seo et~al\mbox{.}}{2018}]%
        {seo2018partially}
\bibfield{author}{\bibinfo{person}{Sungyong Seo}, \bibinfo{person}{Hau Chan},
  \bibinfo{person}{P~Jeffrey Brantingham}, \bibinfo{person}{Jorja Leap},
  \bibinfo{person}{Phebe Vayanos}, \bibinfo{person}{Milind Tambe}, {and}
  \bibinfo{person}{Yan Liu}.} \bibinfo{year}{2018}\natexlab{}.
\newblock \showarticletitle{Partially generative neural networks for gang crime
  classification with partial information}. In
  \bibinfo{booktitle}{\emph{Proceedings of the 2018 AAAI/ACM Conference on AI,
  Ethics, and Society}}. \bibinfo{pages}{257--263}.
\newblock


\bibitem[\protect\citeauthoryear{shafi}{shafi}{2018}]%
        {Disabili59:online}
\bibfield{author}{\bibinfo{person}{Aabid shafi}.}
  \bibinfo{year}{2018}\natexlab{}.
\newblock \bibinfo{title}{Disability rights: Wheelchair users cannot access
  most of Delhi’s buses}.
\newblock
  \bibinfo{howpublished}{\url{https://scroll.in/roving/894005/in-photos-why-wheelchair-users-in-delhi-find-it-difficult-to-use-buses-even-low-floor-ones}}.
\newblock
\newblock
\shownote{(Accessed on 09/25/2020).}


\bibitem[\protect\citeauthoryear{Shah}{Shah}{[n.d.]}]%
        {MissionC54:online}
\bibfield{author}{\bibinfo{person}{Shreya Shah}.}
  \bibinfo{year}{[n.d.]}\natexlab{}.
\newblock \bibinfo{title}{\#MissionCashless: Few use mobiles, fewer know what
  internet is in adivasi belts of Madhya Pradesh}.
\newblock
  \bibinfo{howpublished}{\url{https://scroll.in/article/824882/missioncashless-few-use-mobiles-fewer-know-what-internet-is-in-adivasi-belts-of-madhya-pradesh}}.
\newblock
\newblock
\shownote{(Accessed on 08/14/2020).}


\bibitem[\protect\citeauthoryear{Shankar, Halpern, Breck, Atwood, Wilson, and
  Sculley}{Shankar et~al\mbox{.}}{2017}]%
        {shankar2017no}
\bibfield{author}{\bibinfo{person}{Shreya Shankar}, \bibinfo{person}{Yoni
  Halpern}, \bibinfo{person}{Eric Breck}, \bibinfo{person}{James Atwood},
  \bibinfo{person}{Jimbo Wilson}, {and} \bibinfo{person}{D Sculley}.}
  \bibinfo{year}{2017}\natexlab{}.
\newblock \showarticletitle{No classification without representation: Assessing
  geodiversity issues in open data sets for the developing world}.
\newblock \bibinfo{journal}{\emph{arXiv preprint arXiv:1711.08536}}
  (\bibinfo{year}{2017}).
\newblock


\bibitem[\protect\citeauthoryear{Shanmugavelan}{Shanmugavelan}{2018}]%
        {shanmugavelan}
\bibfield{author}{\bibinfo{person}{Murali Shanmugavelan}.}
  \bibinfo{year}{2018}\natexlab{}.
\newblock \showarticletitle{Everyday Communicative Practices of Arunthathiyars:
  The Contribution of Communication Studies to the Analysis of Caste Exclusion
  and Subordination of a Dalit Community in Tamil Nadu, India}.
\newblock  (\bibinfo{year}{2018}).
\newblock


\bibitem[\protect\citeauthoryear{Shin}{Shin}{2019}]%
        {shin2019toward}
\bibfield{author}{\bibinfo{person}{Donghee~(Don) Shin}.}
  \bibinfo{year}{2019}\natexlab{}.
\newblock \showarticletitle{Toward Fair, Accountable, and Transparent
  Algorithms: Case Studies on Algorithm Initiatives in Korea and China}.
\newblock \bibinfo{journal}{\emph{Javnost - The Public}} \bibinfo{volume}{26},
  \bibinfo{number}{3} (\bibinfo{year}{2019}), \bibinfo{pages}{274--290}.
\newblock
\urldef\tempurl%
\url{https://doi.org/10.1080/13183222.2019.1589249}
\showDOI{\tempurl}
\showeprint{https://doi.org/10.1080/13183222.2019.1589249}


\bibitem[\protect\citeauthoryear{Singh}{Singh}{2018}]%
        {singh2018}
\bibfield{author}{\bibinfo{person}{Ranjit Singh}.}
  \bibinfo{year}{2018}\natexlab{}.
\newblock \showarticletitle{'The Living Dead'}.
\newblock \bibinfo{journal}{\emph{Whispers from the Field: Ethnographic Poetry
  and Creative Prose}} (\bibinfo{year}{2018}), \bibinfo{pages}{29--31}.
\newblock


\bibitem[\protect\citeauthoryear{Singh and Jackson}{Singh and Jackson}{2017}]%
        {singh2017margins}
\bibfield{author}{\bibinfo{person}{Ranjit Singh} {and}
  \bibinfo{person}{Steven~J Jackson}.} \bibinfo{year}{2017}\natexlab{}.
\newblock \showarticletitle{From Margins to Seams: Imbrication, Inclusion, and
  Torque in the Aadhaar Identification Project}. In
  \bibinfo{booktitle}{\emph{Proceedings of the 2017 CHI Conference on Human
  Factors in Computing Systems}}. ACM, \bibinfo{pages}{4776--4824}.
\newblock


\bibitem[\protect\citeauthoryear{Spade}{Spade}{2015}]%
        {spade2015normal}
\bibfield{author}{\bibinfo{person}{Dean Spade}.}
  \bibinfo{year}{2015}\natexlab{}.
\newblock \bibinfo{booktitle}{\emph{Normal life: Administrative violence,
  critical trans politics, and the limits of law}}.
\newblock \bibinfo{publisher}{Duke University Press}.
\newblock


\bibitem[\protect\citeauthoryear{Subramanian}{Subramanian}{2015}]%
        {subramanian2015making}
\bibfield{author}{\bibinfo{person}{Ajantha Subramanian}.}
  \bibinfo{year}{2015}\natexlab{}.
\newblock \showarticletitle{Making merit: The Indian Institutes of Technology
  and the social life of caste}.
\newblock \bibinfo{journal}{\emph{Comparative Studies in Society and History}}
  \bibinfo{volume}{57}, \bibinfo{number}{2} (\bibinfo{year}{2015}),
  \bibinfo{pages}{291}.
\newblock


\bibitem[\protect\citeauthoryear{Sun, Gaut, Tang, Huang, ElSherief, Zhao,
  Mirza, Belding, Chang, and Wang}{Sun et~al\mbox{.}}{2019}]%
        {sun2019mitigating}
\bibfield{author}{\bibinfo{person}{Tony Sun}, \bibinfo{person}{Andrew Gaut},
  \bibinfo{person}{Shirlyn Tang}, \bibinfo{person}{Yuxin Huang},
  \bibinfo{person}{Mai ElSherief}, \bibinfo{person}{Jieyu Zhao},
  \bibinfo{person}{Diba Mirza}, \bibinfo{person}{Elizabeth Belding},
  \bibinfo{person}{Kai-Wei Chang}, {and} \bibinfo{person}{William~Yang Wang}.}
  \bibinfo{year}{2019}\natexlab{}.
\newblock \showarticletitle{Mitigating gender bias in natural language
  processing: Literature review}.
\newblock \bibinfo{journal}{\emph{arXiv preprint arXiv:1906.08976}}
  (\bibinfo{year}{2019}).
\newblock


\bibitem[\protect\citeauthoryear{Swain and Wallentin}{Swain and
  Wallentin}{2009}]%
        {swain2009does}
\bibfield{author}{\bibinfo{person}{Ranjula~Bali Swain} {and}
  \bibinfo{person}{Fan~Yang Wallentin}.} \bibinfo{year}{2009}\natexlab{}.
\newblock \showarticletitle{Does microfinance empower women? Evidence from
  self-help groups in India}.
\newblock \bibinfo{journal}{\emph{International review of applied economics}}
  \bibinfo{volume}{23}, \bibinfo{number}{5} (\bibinfo{year}{2009}),
  \bibinfo{pages}{541--556}.
\newblock


\bibitem[\protect\citeauthoryear{Tamang}{Tamang}{2020}]%
        {tamang2020section}
\bibfield{author}{\bibinfo{person}{Nisha Tamang}.}
  \bibinfo{year}{2020}\natexlab{}.
\newblock \showarticletitle{Section 377: Challenges and Changing Perspectives
  in the Indian Society}.
\newblock \bibinfo{journal}{\emph{Changing Trends in Human Thoughts and
  Perspectives: Science, Humanities and Culture Part I}}
  (\bibinfo{year}{2020}), \bibinfo{pages}{68}.
\newblock


\bibitem[\protect\citeauthoryear{Thakkar, Sambasivan, Kulkarni,
  Kalenahalli~Sudarshan, and Toyama}{Thakkar et~al\mbox{.}}{2018}]%
        {thakkar2018unexpected}
\bibfield{author}{\bibinfo{person}{Divy Thakkar}, \bibinfo{person}{Nithya
  Sambasivan}, \bibinfo{person}{Purva Kulkarni}, \bibinfo{person}{Pratap
  Kalenahalli~Sudarshan}, {and} \bibinfo{person}{Kentaro Toyama}.}
  \bibinfo{year}{2018}\natexlab{}.
\newblock \showarticletitle{The Unexpected Entry and Exodus of Women in
  Computing and HCI in India}. In \bibinfo{booktitle}{\emph{Proceedings of the
  2018 CHI Conference on Human Factors in Computing Systems}}.
  \bibinfo{pages}{1--12}.
\newblock


\bibitem[\protect\citeauthoryear{Thomas}{Thomas}{2006}]%
        {Thomas2006general}
\bibfield{author}{\bibinfo{person}{David~R Thomas}.}
  \bibinfo{year}{2006}\natexlab{}.
\newblock \showarticletitle{A general inductive approach for analyzing
  qualitative evaluation data}.
\newblock \bibinfo{journal}{\emph{American journal of evaluation}}
  \bibinfo{volume}{27}, \bibinfo{number}{2} (\bibinfo{year}{2006}),
  \bibinfo{pages}{237--246}.
\newblock


\bibitem[\protect\citeauthoryear{Thorat and Attewell}{Thorat and
  Attewell}{2007}]%
        {thorat2007legacy}
\bibfield{author}{\bibinfo{person}{Sukhadeo Thorat} {and} \bibinfo{person}{Paul
  Attewell}.} \bibinfo{year}{2007}\natexlab{}.
\newblock \showarticletitle{The legacy of social exclusion: A correspondence
  study of job discrimination in India}.
\newblock \bibinfo{journal}{\emph{Economic and political weekly}}
  (\bibinfo{year}{2007}), \bibinfo{pages}{4141--4145}.
\newblock


\bibitem[\protect\citeauthoryear{Tiwary}{Tiwary}{2015}]%
        {NCRBdata58:online}
\bibfield{author}{\bibinfo{person}{Deeptiman Tiwary}.}
  \bibinfo{year}{2015}\natexlab{}.
\newblock \bibinfo{title}{Almost 68 percent inmates undertrials, 70 per cent of
  convicts illiterate | The Indian Express}.
\newblock
  \bibinfo{howpublished}{\url{https://indianexpress.com/article/india/india-news-india/almost-68-inmates-undertrials-70-of-convicts-illiterate/}}.
\newblock
\newblock
\shownote{(Accessed on 07/28/2020).}


\bibitem[\protect\citeauthoryear{Toyama}{Toyama}{2015}]%
        {toyama2015geek}
\bibfield{author}{\bibinfo{person}{Kentaro Toyama}.}
  \bibinfo{year}{2015}\natexlab{}.
\newblock \bibinfo{booktitle}{\emph{Geek heresy: Rescuing social change from
  the cult of technology}}.
\newblock \bibinfo{publisher}{PublicAffairs}.
\newblock


\bibitem[\protect\citeauthoryear{Tunisia}{Tunisia}{2018}]%
        {TunisiaAI}
\bibfield{author}{\bibinfo{person}{Tunisia}.} \bibinfo{year}{2018}\natexlab{}.
\newblock \showarticletitle{National AI Strategy: Unlocking Tunisia's
  capabilities potential}.
  \bibinfo{howpublished}{\url{http://www.anpr.tn/national-ai-strategy-unlocking-tunisias-capabilities-potential/}}.
  In \bibinfo{booktitle}{\emph{AI workshop.}}
\newblock


\bibitem[\protect\citeauthoryear{Ullah}{Ullah}{[n.d.]}]%
        {Courttol47:online}
\bibfield{author}{\bibinfo{person}{Mazar Ullah}.}
  \bibinfo{year}{[n.d.]}\natexlab{}.
\newblock \bibinfo{title}{Court told design flaws led to Bhopal leak |
  Environment | The Guardian}.
\newblock
  \bibinfo{howpublished}{\url{https://www.theguardian.com/world/2000/jan/12/1}}.
\newblock
\newblock
\shownote{(Accessed on 08/21/2020).}


\bibitem[\protect\citeauthoryear{Upadhya}{Upadhya}{2007}]%
        {upadhya2007employment}
\bibfield{author}{\bibinfo{person}{Carol Upadhya}.}
  \bibinfo{year}{2007}\natexlab{}.
\newblock \showarticletitle{Employment, exclusion and'merit'in the Indian IT
  industry}.
\newblock \bibinfo{journal}{\emph{Economic and Political weekly}}
  (\bibinfo{year}{2007}), \bibinfo{pages}{1863--1868}.
\newblock


\bibitem[\protect\citeauthoryear{Veeraraghavan}{Veeraraghavan}{2013}]%
        {veeraraghavan2013dealing}
\bibfield{author}{\bibinfo{person}{Rajesh Veeraraghavan}.}
  \bibinfo{year}{2013}\natexlab{}.
\newblock \showarticletitle{Dealing with the digital panopticon: the use and
  subversion of ICT in an Indian bureaucracy}. In
  \bibinfo{booktitle}{\emph{Proceedings of the Sixth International Conference
  on Information and Communication Technologies and Development: Full
  Papers-Volume 1}}. \bibinfo{pages}{248--255}.
\newblock


\bibitem[\protect\citeauthoryear{Vivek, Rajendran, Dipanjan, Rajesh, and
  Vibhore}{Vivek et~al\mbox{.}}{2018}]%
        {vivek2018technology}
\bibfield{author}{\bibinfo{person}{Srinivasan Vivek},
  \bibinfo{person}{Narayanan Rajendran}, \bibinfo{person}{Chakraborty
  Dipanjan}, \bibinfo{person}{Veeraraghavan Rajesh}, {and}
  \bibinfo{person}{Vardhan Vibhore}.} \bibinfo{year}{2018}\natexlab{}.
\newblock \showarticletitle{Are technology-enabled cash transfers really
  'direct'?}
\newblock \bibinfo{journal}{\emph{Economic and Political Weekly}}
  \bibinfo{volume}{53}, \bibinfo{number}{30} (\bibinfo{year}{2018}).
\newblock


\bibitem[\protect\citeauthoryear{Wa~Thiong'o}{Wa~Thiong'o}{1992}]%
        {wa1992decolonising}
\bibfield{author}{\bibinfo{person}{Ngugi Wa~Thiong'o}.}
  \bibinfo{year}{1992}\natexlab{}.
\newblock \bibinfo{booktitle}{\emph{Decolonising the mind: The politics of
  language in African literature}}.
\newblock \bibinfo{publisher}{East African Publishers}.
\newblock


\bibitem[\protect\citeauthoryear{Wallerstein}{Wallerstein}{1991}]%
        {wallerstein1991world}
\bibfield{author}{\bibinfo{person}{Immanuel Wallerstein}.}
  \bibinfo{year}{1991}\natexlab{}.
\newblock \showarticletitle{World system versus world-systems: A critique}.
\newblock \bibinfo{journal}{\emph{Critique of Anthropology}}
  \bibinfo{volume}{11}, \bibinfo{number}{2} (\bibinfo{year}{1991}),
  \bibinfo{pages}{189--194}.
\newblock


\bibitem[\protect\citeauthoryear{Wang and Kosinski}{Wang and Kosinski}{2018}]%
        {wang2018deep}
\bibfield{author}{\bibinfo{person}{Yilun Wang} {and} \bibinfo{person}{Michal
  Kosinski}.} \bibinfo{year}{2018}\natexlab{}.
\newblock \showarticletitle{Deep neural networks are more accurate than humans
  at detecting sexual orientation from facial images.}
\newblock \bibinfo{journal}{\emph{Journal of personality and social
  psychology}} \bibinfo{volume}{114}, \bibinfo{number}{2}
  (\bibinfo{year}{2018}), \bibinfo{pages}{246}.
\newblock


\bibitem[\protect\citeauthoryear{website}{website}{2020}]%
        {JayapalJ44:online}
\bibfield{author}{\bibinfo{person}{Jayapal website}.}
  \bibinfo{year}{2020}\natexlab{}.
\newblock \bibinfo{title}{Jayapal Joins Colleagues In Introducing Bicameral
  Legislation to Ban Government Use of Facial Recognition, Other Biometric
  Technology - Congresswoman Pramila Jayapal}.
\newblock
  \bibinfo{howpublished}{\url{https://jayapal.house.gov/2020/06/25/jayapal-joins-rep-pressley-and-senators-markey-and-merkley-to-introduce-legislation-to-ban-government-use-of-facial-recognition-other-biometric-technology/}}.
\newblock
\newblock
\shownote{(Accessed on 07/30/2020).}


\bibitem[\protect\citeauthoryear{Wieringa}{Wieringa}{2020}]%
        {wieringa2020account}
\bibfield{author}{\bibinfo{person}{Maranke Wieringa}.}
  \bibinfo{year}{2020}\natexlab{}.
\newblock \showarticletitle{What to account for when accounting for algorithms:
  a systematic literature review on algorithmic accountability}. In
  \bibinfo{booktitle}{\emph{Proceedings of the 2020 Conference on Fairness,
  Accountability, and Transparency}}. \bibinfo{pages}{1--18}.
\newblock


\bibitem[\protect\citeauthoryear{Xaxa}{Xaxa}{2011}]%
        {xaxa2011tribes}
\bibfield{author}{\bibinfo{person}{Virginius Xaxa}.}
  \bibinfo{year}{2011}\natexlab{}.
\newblock \showarticletitle{Tribes and social exclusion}.
\newblock \bibinfo{journal}{\emph{CSSSC-UNICEF Social Inclusion Cell, An
  Occasional Paper}}  \bibinfo{volume}{2} (\bibinfo{year}{2011}),
  \bibinfo{pages}{1--18}.
\newblock


\bibitem[\protect\citeauthoryear{Xiang and Raji}{Xiang and Raji}{2019}]%
        {xiang2019legal}
\bibfield{author}{\bibinfo{person}{Alice Xiang} {and}
  \bibinfo{person}{Inioluwa~Deborah Raji}.} \bibinfo{year}{2019}\natexlab{}.
\newblock \showarticletitle{On the Legal Compatibility of Fairness
  Definitions}.
\newblock \bibinfo{journal}{\emph{arXiv preprint arXiv:1912.00761}}
  (\bibinfo{year}{2019}).
\newblock


\bibitem[\protect\citeauthoryear{Zevenbergen}{Zevenbergen}{2020}]%
        {zevenbergen2020}
\bibfield{author}{\bibinfo{person}{Bendert Zevenbergen}.}
  \bibinfo{year}{2020}\natexlab{}.
\newblock \emph{\bibinfo{title}{Internet Users as Vulnerable and at-Risk Human
  Subjects: Reviewing Research Ethics Law for Technical Internet Research}}.
\newblock \bibinfo{thesistype}{Ph.D. Dissertation}. \bibinfo{school}{University
  of Oxford}.
\newblock
\newblock
\shownote{Unpublished PhD thesis.}


\bibitem[\protect\citeauthoryear{Zhang, Lemoine, and Mitchell}{Zhang
  et~al\mbox{.}}{2018}]%
        {zhang2018mitigating}
\bibfield{author}{\bibinfo{person}{Brian~Hu Zhang}, \bibinfo{person}{Blake
  Lemoine}, {and} \bibinfo{person}{Margaret Mitchell}.}
  \bibinfo{year}{2018}\natexlab{}.
\newblock \showarticletitle{Mitigating unwanted biases with adversarial
  learning}. In \bibinfo{booktitle}{\emph{Proceedings of the 2018 AAAI/ACM
  Conference on AI, Ethics, and Society}}. \bibinfo{pages}{335--340}.
\newblock


\bibitem[\protect\citeauthoryear{Zhao, Wang, Yatskar, Ordonez, and Chang}{Zhao
  et~al\mbox{.}}{2017}]%
        {zhao2017men}
\bibfield{author}{\bibinfo{person}{Jieyu Zhao}, \bibinfo{person}{Tianlu Wang},
  \bibinfo{person}{Mark Yatskar}, \bibinfo{person}{Vicente Ordonez}, {and}
  \bibinfo{person}{Kai-Wei Chang}.} \bibinfo{year}{2017}\natexlab{}.
\newblock \showarticletitle{Men also like shopping: Reducing gender bias
  amplification using corpus-level constraints}.
\newblock \bibinfo{journal}{\emph{arXiv preprint arXiv:1707.09457}}
  (\bibinfo{year}{2017}).
\newblock


\bibitem[\protect\citeauthoryear{Zmigrod, Mielke, Wallach, and
  Cotterell}{Zmigrod et~al\mbox{.}}{2019}]%
        {zmigrod2019counterfactual}
\bibfield{author}{\bibinfo{person}{Ran Zmigrod}, \bibinfo{person}{Sebastian~J
  Mielke}, \bibinfo{person}{Hanna Wallach}, {and} \bibinfo{person}{Ryan
  Cotterell}.} \bibinfo{year}{2019}\natexlab{}.
\newblock \showarticletitle{Counterfactual Data Augmentation for Mitigating
  Gender Stereotypes in Languages with Rich Morphology}. In
  \bibinfo{booktitle}{\emph{Proceedings of the 57th Annual Meeting of the
  Association for Computational Linguistics}}. \bibinfo{pages}{1651--1661}.
\newblock


\end{thebibliography}
